\documentclass[fleqn,usenatbib]{mnras}

\usepackage{amsmath, amssymb}
\usepackage{newtxtext,newtxmath}
\usepackage[T1]{fontenc}
\DeclareRobustCommand{\VAN}[3]{#2}
\let\VANthebibliography\thebibliography
\def\thebibliography{\DeclareRobustCommand{\VAN}[3]{##3}\VANthebibliography}
\usepackage{graphicx}	
\usepackage{amsmath}	
\usepackage{placeins}
\usepackage{float}
\usepackage[labelfont=bf]{caption}
\usepackage{kotex}
\usepackage[x11names,dvipsnames]{xcolor}
\usepackage[normalem]{ulem}
\usepackage{booktabs}
\usepackage{xcolor}
\usepackage{booktabs}
\usepackage{hyperref}
\usepackage{comment}
\usepackage{array}
\usepackage{dblfloatfix}
\definecolor{darkgreen}{rgb}{0,0.5,0}
\usepackage{scalerel} 
\usepackage{tikz} 

\def\Msun{{\rm M}_{\odot}}

\def \logM {\log M_*\text{[}{\rm M}_{\odot}\text{]}}
\def \logsSFR {\log{\rm sSFR}\text{[}\rm yr^{-1}\text{]}}
\def \logSFR {\log{\rm SFR}\text{[}{\rm M}_{\odot}{\rm yr}^{-1}\text{]}}
\newcommand{\orcidicon}{\includegraphics[width=9pt]{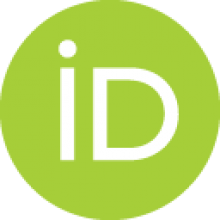}}
\newcommand{\orcid}[1]{\href{https://orcid.org/#1}{\orcidicon}}
\newcommand{\y}[1]{{\textcolor{red}{#1}}}

\newcommand{\yo}[1]{{\textcolor{orange}{#1}}}

\newcommand{\yc}[1]{{\textcolor{cyan}{#1}}}
\newcommand{\yp}[1]{{\textcolor{purple}{#1}}}

\newcounter{subsubsubsection}[subsubsection]

\title[SN~Ia Population Machine. I.]
{SN~Ia Population Machine. I. A Unified Cosmological Simulation--Binary Synthesis Framework Establishing Non-universal Delay-time Distributions and Cosmic Progenitor-channel Dominance Crossover\\
}

\author[S.-J. Yoon, I. Park et al.]
{Suk-Jin Yoon$^{1,2,3,4}$\thanks{E-mail: sjyoon0691@yonsei.ac.kr}\orcid{0000-0002-1842-4325}, 
Inhyuk Park$^{1,2,3}$\orcid{0009-0007-7069-4414},
Woong-Bae G. Zee$^{5}$\orcid{0000-0003-0960-687X},
Chul Chung$^{2,3}$\orcid{0000-0001-6812-4542},
Jun-Sung Moon$^{6}$\orcid{0000-0001-7075-4156},
\newauthor 
Sanjaya Paudel$^{2,3}$\orcid{0000-0003-2922-6866}, 
Kiyun Yun$^{2,3}$\orcid{0000-0003-1636-2455},
Myung-Hun Kim$^{2,3}$\orcid{0009-0008-4928-0599},
and Eun-Taek Gim$^{2,3}$\orcid{0009-0004-2083-3111}
\\
\\
$^{1}$Equal first author \\
$^{2}$Department of Astronomy, Yonsei University, Seoul, 03722, Republic of Korea \\
$^{3}$Center for Galaxy Evolution Research, Yonsei University, Seoul, 03722, Republic of Korea \\
$^{4}$Institute of Natural Sciences, Yonsei University, Seoul, 03722, Republic of Korea \\
$^{5}$School of Liberal Studies, Sejong University, Seoul, 05006, Republic of Korea \\
$^{6}$Institute of Astronomy and Astrophysics, Academia Sinica, No. 1, Sec. 4, Roosevelt Rd., Taipei 106319, Taiwan
}

\begin{document}
\label{firstpage}
\pagerange{\pageref{firstpage}--\pageref{lastpage}}
\maketitle

\begin{abstract}
We present a forward-modeling framework for synthesizing Type~Ia supernova (SN~Ia) populations by coupling cosmological hydrodynamic simulations to binary population synthesis (BPS).
Using {\ttfamily IllustrisTNG} star particles as simple stellar populations, we generate binaries and evolve them with {\ttfamily COMPAS} to produce synthetic SNe~Ia tagged with explosion times and progenitor channels (single- and double-degenerate; SD and DD).
This cosmology--BPS pipeline enables self-consistent, end-to-end tracking of SN~Ia populations from individual galaxies to cosmic scales.
The model reproduces key SN-related observables, including host-galaxy demographics, delay-time distributions (DTDs), SN-rate trends with host properties and redshift, and a progenitor-age `step' implicated by the mass step in Hubble residuals.
Our main findings are as follows.
(1) Contrary to the standard assumption, DTDs appear intrinsically \emph{non-universal}: their form depends on progenitor channel and metallicity, and thus varies systematically across hosts and with redshift.
The commonly adopted DTD is therefore best regarded as a population-averaged approximation rather than a fundamental kernel.
(2) We predict that the dominant SN~Ia progenitor population shifts from SD to DD with cosmic time, with a demographic \emph{crossover} near $z$\,$=$\,0.5 ($\sim$\,5.2~Gyr ago).
This non-monolithic SN~Ia population with a redshift-dependent SD/DD mixture weakens the universality implicit in a single globally calibrated standardization.
Taken together, evolution in both the DTD and the channel mixture can imprint redshift-dependent systematics on SN~Ia luminosities, strengthening the case for jointly inferring progenitor/host-driven effects alongside cosmic acceleration.
The full catalogue and analysis scripts are available via Zenodo.
\end{abstract}


\section{Introduction}
\label{sec:1}

\subsection{SNe~Ia: Cosmological Importance and Progenitor Scenarios}
\label{sec:1.1}

Type~Ia supernovae (SNe~Ia) play a pivotal role in modern astrophysics, particularly in cosmology. 
Their exceptional luminosity and empirically standardizable brightness have made SNe~Ia powerful distance indicators \citep{Phillips1993,Tripp1998}, leading to the discovery of the accelerating expansion of the Universe \citep{Riess_1998, Perlmutter_1999}. 
Despite their importance, the nature of their progenitor systems and explosion mechanisms remains unresolved \citep[see reviews by][]{MaozMannucci2012,WangHang2012,LivioMazzali2018,Liu_2023,RuiterSeitenzahl2025}.
To set the stage for this long-standing progenitor problem, it is useful to briefly recall the key observational and physical properties that define SNe~Ia.
SNe~Ia are characterized by strong Si~II absorption lines and a conspicuous absence of hydrogen and helium features \citep{Filippenko_1997,Parrent_2014,Jha2019,Liu_2023}. 
Their luminosity is thought to arise primarily from the radioactive decay chain ${}^{56}\mathrm{Ni}\rightarrow{}^{56}\mathrm{Co}\rightarrow{}^{56}\mathrm{Fe}$, which produces a $B$-band light curve peaking at ${\rm M}_B\simeq-19.3$~mag and fading over $\sim$100~days \citep{Truran1967, Colgate1969, Arnett1982, Hillebrandt2013}. 
These features imply that SNe~Ia originate from binary systems containing a carbon-oxygen white dwarf (CO-WD) \citep[e.g.,][]{Hoyle1960,HillebrandtNiemeyer2020,Liu_2023,Blondin2024} that accretes material until approaching the Chandrasekhar mass ($M_{\rm Ch}$). 

Observationally, studies based on spectral features, such as the Si~II line velocity, suggest distinct SN~Ia populations associated with different environments and, potentially, with progenitor age and/or metallicity.
For example, high-velocity SNe~Ia preferentially occur in the inner, brighter regions of massive galaxies \citep{Wang2013_HVNV}, suggestive of younger and metal-richer progenitors.
Moreover, high-velocity events show evidence for substantial circumstellar material, inferred from late-time blue light echoes and time-variable Na~I absorption \citep{Wang2019_CSM}, linking at least a subset to SD-like pathways.
In contrast, normal-velocity SNe~Ia tend to exhibit weaker circumstellar material signatures and are often interpreted as more compatible with DD-like origins.
Although the connection between ejecta velocity and progenitor age remains debated \citep{Pan2020_HVSNEnv}, the association of circumstellar material with high-velocity SNe~Ia supports distinct evolutionary routes tied to different progenitor channels. 
Taken together, the diversity of progenitor channels indicates that no single pathway dominates universally. 
If the relative contributions of the different progenitor channels, or the intrinsic properties of their explosions, evolve over cosmic time, such evolution could potentially affect both cosmological measurements based on SNe~Ia as standard candles \citep[e.g.,][]{Rigault_2013, Kim2018, Kang_2020, BroutScolnic2021} and models of galactic chemical enrichment \citep[e.g.,][]{Matteucci2009}. 

Theoretically, SN~Ia progenitors are usually framed in terms of two principal binary channels: the single-degenerate (SD) and double-degenerate (DD) pathways.
In the classical SD channel, a CO-WD accretes hydrogen- or helium-rich material from a non-degenerate companion through binary mass transfer and grows toward $M_{\rm Ch}$ \citep{WhelanIben1973,Nomoto1982a,Nomoto1982b,Nomoto1984}.
Possible donors include a main-sequence star (MS), red- and asymptotic-giant-branch star (GB), and a helium star (He) \citep{Hachisu1999,Han2004,Wang_2009a,Meng2009}.
In the DD channel, two CO-WDs lose angular momentum through gravitational-wave (GW) radiation, spiral inward, and eventually merge \citep{Webbink1984,IbenTutukov1984}.
A further extension of this canonical SD/DD framework is provided by sub-$M_{\rm Ch}$ double-detonation models \citep{Nomoto1982b, wtw86, livn90}, in which helium-shell ignition triggers the detonation of a CO-WD below $M_{\rm Ch}$ \citep{Shen2018sub,Shen2018}.
When the helium donor is non-degenerate, these systems can be viewed phenomenologically as SD-like, whereas helium-WD donors make them DD-like; in some cases, a suitably structured CO-WD may even detonate without additional accretion, triggered solely by its residual helium shell \citep{shen2024}.

Recent studies increasingly suggest that sub-$M_{\rm Ch}$ explosions may account for a large fraction of the SN~Ia population \citep[e.g.,][]{Flors2020,Bravo2022,Kobayashi2020,RuiterSeitenzahl2025}. 
However, current evidence does not support a purely sub-$M_{\rm Ch}$ origin, as several key observables---including stable Ni production, Mn/Fe and Ni/Fe abundance constraints, and some well-observed normal SNe~Ia---still point to a significant contribution from near-$M_{\rm Ch}$ explosions \citep[e.g.,][]{Seitenzahl2013,Yamaguchi2015,Kwok2023,DerKacy2023,Kobayashi2020,Cavichia2024}. 
A balanced interpretation at present is therefore that normal SNe~Ia arise from a mixed population in which sub-$M_{\rm Ch}$ events are likely common, but near-$M_{\rm Ch}$ events remain an essential component \citep[e.g.,][]{Liu_2023,RuiterSeitenzahl2025,Cavichia2024}. 
Overall, the canonical near-$M_{\rm Ch}$ and sub-$M_{\rm Ch}$ scenarios are generally thought to explain most normal SNe~Ia \citep{Mazzali2007, Sim2010, Scalzo2014a, Yamaguchi2015, Das2025}.

\subsection{Delay-time Distributions and SN~Ia/Host Demographics }
\label{sec:1.2}

The delay time of an SN~Ia is the astrophysically fundamental clock: the elapsed time from the star-formation (SF) episode that creates the SN-producing progenitor system to the eventual explosion.
The corresponding delay-time \emph{distribution} (DTD) is a primary population-level diagnostic of SN~Ia progenitor channels and explosion mechanisms \citep[e.g.,][]{WhelanIben1973,IbenTutukov1984,Webbink1984,Scannapieco2005,Sullivan2006,Mannucci2006,Brandt2010,Maoz2010,MaozMannucci2012,Perrett2012,WangHang2012,Graur2013,Graur2014,Maoz2014,Rodney2014,Maoz2017,Friedmann2018,Strolger2020,Wiseman2021,RuiterSeitenzahl2025}.
The ${\rm DTD}(\tau)$ is defined such that, following a single instantaneous star-formation event that forms $1~\Msun$ of stars, \({\rm DTD}(\tau)\) gives the expected number of SNe~Ia that explode at delay $\tau$ \citep{Matteucci_1986, Greggio05, Maoz2014}.
Because its normalization, slope, and short-/long-delay structure are highly sensitive to binary evolution and ignition physics, the observed DTD provides a sensitive empirical discriminant among competing progenitor pathways \citep{Totani2008,Hachisu2008,Mennekens2010,Ruiter2011,Maoz2012,Maoz2014,Claeys_2014}.
Many observational reconstructions find an approximately power-law DTD close to $\propto \tau^{-1}$ over broad delay ranges \citep[e.g.,][]{Totani2008,Maoz2010,Maoz2012,Graur2013,Wiseman2021}, although alternative forms with relatively delayed events have also been reported depending on data/method choices \citep{Strolger2010,Strolger2020}.

The observed SN~Ia birth rate for a given galaxy population can be expressed as the convolution of its star formation history (SFH) with the DTD \citep{Greggio05,Matteucci_1986},
\begin{equation}
\label{eq:SN Ia rate}
R(t) \;=\; \int_0^t \mathrm{SFR}(t-\tau)\,\mathrm{DTD}(\tau)\,d\tau,
\end{equation}
where \(R(t)\) denotes the SN~Ia rate averaged over a galaxy sample or over the Universe at time \(t\), and \(\mathrm{SFR}(t)\) is the corresponding star-formation rate (SFR) at that epoch \citep{Greggio05}.
The SN~Ia rates of individual galaxies collectively build up the cosmic SN~Ia rate, making it a key probe of both binary evolution and cosmology.
This cosmic rate has been modeled by convolving the observed cosmic SFR with an assumed DTD \citep[e.g.,][]{Mannucci2006, Childress2014, Wiseman2021}.
Such an approach enables progenitor-model tests by asking which DTD, when combined with the cosmic SFR, best reproduces the observed SN~Ia rate as a function of redshift \citep{Mannucci2006, Childress2014, Palicio24}.
Comparison with observations therefore provides a cosmic-scale test of our understanding of SN~Ia progenitors.

Given their strong dependence on the underlying stellar population, the properties of SN-host galaxies offer valuable clues to SN~Ia progenitors.
Large observational surveys have established that galaxy properties are tightly correlated with SN~Ia production. 
For instance, the SN~Ia rate tends to be higher in more massive galaxies \citep{Li2011, Brown2019, Wiseman2021} and in galaxies with higher SFRs \citep{Mannucci2005, Sullivan2006}, although the relationships differ between star-forming and quiescent galaxies \citep{Smith2012, Graur2015}. 
SFH-based DTD recovery can couple systematically to the mass-weighted ages of the underlying galaxy population in \texttt{IllustrisTNG} samples \citep{Joshi_2024}.
Together, these trends indicate that galaxy properties (e.g., stellar mass, age, and SFR) trace the SN-progenitor characteristics, particularly the mean progenitor age \citep{Kang_2020,Briday2022,Lee2022}.
Understanding this connection may clarify empirical trends such as the “host-mass step” \citep[e.g.,][]{Kelly2010, Sullivan2010, Gupta_2011, Childress2013}, and is important for improving the precision of SN~Ia cosmology.

\subsection{Toward Improved Forward Modeling: A Cosmological Simulation--Binary Synthesis SN~Ia Population Machine}
\label{sec:1.3}

Recently, forward-modeling approaches have been employed to connect the analytic models of galaxy evolution to the SN~Ia population. 
For instance, \citet{Childress2014} constructed observationally motivated models of galaxy mass assembly histories and convolved them with DTDs, thereby linking host properties to SN~Ia progenitor ages.
\citet{Wiseman2021} and \citet{Wiseman2022} developed a galaxy-based framework relating progenitor ages to light-curve parameters and dust extinction, arguing that complex features of the host mass step, such as its dependence on SN color, may arise from systematic dust-extinction variations correlated with host-galaxy age.
These studies demonstrate the interpretive power of the forward-modeling approach.
However, most of these frameworks rely on semi-empirical or analytical representations of galaxies, or focus on specific observational effects rather than synthesizing the entire SN~Ia population.

Despite extensive observational and theoretical efforts, the relative contributions of the SD and DD channels to the SN~Ia budget---and their evolution over cosmic time---remain highly uncertain \citep{WhelanIben1973,Webbink1984,IbenTutukov1984,MaozMannucci2012,Maoz2014,WangHang2012,LivioMazzali2018,RuiterSeitenzahl2025}.
BPS studies can predict SD and DD rates under different binary-physics assumptions, but they are typically not embedded in a cosmological framework required to track progenitor demographics across redshift \citep{Han2004,Belczynski2005,Ruiter2011,Toonen2012,Bours2013,Claeys_2014}.
Conversely, cosmological hydrodynamic simulations often implement empirical DTDs or analytic prescriptions \citep{Scannapieco2005,Greggio05,Tornatore2007,Wiersma2009,Vogelsberger2013,Schaye2015,Pillepich2018a,Gandhi2022}; without explicit channel-resolved modeling, they are not designed to quantify SD--DD competition directly.
Observational DTD reconstructions provide powerful constraints on the total SN~Ia rate, but they do not by themselves disentangle the underlying progenitor channels \citep{Totani2008,Brandt2010,Maoz2012,Strolger2020,Heringer19,Wiseman2021}.
Taken together, these considerations suggest that a self-consistent, simulation-based quantification of SD and DD evolution with redshift---and a cosmological demonstration of a transition between the two channels---has not yet been established.

In this work, we introdcue a new forward-modeling framework that synthesizes SN~Ia populations directly from a cosmological hydrodynamic simulation by applying BPS to each star particle, treated as a simple stellar population (SSP) characterized by its mass, age, and metallicity.
This approach tracks SD and DD events self-consistently from individual galaxies to cosmic scales.
We thereby show that, contrary to the standard assumption, DTDs are intrinsically \emph{non-universal}: their form depends on progenitor channel and metallicity, and therefore varies systematically among hosts and across redshift.
We further provide the first cosmology-based quantitative determination of the SD and DD contributions over cosmic time, revealing a strong redshift-dependent crossover in which SD dominates at high redshift whereas DD dominates at low redshift.
Taken together, redshift evolution in both the DTD and the channel mixture can induce redshift-dependent systematics on SN~Ia luminosity---an effect long recognized as potentially degenerate with cosmological dimming in Hubble-diagram inference \citep[e.g.,][]{Drell2000,Dominguez2001,Howell2007,Sullivan2010}.

This paper inaugurates our ``SN~Ia Population Machine'' series.
Here, we present a new forward-modeling framework that unifies cosmological hydrodynamic galaxy simulations with binary population synthesis, enabling synthetic SN~Ia populations self-consistently from galactic to cosmic scales.
Section~\ref{sec:2} details the methodology and implementation of the framework.
Sections~\ref{sec:3}--\ref{sec:9} provide a concise preview of the scientific results enabled by the methodology, showcasing representative outputs of the \texttt{SN~Ia Population Machine}.
These sections are deliberately exploratory rather than exhaustive, highlighting key findings that delineate the scope and utility of the framework.
In subsequent papers, we will reorganize and expand this initial results set into several focused science themes, with dedicated analyses and richer theoretical and observational context.
Table~\ref{tab:architecture} summarizes the paper architecture: Part~I (\S\,\ref{sec:2}) introduces the methodology; Part~II (\S\,\ref{sec:3}--\ref{sec:5}) presents SN~Ia host demographics; Part~III (\S\,\ref{sec:6}--\ref{sec:7}) covers SN~Ia progenitor demographics; and Part~IV (\S\,\ref{sec:8}--\ref{sec:9}) describes the cosmic evolution of SN~Ia progenitor demographics.

\begin{table*}
\centering
\caption{Section architecture of the paper.}
\label{tab:architecture}
\footnotesize 
\renewcommand{\arraystretch}{1.08}
\setlength{\tabcolsep}{4pt}
\begin{tabular}{@{}p{0.16\textwidth}p{0.695\textwidth}p{0.110\textwidth}@{}}
\toprule
\toprule
\textbf{Architecture}   & \textbf{Section/Subsection Titles} & \textbf{Comparison to}\\
                        &                   & \textbf{Observations}\\
\midrule
\midrule
\textbf{Introduction} &
\textbf{\ref{sec:1}\ \ Introduction} \\
& \hspace*{0.7em} \ref{sec:1.1} SNe~Ia: Cosmological Importance and Progenitor Scenarios \\
& \hspace*{0.7em} \ref{sec:1.2} Delay-time Distributions and SN~Ia/Host-galaxy Demographics \\
& \hspace*{0.7em} \ref{sec:1.3} Toward Improved Forward Modeling:  \\
& \hspace*{2.2em} A Cosmological Simulation--Binary Synthesis SN~Ia Population Machine \\
\midrule
\textbf{Part I.} &
\textbf{\ref{sec:2}\ \ Methods and Released Data Products} \\
\textbf{Methodology} & \hspace*{0.7em} \ref{sec:2.1}\ \ Cosmological Hydrodynamic Simulations of Galaxies \\
    & \hspace*{0.7em} \ref{sec:2.2}\ \ From Star Particles to Individual Single \& Binary Stars \\
    & \hspace*{0.7em} \ref{sec:2.3}\ \ Binary Population Synthesis: From Binaries to SNe Ia via Single- and Double-degenerate Channels \\
    & \hspace*{0.7em} \ref{sec:2.4}\ \ SNe~Ia in Individual Star Particles: Intrinsic Delay-time Distributions of Simple Stellar Populations \\
    & \hspace*{0.7em} \ref{sec:2.5}\ \ From Star-particle-level SNe~Ia to Subhalo-scale SN~Ia Populations \\
    & \hspace*{0.7em} \ref{sec:2.6}\ \ All-sky Comoving-volume Realization to Build Cosmological SN~Ia Samples \\
    & \hspace*{0.7em} \ref{sec:2.7}\ \ Applying an Observational Time Window to Mock Surveys  & \checkmark~Scalar-level \\
    & \hspace*{0.7em} \ref{sec:2.8}\ \ Simulation--Survey Consistency in the SN~Ia Rate Definition \\
    & \hspace*{0.7em} \ref{sec:2.9}\ \ Observation-driven Redshift Baselines for Global-volume and Local-volume Samples \\
    & \hspace*{0.7em} \ref{sec:2.10}\ \ Final Data Products: Catalogue and Codes \\
\midrule
\textbf{Part II.} &
\textbf{\ref{sec:3}\ \ Anatomy of SN~Ia Populations in Individual Galaxies} \\
\textbf{SN~Ia Hosts} & \hspace*{0.7em} \ref{sec:3.1}\ \ SNe Ia in Representative Galaxies of Different Morphology and Redshift \\
    & \hspace*{0.7em} \ref{sec:3.2}\ \ SNe Ia in Milky Way-like Galaxies at $z=0$  & \checkmark~Scalar-level\\
    & \textbf{\ref{sec:4}\ \ Comparative Demographics of All Galaxies and SN~Ia Host Galaxies} \\
    & \hspace*{0.7em} \ref{sec:4.1}\ \ Global-sample Demographics: All Galaxies versus SN~Ia Hosts  & \checkmark~2D trend \\
    & \hspace*{0.7em} \ref{sec:4.2}\ \ Local and $z\simeq0.5$ Sample Demographics: All Galaxies versus SN~Ia Hosts  & \checkmark~2D trend \\
    & \textbf{\ref{sec:5}\ \ SN~Ia Rate in Galactic Context: Host Dependence}  \\
    & \hspace*{0.7em} \ref{sec:5.1}\ \ Mass-normalized SN~Ia Rate and its Host Dependence  & \checkmark~2D trend \\
    & \hspace*{0.7em} \ref{sec:5.2}\ \ Galaxy-normalized SN~Ia Rate and its Host Dependence  & \checkmark~2D trend \\
\midrule
\textbf{Part III.} &
\textbf{\ref{sec:6}\ \ Delay-Time Distributions of SNe~Ia} \\
\textbf{SN~Ia Progenitors}  & \hspace*{0.7em} \ref{sec:6.1}\ \ Recovered DTDs and Channel-ordered Contributions  & \checkmark~1D distribution \\
    & \hspace*{0.7em} \ref{sec:6.2}\ \ Non-universal DTDs: Progenitor-channel and Metallicity Dependence \\
    & \hspace*{0.7em} \ref{sec:6.3}\ \ Non-universal DTDs: Redshift Evolution \\
& \textbf{\ref{sec:7}\ \ Progenitor Ages of SNe~Ia} \\
    & \hspace*{0.7em} \ref{sec:7.1}\ \ Host-level Mean Progenitor Ages: Host-property and Redshift Dependence\\
    & \hspace*{0.7em} \ref{sec:7.2}\ \ Event-level Individual Progenitor Ages: Progenitor-age Step as an Origin of the Mass/sSFR Steps in HR \\
    & \hspace*{0.7em} \ref{sec:7.3}\ \ Cosmological Implications of Host-dependent SN~Ia Standardization \\
\midrule
\textbf{Part IV.} &
\textbf{\ref{sec:8}\ \ SN~Ia Rates in Cosmic Contexts: Redshift Evolution}   & \checkmark~1D distribution \\
\textbf{SN~Ia Rates} & \textbf{\ref{sec:9}\ \ Cosmic Evolution of SN~Ia Progenitor Channel Mixture}  \\
    & \hspace*{0.7em} \ref{sec:9.1}\ \ SN~Ia Progenitor-channel Dominance Crossover with Cosmic Time \\
    & \hspace*{0.7em} \ref{sec:9.2}\ \ Threefold Drivers of the SD--DD Demographic Mixture Evolution with Cosmic Time \\
    & \hspace*{0.7em} \ref{sec:9.3}\ \ Cosmological Implications of SN~Ia Demographic Transition \\
\midrule
\textbf{Conclusion} &
\textbf{\ref{sec:10}\ \ Conclusion} \\
    & \hspace*{0.7em} \ref{sec:10.1}\ \ Summary \\
    & \hspace*{0.7em} \ref{sec:10.2}\ \ Future Work \\
\bottomrule
\end{tabular}
\end{table*}

\section{Methods and Released Data Products}
\label{sec:2}

\begin{figure*}
\includegraphics[width=0.90\textwidth]{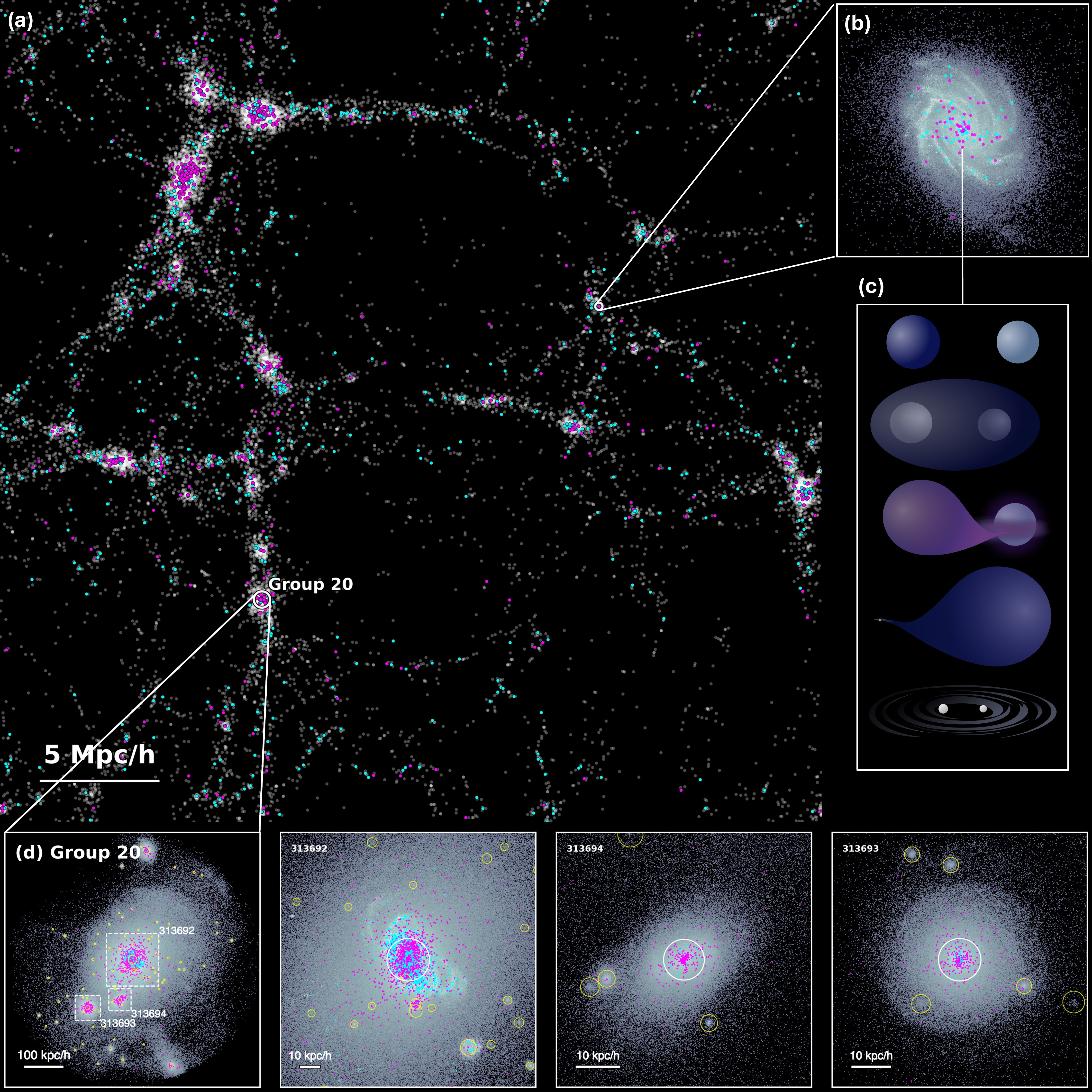}
\caption{
A conceptual overview of our SN~Ia population-synthesis framework, which unified cosmological hydrodynamic galaxy simulations with binary population synthesis to generate SN~Ia populations from individual galaxies to galaxy clusters/groups to cosmic volumes.
\textbf{\textit{(a)}} A slab of the TNG50-1 snapshot at $z=0$.
Subhalos with at least one star particle are shown as white dots with an opacity of 30\,\% and a constant size, without any luminosity weighting.
Cyan and magenta symbols denote SD- and DD-dominated host galaxies, respectively, with SNe~Ia occurring within a mock survey time window of $10^5$~yr (see \S\,\ref{sec:2.7}).
Motivated by the $N_{\rm SD}:N_{\rm DD}\simeq3:7$ ratio at $z=0$ (see \S\,\ref{sec:9}), each subhalo is classified by its internal SN~Ia mixture: subhalos with an SD fraction $>30\,\%$ (i.e., a DD fraction $<70\,\%$) are shown in cyan, whereas those with an SD fraction $<30\,\%$ (i.e., a DD fraction $>70\,\%$) are shown in magenta.
\textbf{\textit{(b)}} Zoom-in view of a galaxy selected from panel \textit{(a)}.
The projected stellar distribution is shown with intensities weighted by the $U$-band flux of star particles.
Star-particle-level stellar-population properties (mass, formation time, metallicity, and phase-space coordinates) are extracted from TNG50-1 snapshots to construct mock galaxies.
Each star particle is treated as a simple stellar population (SSP)---a coeval stellar ensemble with uniform chemical composition---and forms the fundamental unresolved unit of the analysis.
Over sufficiently long observational time windows, a single galaxy can host multiple SNe~Ia (cyan: SD; magenta: DD).
\textbf{\textit{(c)}} Schematic binary evolutionary pathway of a SN~Ia progenitor, generated with the evolution plotter from the \texttt{COMPAS} binary population synthesis suite (see also Fig.\,\ref{fig:Fig4} for examples of SD and DD).
Time increases from top to bottom, and circles denote the primary and secondary stars in each binary.
Star-particle properties, together with the associated binary parameters, are passed to the {\ttfamily COMPAS} engine, which returns SN~Ia events with predicted explosion times and channel classifications.
Applying this procedure to all star particles yields the full SN~Ia population within their host galaxies.
\textbf{\textit{(d)}} Multi-scale zoom-in view of the galaxy cluster labeled `Group~20' (TNG50 halo catalogue index) in panel \textit{(a)}.
The leftmost panel presents the cluster; yellow circles mark subhalos containing at least 30 star particles, and dashed boxes identify the three most massive subhalos selected for closer examination.
The three panels on the right show zoom-in views of these subhalos, arranged from left to right in decreasing stellar mass, using the same visualization scheme as in panel \textit{(b)}.
For completeness, we provide a larger set of analogous views to Appendix\,\ref{appendix:C}, where 10 groups/clusters from panel \textit{(a)} are displayed in the same style as panel \textit{(d)}.
}    
\label{fig:Fig_intro}
\end{figure*}

In this work, we develop an end-to-end framework that integrates cosmological hydrodynamic galaxy simulations with BPS modeling to synthesize SN~Ia populations from galactic to cosmic scales.
Figure\,\ref{fig:Fig_intro} summarizes the workflow.
From \texttt{IllustrisTNG}\footnote{\url{https://www.tng-project.org/}} (panel $a$; \S\S\,\ref{sec:2.1}; \citealt{Pillepich2018b,Springel2018,Nelson2018,Naiman2018,Marinacci2018,Nelson2019}), we extract star-particle-level stellar-population properties---stellar mass, formation time (age), metallicity, and spatial phase-space information---for individual mock galaxies (panel $b$; \S\S\,\ref{sec:2.2}).
These properties, together with the associated initial binary-population parameters (e.g., component masses and orbital separations), are passed to the {\ttfamily COMPAS} BPS engine\footnote{Compact Object Mergers: Population Astrophysics and Statistics; \url{https://compas.science/}} (Team COMPAS; \citealt{COMPASTeam2022a,COMPASTeam2022b,COMPASTeam2025}), which returns SN~Ia events with predicted explosion times and progenitor types (panel $c$; \S\S\,\ref{sec:2.3}--\ref{sec:2.5}).
An all-sky cosmic-volume realization of the \texttt{IllustrisTNG} snapshots (panel $a$) then embeds the full galaxy population, including SN~Ia hosts (panels $b$ and $d$), in a cosmological setting (\S\S\,\ref{sec:2.6}).
To enable survey-style comparisons, we impose an observational time window emulating real temporal baselines (\S\S\,\ref{sec:2.7}) and adopt consistent definitions of the SN~Ia rate and redshift baseline (\S\S\,\ref{sec:2.8}--\ref{sec:2.9}).
Together, these steps yield a final mock SN~Ia catalogue that captures the distribution of events across cosmic volume and time (\S\S\,\ref{sec:2.10}) and constitutes the dataset analyzed throughout the paper.
This catalogue provides a self-consistent basis for connecting SN~Ia populations to host-galaxy properties and tracking those connections across cosmic time (\S\,\ref{sec:3}--\ref{sec:9}).

We emphasize that, more broadly, the same framework naturally enables analogous forward modeling of other stellar transients in a fully cosmological setting: from Cepheid variables anchoring the Cepheid--Hubble flow calibration, to core-collapse SNe driving galactic and cosmic chemical evolution, to compact-object mergers involving neutron stars (NSs) and black holes (BHs)---namely NS+NS, NS+BH, and BH+BH systems---that are directly relevant to GW studies.
Indeed, cosmological-simulation--informed BPS frameworks have already been successfully deployed in compact-object merger studies \citep[e.g.,][]{Mapelli2017,Lamberts2019}.
To our knowledge, the present work is the first to apply an analogous \texttt{IllustrisTNG}+\texttt{COMPAS} strategy in a systematic way to galactic and cosmic SN~Ia population synthesis and to the coupled evolution of the SD and DD channels.

For ease of navigation, we provide a consolidated notation guide in Appendix\,\ref{appendix:A}. 
In particular, Appendix\,\ref{appendix:A} collects the acronyms and survey/simulation/code labels (Table~\ref{tab:acronyms}) and the parameters and variables (Table~\ref{tab:variables}) used throughout the paper, each with a one-line definition and alphabetical ordering for quick lookup. 
The tables are intended to reduce ambiguity in terminology and notation and to streamline cross-referencing across the cosmological-simulation, BPS, observational, and cosmology components of our analysis.

For clarity, we define here the notation used for host-level and SN-progenitor-level quantities.
For each host galaxy, we denote the host-level mean stellar age and metallicity by $T_*$ and $Z_*$, respectively. For the SN~Ia progenitor population associated with that host, we define the host-level mean progenitor age and metallicity by $T_{\rm pro}$ and $Z_{\rm pro}$, respectively, such that
\begin{equation}
\begin{gathered}
T_{\rm pro}=\frac{1}{N_{\rm SN}}\sum_i T_{{\rm pro},i}~~\&~~Z_{\rm pro}=\frac{1}{N_{\rm SN}}\sum_i Z_{{\rm pro},i}~,
\end{gathered}
\label{eq:TproZpro}
\end{equation}
where $T_{{\rm pro},i}$ and $Z_{{\rm pro},i}$ are the age and metallicity of the individual progenitor corresponding to the $i$th SN~Ia event.
$T_{{\rm pro},i}$ is equivalent to the conventional delay time, $\tau$.

\subsection{Cosmological Hydrodynamic Simulations of Galaxies}
\label{sec:2.1}

Our model employs the mock galaxy database from the \texttt{IllustrisTNG} project (\citealt{Pillepich2018b,Springel2018,Nelson2018,Naiman2018,Marinacci2018,Nelson2019}), produced with the moving-mesh code \textsc{AREPO} \citep{Springel2010}.
We adopt TNG100-1 (hereafter TNG100) as our fiducial cosmological realization because it provides a pragmatic balance between volume and resolution for predicting SN~Ia populations.
TNG50 achieves superior resolution but is limited by its small volume, reducing the sampling of rare systems and increasing finite-volume and cosmic-variance uncertainties; conversely, TNG300 offers a much larger volume but at substantially lower resolution, which can compromise host-property inference and the modeling of sub-grid-dependent processes relevant to SN progenitor demographics.

The TNG suite provides snapshot outputs spanning $z$\,=\,20.05 to 0.0 (100 snapshots in total: 20 ``full'' and 80 ``mini'').
We run the full simulation over $0 \leq z \leq 5$, constructing samples across this range for (i) all galaxies and (ii) SNe~Ia and their host galaxies (see Appendices~\ref{appendix:D} and \ref{appendix:E}).
This interval is motivated both practically and conceptually: the range $0 \leq z \leq 5$ effectively includes nearly the full galaxy population relevant for this work, except for a small minority of the earliest systems at the highest redshifts \citep[e.g.,][]{Madau2014,Behroozi2019,Forster-Schreiber2020,Robertson2022}.
For the fiducial analyses in this paper, however, we restrict attention to two explicitly defined subsets: the ``global-volume'' and ``local-volume'' galaxy samples, spanning $0 \leq z \leq 3$ and $0 \leq z \leq 0.1$, respectively (see \S\S\,\ref{sec:2.9}).
For SN~Ia cosmology, the ``global-volume'' subset is taken to represent the full SN~Ia and host-galaxy populations, while the ``local-volume'' subset provides the nearby reference frame for interpreting redshift evolution in the global-volume sample.

The TNG100 evolves a comoving volume of $75^3~[h^{-1}{\rm {cMpc}}]^3$ ($h = 0.6774$), modeling the coupled dynamics of dark matter, baryons, and supermassive black holes.
The simulation includes an advanced treatment of the key physical processes driving galaxy formation and evolution, such as radiative cooling, SF, chemical enrichment, and feedback from both SNe and AGNs. 
Compared to the original Illustris project \citep{Vogelsberger2014a,Vogelsberger2014b}, the TNG100 employs improved subgrid models for feedback and chemical evolution \citep{Weinberger2017,Pillepich2018a}, producing more realistic galaxy populations across cosmic time. 
The TNG100 provides high mass resolution ($9.4395\,\times\,10^{-5}\,\text{}[10^{10}\,h^{-1}\,\Msun\text{]}\,\simeq\,1.4\,\times\,10^6\,\text{[}\Msun\text{]}$ per baryonic particle) and tracks individual elemental abundances (e.g., C, O, and Fe), enabling metallicity-dependent modeling of stellar populations.
Combined with physically motivated prescriptions for stellar and AGN feedback, these features yield realistic SFHs and chemical enrichment patterns---key inputs for our SN~Ia population-synthesis framework.
For a subset of the analysis in \S\,\ref{sec:3}, we additionally exploit the higher-resolution TNG50-1 simulation \citep{Nelson2019b, Pillepich2019}. 
This allows us to characterize better the spatial distribution of SNe~Ia within mock host galaxies and to assess its impact on SN~Ia properties.
Unless otherwise noted, however, all results in this paper are based on the TNG100 simulation.

In each snapshot of TNG100, galaxies are identified as self-bound subhalos using the \textsc{SUBFIND} algorithm \citep{Springel2001}.
We consider only well-defined subhalos with \texttt{SubhaloFlag} $\neq 0$ and compute their stellar masses by summing the masses of star particles with \texttt{GFM\_StellarFormationTime} $> 0$ (thereby excluding wind particles with negative formation times).
We then retain galaxies satisfying $\logM \geq 8.0$.
For TNG100, the target baryonic mass resolution is $m_{\rm baryon} \simeq 1.4\times10^{6}\,\Msun$, which sets the typical initial mass of a star particle; therefore, the stellar-mass threshold $\logM=8.0$ corresponds to $\gtrsim 70$ star particles.

A collection of star particles within each subhalo serves as the building block of the unresolved stellar population. 
Each particle represents a SSP---a coeval group of stars with a common chemical composition.
Across our sample, mock galaxies spanning $M_*\simeq10^{8.0}~\Msun$ to $10^{12.5}~\Msun$ contain $\sim$\,70 to $\sim\!4\,\times\,10^{6}$ star particles, with a mean and a median of $\sim\!7\times10^{3}$ and $\sim\!5\times10^{2}$, respectively.
We use the galaxy-scale properties provided by the TNG mock galaxy catalogue, including stellar mass, SFR, and mass-weighted mean stellar metallicity.
In that catalogue, these integrated quantities are defined within twice the stellar half-mass comoving radius, $2R_\mathrm{half}$.
Since the mass-weighted mean stellar age is not provided, we compute it separately using the same aperture, $2R_\mathrm{half}$.
For the SN~Ia calculation, by contrast, we consider all star particles bound to each subhalo, including those at radii beyond $2R_\mathrm{half}$, to capture events occurring in the galactic outskirts.
This more extended selection, however, has a negligible impact on the inferred SN~Ia statistics, because the vast majority of bound star particles lie within $2R_\mathrm{half}$.

In constructing our TNG100 galaxy sample, we identify a rare group of unusually compact subhalos that exhibit implausibly young and metal-rich stellar populations.
Specifically, these systems have $R_{\rm half}<1~h^{-1}\rm{cKpc}$, stellar age $T_*<1~\mathrm{Gyr}$, and stellar metallicity $Z_*$ exceeding the fiducial mass--metallicity sequence by $>0.5$~dex.
Although they are not marked as spurious (i.e., ${\tt SubhaloFlag}\neq 0$) and can still contain substantial numbers ($>70$) of star particles, their sizes approach the numerical resolution limit.
At $z=0$, the stellar gravitational softening length in TNG100 is $\epsilon_*\simeq 0.74~\mathrm{Kpc}$, so $R_{\rm half}\lesssim 2\epsilon_*$ places these systems in a barely resolved regime.
In this regime, over-cooling can produce overly concentrated SF and inefficient metal mixing, leading to runaway self-enrichment that yields artificially young and metal-rich stellar populations (e.g., \citealt{Pillepich2018a,Genel2018}).
We therefore conservatively exclude subhalos with $R_{\rm half}<1~h^{-1}\rm{cKpc}$ from all subsequent analyses.

In this work, galaxy mass and SFR are the two TNG100 galaxy properties most frequently discussed. 
We convert all masses from the native \texttt{IllustrisTNG} unit, $10^{10}~h^{-1}\Msun$, to physical masses in $\Msun$ assuming $h=0.6774$.
For SFR, we adopt the \texttt{IllustrisTNG} ``instantaneous'' value as our fiducial measure. 
Using star-particle formation times, \citet{Donnari2019} constructed time-averaged SFR estimators over fixed look-back windows of $10$\,--\,$10^{3}$\,Myr and showed that the instantaneous SFR is broadly consistent with the short-timescale estimator, while the $10^{3}$\,Myr-averaged SFR is only modestly lower by $\sim0.1$~dex at $z=0$ and by $\gtrsim0.2$~dex at $z=2$. 
Given these small offsets, and for simplicity and consistency with the public TNG catalogues, we use the instantaneous SFR throughout this paper. 
In addition, for time-related quantities stored in TNG snapshots as scale factors (e.g., the snapshot time and stellar formation time), we convert them to cosmic times in physical Gyr using the adopted TNG cosmological parameters.

Throughout this work, we adopt the \citet{Planck2015} cosmological parameters assumed in the \texttt{IllustrisTNG} suite to ensure self-consistency.
Specifically, we assume a spatially flat $\Lambda$CDM cosmology with
$\Omega_{\rm m,0}=0.3089$, $\Omega_{\Lambda,0}=0.6911$, $\Omega_{\rm b,0}=0.0486$,
$\sigma_8=0.8159$, $n_s=0.9667$, and $h=0.6774$ (i.e., $H_0=67.74~{\rm km~s^{-1}~Mpc^{-1}}$).
These parameters imply a present-day cosmic age of $t(z=0)=13.8027~{\rm Gyr}$,
equivalently a look-back time of $\Delta t_{\rm lb}(z\rightarrow\infty)=13.8027~{\rm Gyr}$.

\subsection{From Star Particles to Individual Single \& Binary Stars}
\label{sec:2.2}

SN~Ia progenitors are binary systems, so BPS predictions for SNe~Ia depend fundamentally on the assumed initial binary demographics, including the fraction of stars born in binaries and the distributions of stellar masses and orbital properties \citep[e.g.,][]{Raghavan2010,DucheneKraus2013,SanaEtAl2012}.
Any BPS treatment of SNe~Ia therefore requires explicit assumptions about the initial binary pairing function, mass ratios, and orbital separations of primary and secondary stars.
For each star particle extracted from TNG100, we first populate stars according to the Chabrier initial mass function (IMF; \citealt{Chabrier2003})\footnote{\url{https://github.com/keflavich/imf}} for zero-age main-sequence (ZAMS) stars.
We then assemble these stars into binary systems following the observationally motivated prescriptions of \citet{MoeDiStefano2017}, which specify the correlated probabilities of the binary fraction, primary mass ($M_1$), mass ratio ($q=M_2/M_1$), orbital period ($P_{\rm orb}$), and eccentricity.

We initialize the binary grids with primary masses $M_1 \ge 0.7~{\rm M}_{\odot}$, secondary masses $M_2 \ge 0.08~{\rm M}_{\odot}$, and initial orbital periods $P_{\rm orb} \in [1, 10^5]$ days. 
The lower limit for the primary mass\footnote{The lower ZAMS mass required for a star to reach the WD stage by the present epoch is metallicity dependent. 
From the analytic stellar-lifetime prescriptions of \citet{Hurley2000}, a $0.8\,\Msun$ star, for instance, has a main-sequence lifetime of $\sim$\,17--20~Gyr at near-solar metallicity, exceeding the age of the Universe. 
But this decreases to $\sim$\,11--12~Gyr at very low metallicity ($Z \simeq 10^{-4}$), so such stars can in principle form WDs within a Hubble time in metal-poor environments. 
In contrast, a $0.7\,\Msun$ star remains too long-lived even at $Z \simeq 10^{-4}$, with a lifetime of $\sim$\,15--18~Gyr, and thus cannot produce a WD by the present epoch. 
We therefore adopt $M_1 \ge 0.7\,\Msun$ as a conservative lower limit on the primary mass.} 
ensures that the stars can evolve into WDs within a Hubble time \citep{IbenTutukov1984,Hurley2000}, while the secondary-mass cutoff corresponds to the hydrogen-burning limit and the validity range of the binary star evolution fitting formulae of \citet{Hurley2002}. 
The orbital period range is chosen to encompass all evolutionary channels that experience binary interactions; systems with $P_{\rm orb} < 1$ day typically experience ZAMS contact and premature mergers\footnote{Even among binaries with $P_{\rm orb} \ge 1$ day, systems that are already in Roche-lobe contact at initialization are reclassified as merged single stars to avoid potential numerical issues in \texttt{COMPAS}.} \citep{MoeDiStefano2017}, whereas systems with $P_{\rm orb} > 10^5$ days are too wide to undergo Roche-lobe overflow and effectively evolve as single stars\footnote{We set the upper limit of the initial orbital period to $10^5$ days. 
Following the definition by \citet{PostnovYungelson2014}, the evolutionary trajectory of a binary system deviates from that of single stars only if it undergoes Roche-lobe overflow. 
Since the maximum stellar radius during the extreme supergiant or AGB phase rarely exceeds $\sim$\,2000\,${\rm R}_\odot$, any binary system with $P_{\rm orb}$\,$>$\,$10^5$ days (corresponding to an orbital separation $a \gtrsim 6000\,{\rm R}_\odot$) will never experience Roche-lobe overflow and effectively evolves as two independent single stars. 
Thus, our period range fully encompasses all possible interacting binaries.} \citep[e.g.,][]{Paczynski1976,Hurley2002, Ivanova2013,PostnovYungelson2014,DeMarcoIzzard2017}. 
Unlike $M_1$, $M_2$, and $P_{\rm orb}$, we impose no explicit restriction on the initial eccentricity.

\begin{figure}
\includegraphics[width=0.95\columnwidth]{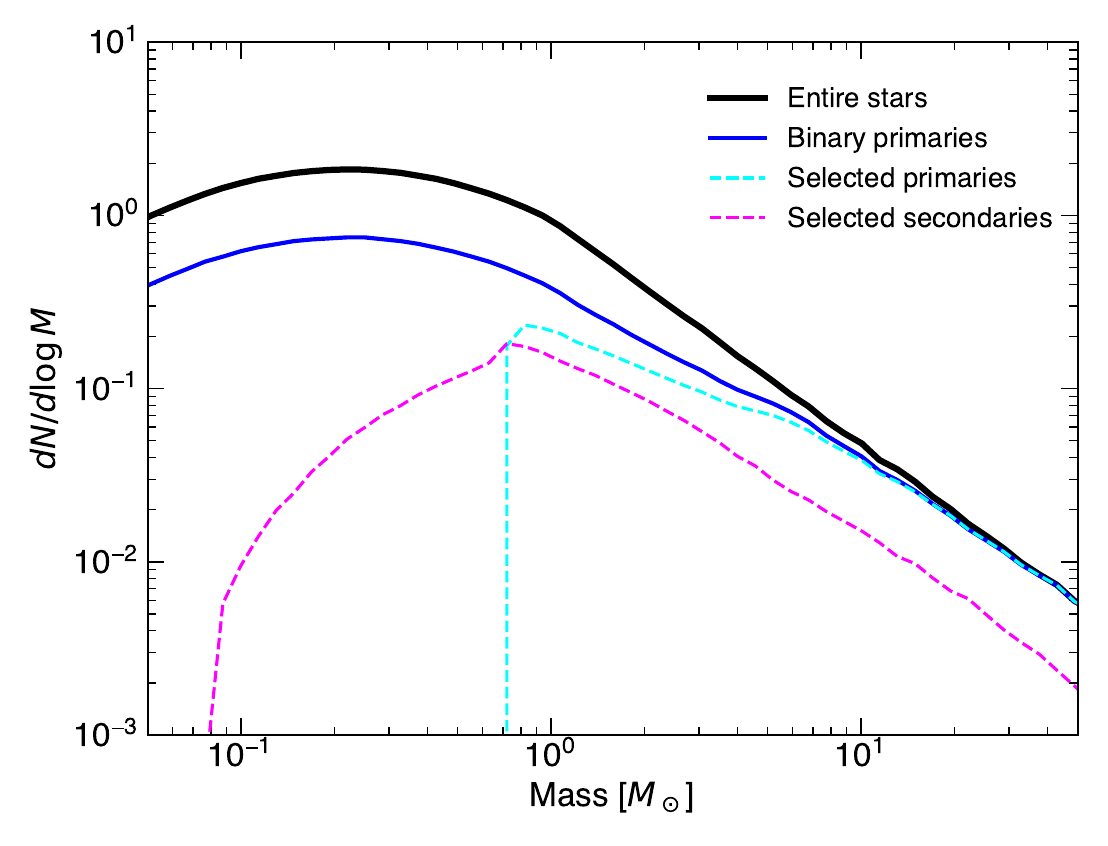}
\caption{
The initial mass functions for all stars and binary components, generated from a reference star particle with $4\times10^6~\Msun$. 
We adopt the canonical \citet{Chabrier2003} initial mass function with a high-mass power-law slope of $-2.3$, producing $\sim$\,6.3 million stars (thick black solid line). 
Next, we combine these stars into binary systems using the joint probability distribution of \citet{MoeDiStefano2017}, which provides correlated probabilities of the binary fraction, primary mass ($M_1$), mass ratio ($q=M_2/M_1$), orbital period ($P_{\rm orb}$), and eccentricity. 
Blue solid line shows the initial mass functions of the primary stars in $\sim$\,2.6 million binary pairs. 
The cyan and magenta dashed lines denote the IMFs of the primary and secondary stars, respectively, in $\sim\,$0.4 million binary pairs that satisfy our selection criteria: $M_1 \ge 0.7\,\Msun$, $M_2 \ge 0.08\,\Msun$, and $P_{\rm orb} \in [1, 10^5]$ days.
}
\label{fig:Fig2}
\end{figure}

Figure\,\ref{fig:Fig2} shows the resulting IMFs for all (single and binary) stars and for binary components.
These IMFs are generated assuming a star particle with $4\times10^6~\Msun$, which we adopt as a conservative upper-envelope mass that brackets the maximum star-particle mass expected in TNG100 (typically $\lesssim 2.8\times10^6~\Msun$).
In other words, $4\times10^6~\Msun$ is a sufficiently large reference mass, ensuring that our IMF sampling covers the full mass range of star particles.
For the reference mass, sampling the Chabrier IMF yields $\sim$\,6.3 million stars, whose mass distribution is shown by the thick black curve.
From them, we draw $\sim$\,2.6 million binary pairs, whose primary-mass distribution is indicated in blue.
Imposing our adopted mass and orbital-period cuts isolates the subsets of primaries and secondaries in $\sim\,$0.4 million binary pairs, shown in cyan and magenta, respectively.

\subsection{Binary Population Synthesis: From Binaries to SNe Ia via Single- and Double-degenerate Channels}
\label{sec:2.3}

To model the binary evolution leading to SNe~Ia, we employ the open-source rapid BPS code \texttt{COMPAS} (Team COMPAS; \citealt{COMPASTeam2022a,COMPASTeam2022b,COMPASTeam2025}).
Several BPS frameworks have been developed to study compact objects and SN progenitors, including \texttt{StarTrack} \citep{Belczynski2008}, \texttt{SeBa} \citep{Toonen2012}, \texttt{BSE/binary\_c} \citep{Hurley2002, Izzard2004}, the Brussels code \citep{Mennekens2010}, \texttt{BPASS} \citep{EldridgeStanway2016,Eldridge2017}, \texttt{COSMIC/MOBSE} \citep{Breivik2020, Giacobbo2018}, and \texttt{POSYDON} \citep{Fragos2023}.
Among these, \texttt{COMPAS} provides a computationally efficient platform for evolving large ensembles of binaries, while retaining detailed treatments of mass transfer, common-envelope (CE) evolution, and GW-driven orbital decay.
The code has been extensively validated in studies of compact objects and GW sources \citep[e.g.,][]{Stevenson2017,VignaGomez2018}.
\texttt{COMPAS} was built on the stellar- and binary-evolution formalism\footnote{\texttt{COMPAS}, like other rapid BPS frameworks, remains prescription-based: uncertain phases such as mass retention on accreting WDs, CE evolution, shell ignition, and the mapping from binary state to explosion outcome are not derived from detailed multidimensional hydrodynamic calculations, and the sub-$M_{\rm Ch}$ channel in particular is represented through simplified criteria rather than a complete physical census.} of \citet{Hurley2000} and \citet{Hurley2002}, but has been substantially updated with more modern treatments of binary interactions and accreting WD systems (Team COMPAS; \citealt{COMPASTeam2025}).
These features make \texttt{COMPAS} well suited to our goal of generating large, internally consistent SN~Ia populations, and then coupling them to cosmological mock galaxies to construct a cosmic-scale SN~Ia population model.

Most traditional rapid-BPS frameworks were originally developed with canonical near-$M_{\rm Ch}$ pathways as their primary focus.
In contrast, the treatment of sub-$M_{\rm Ch}$ explosions remains considerably less uniform across codes.
Consequently, whether a given BPS prediction is considered to include sub-$M_{\rm Ch}$ SNe~Ia depends sensitively on the specific code version and on the adopted prescriptions for WD accretion, helium retention, and shell ignition.
Among the widely used public frameworks, \texttt{StarTrack} \citep{Belczynski2008} and, more recently, \texttt{COMPAS} \citep{COMPASTeam2025} implement prescription-based sub-$M_{\rm Ch}$ channels linked to helium-accretion physics and double-detonation criteria.
In many other codes, however, sub-$M_{\rm Ch}$ outcomes are either not included, only approximately treated, or inferred in post-processing from progenitor populations rather than generated by a native explosion module.
Taken together, this work uses \texttt{COMPAS} as the BPS engine of our \texttt{SN~Ia Population Machine} to generate WD systems that can produce SNe~Ia across cosmic time.


\begin{table}
\setlength{\tabcolsep}{3pt}
\caption{Mapping {\ttfamily COMPAS} ``stellar phases'' to ``stellar types'' used in our simulation. We group the 16 \texttt{COMPAS} ``stellar phases'' into nine ``stellar types''.}
\label{tab:sttypes}
\footnotesize
\begin{tabular}{>{\raggedright\arraybackslash}p{0.0945\textwidth} >{\raggedright\arraybackslash}p{0.28\textwidth} >{\raggedright\arraybackslash}p{0.07\textwidth}}
\midrule
\midrule
\textbf{Stellar}    & \textbf{Meaning}  & \textbf{Stellar} \\
\textbf{Phase}      &                   & \textbf{Type} \\
\midrule
MS $\leq0.7$ & Main-sequence star, mass $\leq0.7~\Msun$ & MS\\
MS $>0.7$ & Main-sequence star, mass $>0.7~\Msun$ & MS\\
HG & Hertzsprung gap & MS\\
FGB & First giant-branch star & GB\\
CHeB$^\dagger$ & (Hydrogen-rich) Core-helium-burning star & GB\\
EAGB & Early asymptotic-giant-branch star & GB\\
TPAGB & Thermally pulsing AGB star & GB\\
HeMS & Helium main-sequence star & He\\
HeHG & Helium Hertzsprung gap & He\\
HeGB & Helium giant-branch star & He\\
HeWD & Helium white dwarf & He-WD\\
COWD & Carbon--Oxygen white dwarf & CO-WD\\
ONeWD & Oxygen--Neon white dwarf & ONe-WD\\
NS & Neutron star & NS\\
BH & Black hole & BH\\
MR & Massless remnant & MR\\
\midrule
\end{tabular}
\begin{minipage}{\columnwidth}
\footnotesize
\textit{Note.} 
$^\dagger$Within the stellar-evolution prescription adopted by \texttt{COMPAS}, CHeB stars generally lie in the horizontal-branch, ``giant'' regime, although sufficiently massive stars at low metallicity may enter CHeB before reaching the GB \citep{Hurley2000,COMPASTeam2022a}. 
Motivated by the internal \texttt{COMPAS} treatment, which groups CHeB stars with other giant-like post-MS donors in parts of its binary-interaction machinery, we adopt the simpler working classification and include CHeB companions in the WD+GB category \citep{COMPASTeam2022a}.
\end{minipage}
\end{table}

\begin{table*}
\centering
\caption{Principal selection criteria used to define the four SN~Ia progenitor categories in this study.}
\label{tab:snia_category}
\footnotesize
\setlength{\tabcolsep}{4pt}
\renewcommand{\arraystretch}{1.15}
\begin{tabular}{>{\raggedright\arraybackslash}p{0.11\textwidth} >{\raggedright\arraybackslash}p{0.36\textwidth} >{\raggedright\arraybackslash}p{0.41\textwidth}}
\midrule
\midrule
\textbf{Category}$^\text{a}$ & \textbf{SD} & \textbf{DD}\\
\midrule
\textbf{Near-$M_{\rm Ch}$} 
&
$\blacktriangleright$ \texttt{COMPAS}$^\text{b}$-generated binary-evolution tracks \newline
\hspace*{0.7em} $\vartriangleright$ CO-WD + non-degenerate (MS, GB, and He) \newline
$\blacktriangleright$ External, pre-computed parameter grids$^\text{c}$ \newline
\hspace*{0.7em} $\vartriangleright$ $M_1^{\rm WD} > 1.378~\Msun$
&
$\blacktriangleright$ \texttt{COMPAS}$^\text{b}$ output \newline
\hspace*{0.7em} $\vartriangleright$ CO-WD + CO-WD \newline
\scriptsize 
\hspace*{2.4em} $\bullet$ \texttt{event\_kind} = \texttt{`dco\_merge'} in \texttt{BSE\_Double\_Compact\_Objects}$^\text{d}$ \newline
\footnotesize
$\blacktriangleright$ Post-processing \newline
\hspace*{0.7em} $\vartriangleright$ $M_{\rm tot}~(=M_1^{\rm WD}+M_2^{\rm WD}) > 1.378~\Msun$ at DCO formation 
\\
\midrule
\textbf{Sub-$M_{\rm Ch}$} 
&
$\blacktriangleright$ \texttt{COMPAS}$^\text{b}$ output \newline
\hspace*{0.7em} $\vartriangleright$ CO-WD + non-degenerate He star \newline
\scriptsize 
\hspace*{2.4em} $\bullet$ \texttt{event\_kind} = \texttt{`sn\_event'} in \texttt{BSE\_Supernovae}$^\text{d}$ \newline
\hspace*{2.4em} $\bullet$ \texttt{SN\_Type} = \texttt{`HeSD'} (He-shell detonation) in \texttt{BSE\_Supernovae}$^\text{d}$ \newline
\footnotesize
&
$\blacktriangleright$ \texttt{COMPAS}$^\text{b}$ output \newline
\hspace*{0.7em} $\vartriangleright$ CO-WD + He-WD \newline
\scriptsize 
\hspace*{2.4em} $\bullet$ \texttt{event\_kind} = \texttt{`dco\_merge'} in \texttt{BSE\_Double\_Compact\_Objects}$^\text{d}$ \newline
\footnotesize
$\blacktriangleright$ Post-processing \newline
\hspace*{0.7em} $\vartriangleright$ $M_{\rm tot}~(=M_1^{\rm WD}+M_2^{\rm WD}) < 1.378~\Msun$ at DCO formation \newline 
\hspace*{0.7em} $\vartriangleright$ $M_{\rm p}~(=\max(M_1^{\rm WD},M_2^{\rm WD})) > 0.8~\Msun$ at DCO formation$^\text{e}$
\\
\midrule
\end{tabular}
\begin{minipage}{1.00\textwidth}
\vspace{1mm}
\footnotesize
\textit{Notes.}
$^\text{a}$ Category: The SD/DD class is defined by whether the companion is non-degenerate or degenerate at explosion or merger/ignition, and the near-$M_{\rm Ch}$/sub-$M_{\rm Ch}$ class by whether the explosion mass is above or below $1.378~\Msun$. \citep[e.g.,][]{WhelanIben1973,IbenTutukov1984,Webbink1984,Nomoto1984,Sim2010,WangLiHan2010,WangHan2010a,WangHan2010b,Maoz2014,Ruiter2014,Meng2017,Shen2018sub}. 
~$^\text{b}$ \texttt{COMPAS}: \citet{COMPASTeam2022a,COMPASTeam2025}.
~$^\text{c}$ External grids: \citet{WangLiHan2010,WangHan2010a,WangHan2010b,Meng2017}.
~$^\text{d}$ The labels and event flags used in the \texttt{COMPAS} output.
~$^\text{e}$ Sub-$M_{\rm Ch}$ DD $M_{\rm p}$: \citet{Ruiter2011} \citep[see also][]{Sim2010,Ruiter2014,Shen2018sub,Polin_2019}.
\end{minipage}
\end{table*}

\subsubsection{Single-degenerate SNe~Ia}
\label{sec:2.3.1}

In this work, we use \texttt{COMPAS} to evolve binary populations and generate the statistical ensemble of WD binaries that can potentially produce SNe~Ia.
In the SD scenario, a CO-WD accretes hydrogen- or helium-rich material from a non-degenerate companion and grows toward thermonuclear ignition; possible donors include MS, GB, and He companions (Table~\ref{tab:sttypes} and Appendix\,\ref{appendix:B}).
Because the microphysics of stable WD accretion---including mass-retention efficiency, wind mass loss, disk instability, and shell ignition---remains challenging to capture robustly in rapid-BPS formalisms, we adopt a hybrid treatment of the SD channel.
Specifically, we retain \texttt{COMPAS}-native sub-$M_{\rm Ch}$ SD events produced by helium-rich accretion onto CO-WDs through helium-shell detonation, while identifying canonical near-$M_{\rm Ch}$ SD progenitors by mapping WD + donor pairs in \texttt{COMPAS} onto externally computed explosion conditions \citep{WangLiHan2010,WangHan2010a,WangHan2010b,Meng2017} in the $M_1^{\rm WD}-M_2-\log P_{\rm orb}$ space.
The externally calibrated grids were derived from detailed binary-evolution calculations that explicitly follow the subsequent accretion and WD-growth phase, whereas rapid-BPS frameworks necessarily approximate the same phase with simplified analytic prescriptions.\footnote{More precisely, \texttt{COMPAS} is used in our hybrid framework as a binary-population generator, whereas the externally adopted SD progenitor regions serve as calibrators of the accreting-WD phase that determines the SD explosion criterion. In that limited sense, externally precomputed detailed grids may provide higher physical fidelity than native rapid-BPS prescriptions for the semidetached accretion phase, although that fidelity is conditional on the adopted SD microphysics and should not be interpreted as establishing the truth of the SD channel itself.}
The grids therefore provide a physically motivated summary of the channel-specific conditions under which an accreting CO-WD can grow to explosion, subject to the assumptions of the underlying calculations.\footnote{We note that the external grids cited here are not fully homogeneous: the \citet{WangLiHan2010} and \citet{WangHan2010a,WangHan2010b} contours are based on more traditional optically-thick-wind-type assumptions, whereas \citet{Meng2017} presents an alternative common-envelope-wind-based treatment, particularly for the WD+MS channel. 
Accordingly, these works are grouped here only in the broader sense that they provide SD progenitor regions derived from detailed binary-evolution calculations, rather than as a single unified parameter set obtained under one identical physical prescription.}
The principal selection criteria used to define the near- and sub-$M_{\rm Ch}$ SD progenitor categories in this study are summarized in Table~\ref{tab:snia_category}.

Operationally, we combine the two SD prescriptions for sub- and near-$M_{\rm Ch}$ channels through a hierarchical classification scheme.
The sub-$M_{\rm Ch}$ progenitors are first selected by \texttt{COMPAS} when binaries meet the \texttt{COMPAS}-native sub-$M_{\rm Ch}$ double-detonation criterion for helium-rich accretion onto CO-WDs (i.e., $M_1^{\rm WD} \ge 0.9~\Msun$ and $M_{\rm He\text{-}shell} \ge 0.05~\Msun$), which can trigger helium-shell detonations before the accreting WD reaches $M_{\rm Ch}$.
The external near-$M_{\rm Ch}$ progenitor regions are then applied to binaries that do not satisfy the above sub-$M_{\rm Ch}$ condition.
This ordering assigns temporal priority to pre-emptive sub-$M_{\rm Ch}$ SD detonations and thus avoids artificial suppression of the sub-$M_{\rm Ch}$ channel; the near-$M_{\rm Ch}$ SD catalogue correspondingly contains only those systems that avoid an earlier sub-$M_{\rm Ch}$ detonation and continue to grow toward $M_{\rm Ch}$.
Although this procedure does not fully exploit the \texttt{COMPAS}-native SD treatment, it anchors the most uncertain phase of SD evolution to detailed calculations rather than to a single rapid-code implementation.

\begin{figure*}
\includegraphics[width=0.8\textwidth]{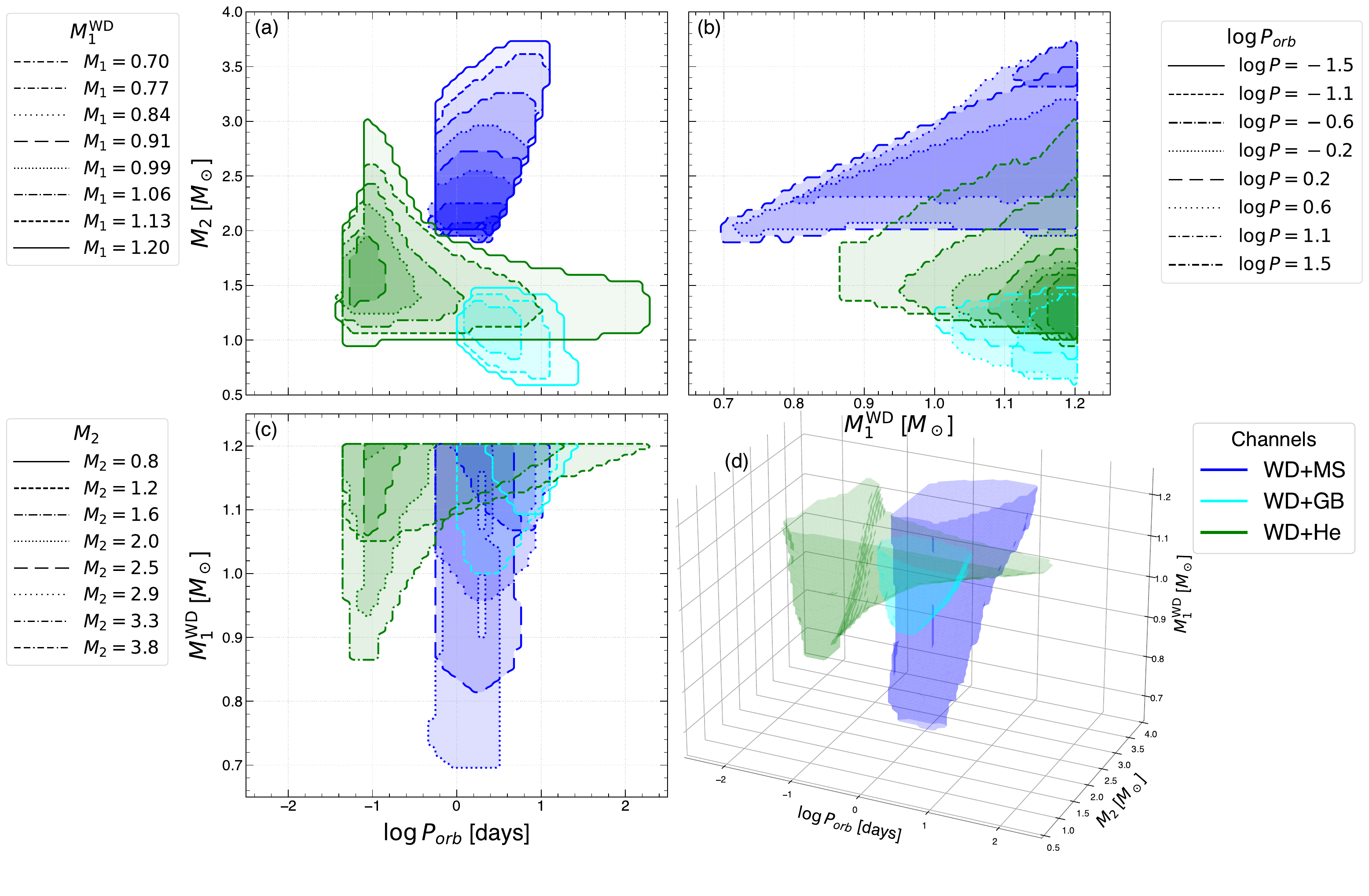}
\caption{
Three 2D projections (panels $a$--$c$) of our adopted 3D $M_{1}^{\rm WD}$--$M_2$--$\log P_{\rm orb}$ parameter-space polyhedra (panel $d$) for the single-degenerate SN~Ia channels (WD+MS, WD+GB, and WD+He).
The blue, cyan, and green regions denote the WD+MS \citep{Meng2017}, WD+GB \citep{WangLiHan2010}, and WD+He \citep{WangHan2010a} channels, respectively.
Contours with different line types indicate the regions associated with the corresponding binary parameters ($M_{1}^{\rm WD}$, $M_2$, and $\log P_{\rm orb}$), as labeled in the legends.
As the binary stellar and dynamical evolution proceeds, we record the companion's evolutionary stage (i.e., MS, GB, or He) at each time step (Table~\ref{tab:sttypes}).
When a binary simultaneously satisfies the companion-type and the $M_{1}^{\rm WD}$--$M_2$--$\log P_{\rm orb}$ conditions, we identify it as an SN~Ia event in the corresponding single-degenerate channel.
}
\label{fig:Fig3}
\end{figure*}

Figure\,\ref{fig:Fig3} visualizes the external grids for near-$M_{\rm Ch}$ SD progenitors adopted in this study: three 2D projections (panels $a$--$c$) of the 3D $M_{1}^{\rm WD}$--$M_2$--$\log P_{\rm orb}$ parameter-space polyhedra (panel $d$) for WD+MS \citep{Meng2017}, WD+GB \citep{WangLiHan2010}, and WD+He \citep{WangHan2010b} channels.
Panels ($a$--$c$) show how the allowed regions vary with $M_{1}^{\rm WD}$, $\log P_{\rm orb}$, and $M_2$, respectively.
The sizes and boundaries of these polyhedra are physically motivated.
For instance, panel ($a$) shows that more massive WDs generally occupy larger regions of parameter space, implying a higher probability of SN~Ia formation because less additional mass is required to reach $M_{\rm Ch}$.
The left and right boundaries correspond to the ZAMS stellar radii and the onset of dynamical instability, respectively, while the upper and lower edges reflect constraints from stable mass transfer and critical mass-transfer rates.

As the binary stellar and orbital evolution proceeds, we record the companion's evolutionary stage (i.e., MS, GB, or He) at each time step (Table~\ref{tab:sttypes}).
When a binary simultaneously satisfies the companion-type and the $M_{1}^{\rm WD}$--$M_2$--$\log P_{\rm orb}$ conditions, we classify it as an SN~Ia event in the corresponding near-$M_{\rm Ch}$ SD channel.
In practice, we continuously scan the post-CO-WD evolutionary track from \texttt{COMPAS} and identify an SN Ia event whenever the binary satisfies the companion stellar type criterion and its trajectory intersects the corresponding external parameter-space region.
Because {\ttfamily COMPAS} outputs discrete time steps in $M^{\rm WD}_1$, $M_2$, and $\log P_{\rm orb}$, some systems can step over narrow valid regions and thus be missed.
We therefore interpolate the evolutionary tracks on a progressively refined temporal grid until the sampling of the valid regions converges, thereby reducing such losses.
Given the lack of a clear physical or numerical justification for higher-order schemes, we adopt linear interpolation between adjacent {\ttfamily COMPAS} temporal grids and their relevant parameters.

For the near-$M_{\rm Ch}$ SD channel, we take the delay time to be the interval from progenitor-binary birth at the SF epoch to the system’s first entry into the SN-progenitor parameter space (i.e., the onset of stable burning and WD mass growth), rather than to the eventual explosion epoch after the terminal mass-accretion phase.
Because the adopted SD parameter spaces are constructed to produce SNe~Ia through stable accretion, the remaining growth to $M_{\rm Ch}$ is rapid.
For example, \citet{Chomiuk2021} showed that stable hydrogen-burning occurs at $\dot{M} \simeq$~1\,--\,5\,$\times 10^{-7}~\Msun\,{\rm yr^{-1}}$, depending on $M_{\rm WD}$, with the threshold values, 4\,$\times$\,$10^{-8}$\,--\,4\,$\times$\,$10^{-7}$\,$\Msun\,{\rm yr}^{-1}$, increasing with $M_{\rm WD}$.
This implies a remaining growth time of only a few to a few tens of Myr, depending on the mass-retention efficiency.
To account for this terminal accretion phase, we add a fixed 30 Myr to the time of the system's first entry into the SN-progenitor parameter space.
This addition is negligible compared to the 0.1--10~Gyr timescales of stellar evolution and binary interaction and has little impact on the global DTD shape, although it can slightly affect the youngest end of the progenitor-age distribution.

\subsubsection{Double-degenerate SNe~Ia}
\label{sec:2.3.2}

In the DD realization, we continue to use \texttt{COMPAS} to follow the formation of double-WD binaries and their subsequent orbital evolution.
In this channel, WD+WD systems are produced by binary evolution, lose orbital angular momentum through GW radiation, and may eventually merge, triggering thermonuclear ignition that can lead to SNe~Ia.
The total delay time therefore consists of two components: the time required to form the double-WD binary and the subsequent GW-driven inspiral time.
\texttt{COMPAS} computes the inspiral timescale as
\begin{equation}
\label{eq:gw radiation}
t_{\text{GW}} = \frac{5}{256} \, \frac{1}{3.15 \times 10^{13}} \, \frac{c^5}{G^3} \, \frac{a^4}{M_{\rm p} \, M_{\rm s} \,(M_{\rm p} + M_{\rm s})}\,\,[\text{Myr}],
\end{equation}
where $c$ is the speed of light, $G$ is the gravitational constant, $a$ is the orbital semi-major axis, and $M_{\rm p}$ and $M_{\rm s}$ denote the primary and secondary masses, respectively \citep{Peters1964}.
Owing to its steep dependence on orbital separation and component masses, $t_{\rm GW}$ spans an extremely broad range, from Myr for the tightest binaries to many Gyr for wider ones, with some systems remaining unmerged beyond a Hubble time.

WD mergers span several compositions, including He-, CO-, and ONe-WDs (Table~\ref{tab:sttypes} and Appendix\,\ref{appendix:B}), but not all such systems are regarded as DD SN~Ia progenitors.
He+He WD mergers are generally too low in mass to reach $M_{\rm Ch}$ and are expected to produce spectra distinct from those of normal SNe~Ia, and are therefore excluded from the DD population.
CO+CO WD mergers are typically taken as the fiducial DD progenitors.
CO+He WD mergers are treated separately as candidate sub-$M_{\rm Ch}$ DD systems, motivated by the viability of thin helium-shell double detonations for a subset of normal SNe~Ia \citep{Bildsten2007,Shen2009,Woosley2011}.
ONe-WDs (typically with initial masses $\gtrsim 8~\Msun$) are more likely to undergo accretion-induced collapse to neutron stars than explosion as SNe~Ia \citep{Saio1985, Nomoto1991, Gutierrez1996}, and are therefore excluded from the DD population.
Our choice retains the bulk of the DD SN~Ia population, since BPS calculations show that CO+CO WD pairs dominate (88 -- 100\,\%) the DD population of WD$+$WD mergers with combined mass above $M_{\rm Ch}$ \citep{Belczynski2005, Ruiter2009}.

Depending on the component masses, compositions, and merger dynamics, WD mergers may produce either near-$M_{\rm Ch}$ explosions or sub-$M_{\rm Ch}$ detonations.
Early studies often assumed that an SN~Ia occurs whenever the combined WD mass exceeds $M_{\rm Ch}$ \citep[e.g.,][]{Belczynski2005, Ruiter2009}, but later work emphasized the importance of the mass ratio, with DD SNe~Ia preferentially expected for near-equal-mass systems \citep{Pakmor2010, Pakmor2011}. 
\citet{Ruiter2013} further introduced a critical mass-ratio threshold that depends on both the primary mass and the mass ratio.
Given this diversity of proposed DD explosion criteria, we adopt an inclusive but structured classification scheme for the DD populations. 
Systems that merge as CO+CO WDs are treated as near-$M_{\rm Ch}$ DD candidates and are required to satisfy the total mass $M_{\rm tot}\,(=M_{1}^{\rm WD}+M_{2}^{\rm WD}) > 1.378~\Msun$. 
In contrast, systems that merge as CO+He WDs are treated as sub-$M_{\rm Ch}$ DD candidates and are required to satisfy $M_{\rm tot} < 1.378~\Msun$ and the primary mass $M_{\rm p}~(=\max(M_{1}^{\rm WD},M_{2}^{\rm WD})) > 0.8~\Msun$ \citep{Ruiter2011}.
Because the near- and sub-$M_{\rm Ch}$ channels compete for the same underlying progenitor population, their inferred relative fractions are contingent on the adopted selection criteria.
The principal selection criteria used to define the near- and sub-$M_{\rm Ch}$ DD progenitor categories in this study are summarized in Table~\ref{tab:snia_category}.

\begin{figure*}
\includegraphics[width=\textwidth]{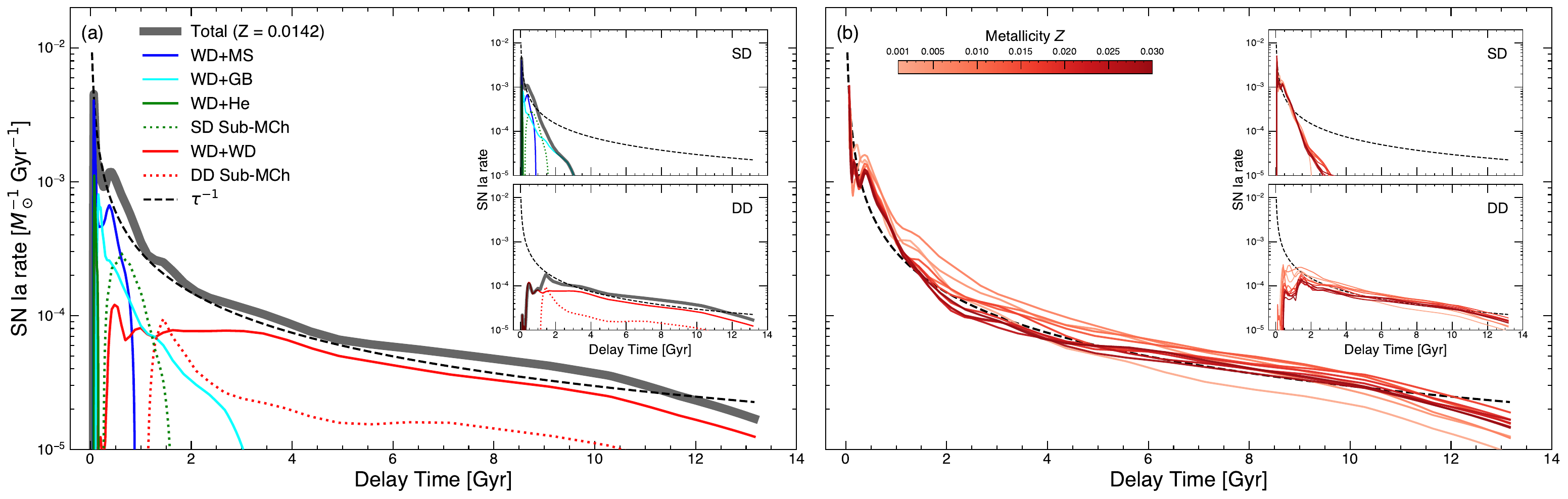}
\caption{
The star-particle-level delay-time distributions: the SN~Ia rate as a function of delay time ($\tau$ in units of Gyr) following a single star-burst of $1~\Msun$ simple-stellar-population formation.
\textbf{\textit{(a)}} Star-particle-level SSP DTDs for Z = 0.0142 (the adopted solar metallicity in \texttt{COMPAS}).
Blue, cyan, and green solid curves show the SD near-$M_{\rm Ch}$ channels (WD+MS, WD+GB, and WD+He), and the red solid curve shows the DD near-$M_{\rm Ch}$ channel (WD+WD). 
Green and red dotted curves denote sub-$M_{\rm Ch}$ progenitors for SD and DD channels, respectively.
The thick grey curve is the all-channel sum, and the black dashed curve marks $\tau^{-1}$.
Insets separate SD and DD contributions, highlighting a temporal dichotomy: SD events cluster at prompt delays ($<1.5$~Gyr), whereas DD mergers dominate at tardy delays ($>1.5$~Gyr).
The SD and DD components connect smoothly near 1.5\,Gyr without any explicit normalization adjustment; the apparent continuity is coincidental rather than imposed.
Overall, the combined DTD is broadly consistent with the commonly adopted $\tau^{-1}$ form.
\textbf{\textit{(b)}}
The full set of star-particle-level SSP DTDs across the 11 \texttt{COMPAS} progenitor metallicities, color-coded from metal-poor (light red) to metal-rich (dark red) as indicated by the color bar.
The main panel shows the total SN~Ia rate, and the insets isolate the SD (top) and DD (bottom) contributions.
The progenitor-metallicity dependence is strongly channel-specific: the SD DTD is nearly invariant, while the DD DTD progressively flattens toward higher metallicity.
For the effect of the $Z=0.001$ outlier, see the text.
This behaviour argues against a single universal DTD kernel, instead favoring an environment-conditioned response function whose effective shape depends on progenitor metallicity.
}
\label{fig:Fig5}
\end{figure*}

\begin{figure}
\includegraphics[width=\columnwidth]{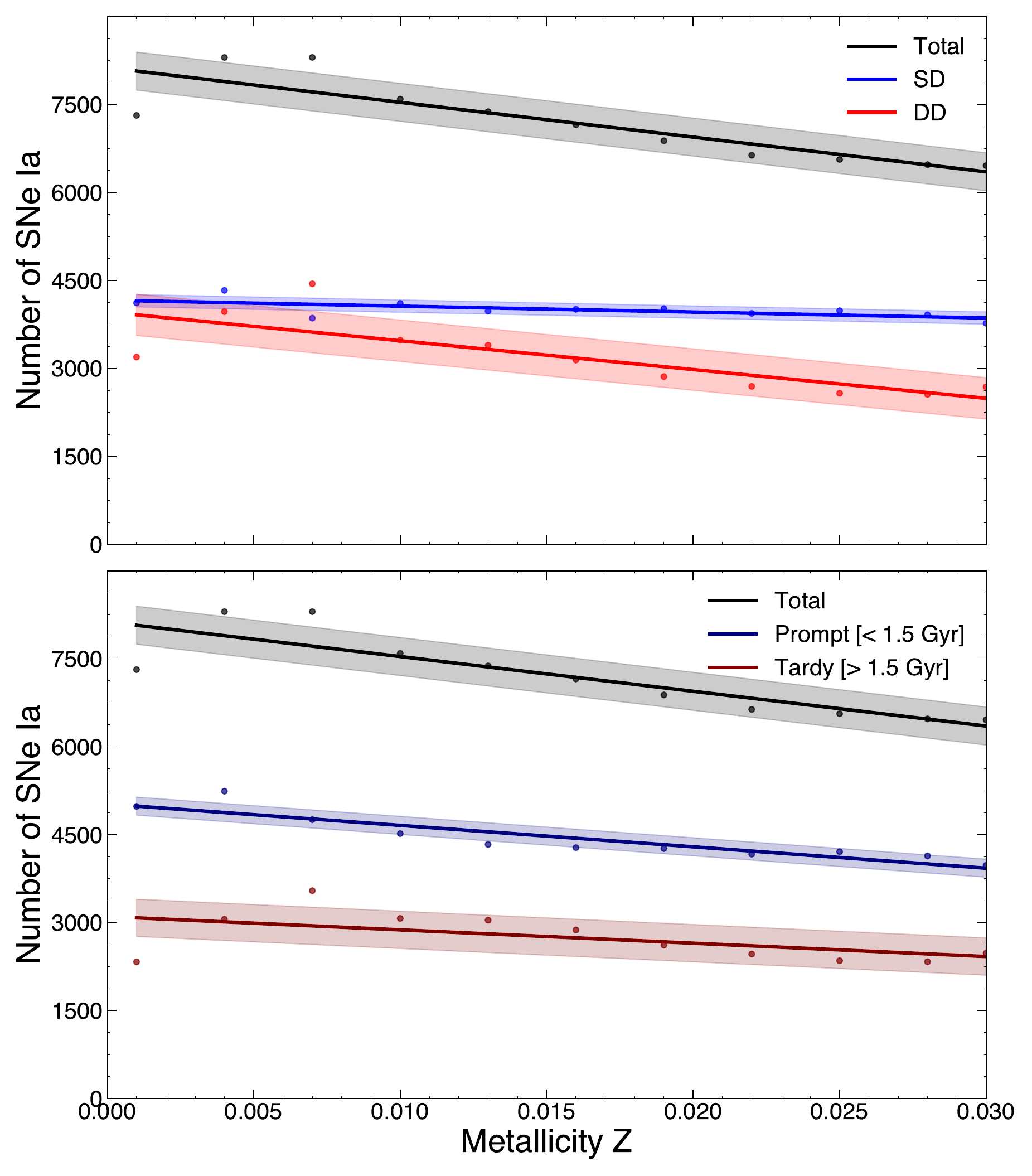}
\caption{
Total SN~Ia productions (yields) over a Hubble time for $4\times10^6\ \Msun$ star particles evaluated at the 11 \texttt{COMPAS} metallicities.
To quantify how metallicity-dependent DTD shapes translate into SN~Ia productions, the yields are computed by integrating the DTDs in Fig.\,\ref{fig:Fig5}($b$) over delay time.
\textbf{\textit{(Upper)}} Channel-separated yields for the SD (blue) and DD (red).
Points show the measurements, solid lines give the best-fit linear trends, and shaded bands mark the $1\sigma$ regression uncertainties.
Consistent with the progenitor-metallicity dependence shown in Fig.\,\ref{fig:Fig5}($b$), the SD yield is nearly metallicity-invariant, while the DD yield declines toward higher metallicity.
\textbf{\textit{(Lower)}} Same as the upper panel, but with yields split into prompt ($\le 1.5$~Gyr) and tardy ($>1.5$~Gyr) populations.
The prompt yield depends only weakly on progenitor metallicity, indicating that the prompt population is a mixture of SD events and early-time DD mergers.
}
\label{fig:Fig6}
\end{figure}

\subsection{SNe~Ia in Individual Star Particles: Intrinsic Delay-time Distributions of Simple Stellar Populations}
\label{sec:2.4}

In TNG100, individual star particles serve as the fundamental building blocks of galactic stellar populations.
Using each star particle as an SSP, we generate binaries and pass them to \texttt{COMPAS} for BPS calculations.
For each particle, \texttt{COMPAS} returns a population of SNe~Ia, each tagged with a delay time (i.e., progenitor age); the resulting ensemble of delays defines an intrinsic star-particle-level SSP DTD, which subsequently assembles into host-specific and, ultimately, cosmic DTDs.
To examine the shape of these DTDs, we perform an \emph{idealized} experiment in which SNe~Ia are generated within individual star particles.
Specifically, we consider 11 star particles with $4\times10^6\ \Msun$ (our reference mass), assign each to one of the 11 \texttt{COMPAS} progenitor metallicity grids\footnote{Throughout this work, we discretize $Z=0.001$--$0.03$ into 1000 logarithmically spaced grid points and evaluate any $Z$-dependent quantity by linear interpolation in $Z$ between adjacent grid points.} spanning $Z=0.001$--$0.03$, and set all SF times to $t=0$.
Because the SFH of each particle is a temporal $\delta$-function burst, the resulting DTDs correspond to the \emph{intrinsic} response functions of the BPS model itself.
They can therefore be interpreted directly, without the need to deconvolve extended and complex SFHs.

Figure\,\ref{fig:Fig5} presents the star-particle-level single-burst SSP DTDs, namely the SN~Ia rate as a function of delay time ($\tau$ in units of Gyr) following a single burst of $1~\Msun$ SSP formation.
Panel ($a$) shows the star-particle-level DTD at $Z = 0.0142$, the solar metallicity adopted in \texttt{COMPAS}.
The inset panels decompose the total DTD into its SD and DD components and highlight a clear temporal dichotomy: SD events are concentrated at prompt delays ($<1.5$~Gyr), whereas DD mergers dominate the tardy regime ($>1.5$~Gyr).
The DD channel is, however, not confined to old populations: compact double-WDs can also produce prompt events, while the broad GW inspiral-time distribution generates tardy events at late times. 
In the main panel, the total SD+DD DTD exhibits a steep prompt rise followed by a long tail, reflecting the time-ordered emergence of distinct progenitor channels.
For the SD channel, rapidly evolving WD+MS systems generate the earliest sharp peak at $< 0.1~{\rm Gyr}$, followed sequentially by contributions from near-$M_{\rm Ch}$ WD+He, WD+GB, and sub-$M_{\rm Ch}$ WD+He systems.
For the DD channel, the near-$M_{\rm Ch}$ and sub-$M_{\rm Ch}$ progenitors peak at $\sim$\,0.4~Gyr and $\sim$\,1.4~Gyr, respectively, and then decline more gradually.
Overall, the SD and DD components join smoothly around 1.5~Gyr without any explicit re-normalization.
The combined DTD is broadly consistent with the commonly adopted analytic form $\tau^{-1}$.

Panel ($b$) generalizes this analysis to the full set of star-particle-level DTDs spanning the 11 \texttt{COMPAS} progenitor metallicities, with the SD and DD components shown separately in the inset panels.
These intrinsic SSP DTDs of individual star particles reveal a clear metallicity dependence that is strongly channel-specific: the SD DTD remains nearly unchanged across metallicity, whereas the DD DTD systematically flattens, with its power-law slope becoming progressively shallower at higher metallicity.
Accordingly, the overall metallicity dependence of the total DTD shape is driven primarily by changes in the DD component rather than by variations in the SD contribution.
This behaviour implies that the SN~Ia DTD is not a single universal response function whose effective shape (i.e., normalization and power-law slope) varies with progenitor metallicity.

Figure\,\ref{fig:Fig6} summarizes the total SN~Ia yield over a Hubble time for $4\times10^6\ \Msun$ star particles evaluated across the 11 \texttt{COMPAS} metallicity grids.
To quantify how metallicity-dependent SSP DTDs map into the SN~Ia production, we integrate each DTD in Fig.\,\ref{fig:Fig5} over the full delay-time domain.
The upper panel shows the resulting yields decomposed by the progenitor channel.
In direct agreement with the channel-specific metallicity trends in Fig.\,\ref{fig:Fig5}($b$), the SD yield remains nearly unchanged with metallicity, whereas the DD yield decreases systematically toward higher metallicity.
The metallicity dependence of the total SN~Ia yield is therefore set primarily by the DD component, not by the SD contribution (see \S\,\ref{sec:6}).
The lower panel recasts the same events by delay time into prompt ($<1.5$~Gyr) and tardy ($>1.5$~Gyr) populations, with 1.5~Gyr approximately marking the crossover between the SD and DD DTDs.
The prompt component comprises both SD events and early-time DD events.
Because it includes the metallicity-sensitive 0.5\,$<$\,$\tau$\,$<$1.5\,Gyr segment of the DD DTD, the prompt component exhibits a stronger metallicity dependence than the SD population alone.
By contrast, the tardy component excludes this most metallicity-sensitive DD interval, 0.5\,$<$\,$\tau$\,$<$1.5\,Gyr, and therefore shows a weaker metallicity dependence than the DD population as a whole.

Taken together, Figures\,\ref{fig:Fig5} and \ref{fig:Fig6} demonstrate that metallicity acts in a channel-dependent manner, primarily modulating the subset of binaries that evolve into DD systems rather than the full SN~Ia population.
For the DD channel, higher metallicity enhances wind-driven mass loss during the giant phases, producing smaller final stellar masses \citep[e.g.,][]{Weiss2009, Doherty2014, Romero2015}.
At the same time, higher metallicity can shift the DD contribution toward longer delay times by altering both the pre-WD evolutionary timescales and the remnant masses of the resulting binaries, which in turn can modify their subsequent GW-driven merger times \citep[e.g.,][]{Hurley2000,Toonen2012,Doherty2014,Romero2015}.
Consequently, these effects reduce the incidence of DD SNe~Ia at higher metallicity \citep{Gandhi2022}.
Our results indicate that the SN~Ia DTD is not a single universal kernel, but rather an environment-conditioned response function, ${\rm DTD}(\tau;Z)$, whose normalization (i.e., total yield) and effective power-law slope vary systematically with progenitor metallicity.
Such metallicity dependence can propagate into host- and redshift-dependent signatures in SN~Ia populations.
We examine the DTD non-universality on galaxy and cosmological scales in \S\,\ref{sec:6}.

\subsection{From Star-particle-level SNe~Ia to Subhalo-scale SN~Ia Populations}
\label{sec:2.5}

With the algorithm that generates star-particle-level SN~Ia populations, we assemble subhalo-scale (i.e., galaxy-scale) SN~Ia populations by summing the contributions from all member star particles in each subhalo.
For a given subhalo $g$, we consider its constituent star particles $i$, characterized by stellar mass at birth $M_{{\rm particle},i}^{\rm birth}$, birth time $t_{{\rm particle},i}^{\rm birth}$, metallicity $Z_{{\rm particle},i}$, and phase-space coordinates ${\bf x}_{{\rm particle},i}(t)$ and ${\bf v}_{{\rm particle},i}(t)$ in the simulation.
Each particle $i$ yields a catalogue of SN~Ia events, indexed by $k$, with delay times $\tau_{{\rm particle},i,k}$ and progenitor-channel labels drawn from the BPS realization appropriate for $Z_{{\rm particle},i}$ (\S\S\,\ref{sec:2.4}).
Accordingly, the assembled subhalo population is built from metallicity-assigned DTDs and naturally reflects metallicity-dependent variations in both SN~Ia productivity and progenitor-channel mix.

For each SN~Ia event $(i,k)$, we compute the explosion time
\begin{equation}
t_{{\rm particle},i,k}^{\rm explosion} \equiv t_{{\rm particle},i}^{\rm birth} + \tau_{{\rm particle},i,k}~,
\end{equation}
which combines the SSP-level delay times with the galaxy’s built-up SFH.
We define the host as the subhalo that contains the parent star particle at $t_{{\rm particle},i,k}^{\rm explosion}$ (rather than at $t_{{\rm particle},i}^{\rm birth}$), so that each SN~Ia is linked to the galaxy within which it actually explodes.
This definition is essential in the presence of galactic interactions (e.g., mergers and satellite accretion) that can relocate stellar populations between formation and explosion.
To preserve mass-consistent normalization across particles, we scale the expected number of SN~Ia events (i.e., the Poisson expectation value) for each particle $i$ by the mass ratio given by
\begin{equation}
W_{{\rm particle},i} \equiv \frac{M_{{\rm particle},i}^{\rm birth}}{M_{\rm ref}}~,
\end{equation}
where the star-particle-level SN~Ia populations are normalized to a reference SSP mass $M_{\rm ref}$ ($4\times10^6~\Msun$ in our case).

We assign host properties (e.g., $T_*$, $Z_*$, $M_*$, ${\rm SFR}$, and ${\rm sSFR}$) at the explosion epoch to each event.
We also retain kinematic and spatial information by inheriting the parent particle phase-space coordinates at $t_{{\rm particle},i,k}^{\rm explosion}$, ${\bf x}_{{\rm particle},i}(t_{{\rm particle},i,k}^{\rm explosion})$ and ${\bf v}_{{\rm particle},i}(t_{{\rm particle},i,k}^{\rm explosion})$, enabling direct galaxy-scale analyses of SN~Ia locations and velocities.
In this construction, each subhalo-scale SN~Ia population is a star-particle-mass-weighted superposition of metallicity-dependent SSP DTDs, encoding the subhalo’s assembled SF and enrichment history.

\subsection{All-sky Comoving-volume Realization to Build Cosmological SN~Ia Samples}
\label{sec:2.6}

In this section, we describe how we embed the subhalo (galaxy) populations in a cosmological setting and construct a single, internally consistent SN~Ia catalogue.
To map the discrete snapshot outputs of the TNG100 simulation\footnote{The TNG suite provides two snapshot types---20 ``full'' and 80 ``mini''---spanning $z$\,=\,20.05 to 0.0.
Our fiducial analysis exploits 83 snapshots (13 full and 70 mini) over $0 \leq z \leq 5$.
For reference, the $0 \leq z \leq 3$ and $0 \leq z \leq 0.1$ subsamples use 75 snapshots (11 full and 64 mini) and 9 snapshots (2 full and 7 mini), respectively.} onto the real Universe, we carry out an explicit all-sky comoving-volume realization.
Specifically, the number of simulated galaxies and SNe~Ia in each redshift shell is set by the corresponding differential comoving volume element, so that the resulting mock galaxy/SN catalogue is defined on the same volumetric basis as observations.
We adopt an idealized, isotropic full-sky survey and neglect directional and kinematic effects, as these are irrelevant to the count-based volume bookkeeping considered here.
In this sense, our construction represents a minimal mock light cone: it maps comoving volume elements to discrete simulation snapshots, without modeling cube rotation, observer motion, or redshift-space distortions (see, e.g., \citealt{Blaizot2005, Merson2013} for related mock light-cone concepts).

\smallskip
\noindent\textit{(i) Comoving distances and volume elements in FLRW cosmology.} --- For a homogeneous and isotropic Friedmann--Lema\^{i}tre--Robertson--Walker cosmology, the Hubble expansion rate is
\begin{equation}
\begin{aligned}
H(z) = H_0\,\left[\Omega_{\rm m}(1+z)^3 + \Omega_{\rm k}(1+z)^2 + \Omega_\Lambda\right]^{1/2},
\end{aligned}
\label{eq:Hz}
\end{equation}
where $H_0$, $\Omega_{\rm m}$, $\Omega_{\rm k}$, and $\Omega_\Lambda$ are the Hubble parameter at $z=0$ (Hubble constant), matter density parameter, curvature density parameter, dark-energy density parameter (cosmological constant), respectively.
The line-of-sight comoving distance is
\begin{equation}
\chi(z) = c \int_{0}^{z}\frac{{\rm d}z'}{H(z')}~.
\label{eq:chi_of_z}
\end{equation}
Assuming full-sky coverage, the differential comoving volume element is
\begin{equation}
\frac{{\rm d}V_{\rm c}}{{\rm d}z} = 4\pi\,D_{\rm M}^2(z)\,\frac{c}{H(z)}~,
\label{eq:dVdz}
\end{equation}
where, in a spatially flat universe ($\Omega_{\rm k}=0$), the transverse (proper-motion) comoving distance is given by $D_{\rm M}(z)\equiv\chi(z)$.
Equivalently, the comoving volume between two comoving radii $\chi_0$ and $\chi_1$ is
\begin{equation}
V_{\rm shell}(\chi_0,\chi_1)=\int_{\chi_0}^{\chi_1}4\pi \chi^2\,{\rm d}\chi
=\frac{4\pi}{3}\left(\chi_1^3-\chi_0^3\right).
\label{eq:Vshell}
\end{equation}
These relations define the full-sky comoving shell volume that must be represented when using a single periodic box as the fundamental volume unit.

\smallskip
\noindent\textit{(ii) Territories of snapshots.} --- 
We let snapshot $i$ correspond to a representative redshift $z_i$ and comoving distance $\chi_i\equiv \chi(z_i)$. 
We assign to snapshot $i$ the radial territory bounded by the midpoints of adjacent snapshots in comoving distance,
\begin{equation}
\begin{gathered}
\chi_{i-1/2}\equiv \frac{1}{2}\left(\chi_{i-1}+\chi_i\right)~~\&~~\chi_{i+1/2}\equiv \frac{1}{2}\left(\chi_i+\chi_{i+1}\right),
\end{gathered}
\label{eq:midpoint_bounds}
\end{equation}
thereby partitioning the light-cone volume into non-overlapping comoving radial shells. 
Within each territory, we treat the snapshot as the discrete representation of galaxy/SN properties across that comoving interval.

\smallskip
\noindent\textit{(iii) Box-length tiling and weights.} --- 
We denote the comoving side length and comoving volume of the simulation cube by $L_{\rm box}$ and $V_{\rm box}=L_{\rm box}^3$. Within the territory $[\chi_{i-1/2},\,\chi_{i+1/2}]$, we subdivide the radial interval into consecutive segments of length $L_{\rm box}$, starting at $\chi_{i-1/2}$ and stepping outward in increments of $L_{\rm box}$. 
If the remaining thickness is $<L_{\rm box}$---a common situation at low redshift where adjacent snapshots are closely spaced in $\chi$---the last segment is \emph{shortened} so that its outer boundary coincides exactly with $\chi_{i+1/2}$. 
This construction yields a set of tiles $t$ with bounds $[\chi_{t,0},\,\chi_{t,1}]$, including partial tiles with $\chi_{t,1}-\chi_{t,0}<L_{\rm box}$.
For each tile, we assign a shell-volume weight equal to the shell volume in units of the box volume,
\begin{equation}
w_{i,t} \equiv \frac{V_{\rm shell}(\chi_{t,0},\chi_{t,1})}{V_{\rm box}}
= \frac{4\pi}{3} \frac{\left(\chi_{t,1}^3-\chi_{t,0}^3\right)}{V_{\rm box}}~,
\label{eq:tile_weight}
\end{equation}
which represents the effective number of box volumes required to fill that full-sky comoving shell.
Number counts and any number-weighted mean quantities are then obtained by summing over galaxies (or SN events) in each tile, weighted by $w_{i,t}$:
\begin{equation}
\begin{aligned}
& N = \sum_i \sum_{t\in i} w_{i,t}\, N_{i,t}~~\&~~\langle X \rangle = \frac{\sum_i \sum_{t\in i} w_{i,t}\sum_{j\in(i,t)} X_j}{N},
\end{aligned}
\end{equation}
where $N_{i,t}$ is the number of galaxies (or SN events) in tile $t$ of snapshot $i$, and $X_j$ denotes the corresponding property of galaxy $j$ (or SN event $j$).

\smallskip
\noindent\textit{(iv) Redshift smearing within tiles.} --- 
When an explicit redshift coordinate is required (e.g., for panels with redshift on the $x$-axis), we smear each tile’s contribution by sampling uniformly in comoving volume. 
Specifically, we draw $u \simeq \mathcal{U}(\chi_{t,0}^3, \chi_{t,1}^3)$ and then set
\begin{equation}
\begin{aligned}
& \chi_{\rm sm} = u^{1/3}~~\&~~z_{\rm sm} = \chi^{-1}(\chi_{\rm sm})~.
\end{aligned}
\label{eq:z_smearing}
\end{equation}
This procedure yields a redshift distribution consistent with the shell-volume weighting implied by Eq.~\ref{eq:Vshell}.
Each smeared contribution retains the tile weight $w_{i,t}$ from Eq.~\ref{eq:tile_weight}. 

\subsection{Applying an Observational Time Window to Mock Surveys}
\label{sec:2.7}

\subsubsection{Definition of source-frame and observer-frame catalogues}
\label{sec:2.7.1}

Although our cosmological simulation $+$ BPS framework produces SNe~Ia continuously throughout cosmic time, comparison with observations requires constructing a mock survey with a fixed observational time window, $\Delta t$, designed to mimic the finite temporal baselines of real SN surveys.
Because of cosmological time dilation, an interval measured in the observer frame corresponds to a shorter interval in the source frame by a factor of $1/(1+z)$.
Each SN~Ia event is specified by a source-frame explosion time $t_{\rm explosion}=t_0-\Delta t_{\rm lb}(z)$, where $t_0$ is the present epoch and $\Delta t_{\rm lb}(z)$ is the lookback time at $z$.
Photons emitted toward the observer propagate along the past light cone and are received at an observer-frame arrival time $t_{\rm arrival}$; events observed at the present epoch satisfy $t_{\rm arrival}=t_0$.
The mock-observed catalogue is therefore defined by the observer-frame interval $t_0-\Delta t \le t_{\rm arrival} \le t_0$.
At fixed $z$, this is equivalent to selecting explosions in the source-frame interval
$t_0-\Delta t_{\rm lb}(z)-\Delta t/(1+z) \le t_{\rm explosion} \le t_0-\Delta t_{\rm lb}(z)$.
Applying the $1/(1+z)$ factor therefore yields the observer-frame catalogue, whereas omitting it yields the corresponding source-frame catalogue.

In this paper, we adopt the \emph{source frame} as our default reference frame in order to present the intrinsic predictions of the model; accordingly, all figures and discussion are based on source-frame quantities.
Operationally, the standard catalogue is constructed without applying the $1/(1+z)$ contraction of the observational time window, so our default results correspond to quantities corrected for cosmological time dilation.
This follows standard observational practice, in which SN-rate-related inference products are generally reported in the source frame after applying the time-dilation correction.
Also, the host-galaxy demographics are presented in the source frame in \S\,\ref{sec:4} and Appendix\,\ref{appendix:D}.
Given that observed host-demographic samples are often shown without an explicit time-dilation correction, we also provide the model host demographics in the observer frame in Appendix\,\ref{appendix:E}, which are more directly comparable to observations.
Figure\,\ref{fig:corner_host_observerframe} shows that the host distribution changes only marginally between the source and observer frames.
Our full catalogue of simulated SNe~Ia (Table~\ref{tab:snpm_catalogue}) lists the quantities required to transform the intrinsic source frame into the observer frame..

\begin{figure}
\includegraphics[width=0.95\columnwidth]{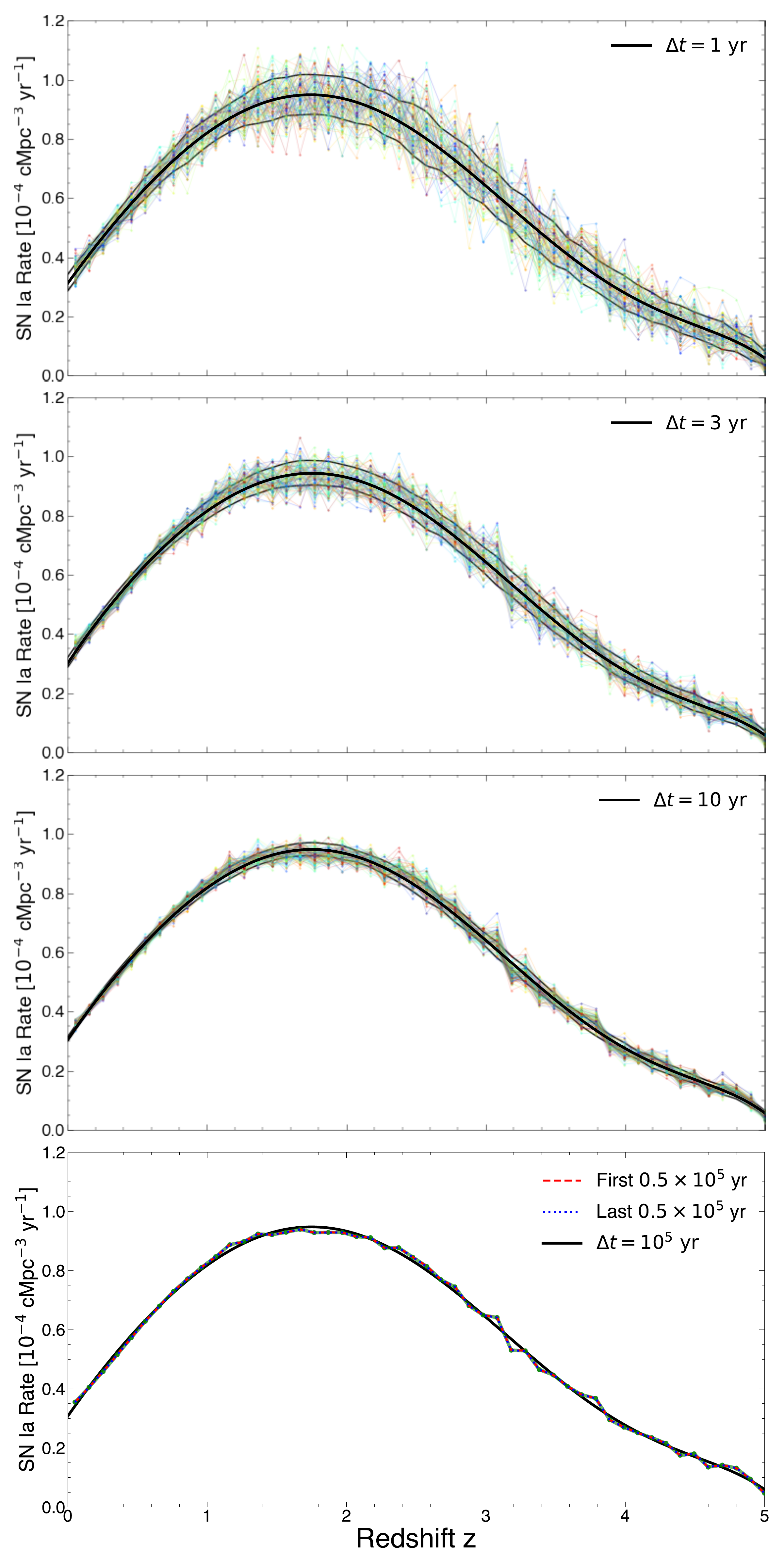}
\caption{
Cosmic volume-normalized SN~Ia rate ($R_{\rm vol}$) as a function of redshift for different observational time windows ($\Delta t$). 
To quantify stochasticity, we consider four observational time windows, $\Delta t=$ 1, 3, 10, and $10^5$~yr (top to bottom), and partition $0 \le z \le 5$ into 50 bins.
For each $\Delta t$, we generate 100 independent realizations, plotted as colored points connected by lines.
In each redshift bin, we compute the mean and 1$\sigma$ scatter across the 100 runs.
Thick curve is a sixth-order polynomial fit to the 50-bin mean points, with the order chosen for consistency with Fig.\,\ref{fig:Cosmic_crossover}.
Thin curves bracketing the fit indicate the bin-wise 1$\sigma$ uncertainty.
The upper three panels show substantial run-to-run scatter, demonstrating that short $\Delta t$ intervals suffer stochastic noise. 
In contrast, the bottom panel shows our fiducial model of $\Delta t=10^5$~yr, which yields a statistically ideal trend. 
Because the $10^5$~yr model has, by construction, only a single realization (green points connected by lines), any residual variation reflects model fluctuations (systematic uncertainty) rather than random uncertainty. 
As a robustness check on the adopted window length, we split the $10^{5}$~yr interval into two equal halves and compute $R_{\rm vol}$ separately for the first and last $0.5\times10^{5}$~yr (red dashed and blue dotted lines, respectively). 
The two curves are indistinguishable, confirming that our choice of $\Delta t=10^{5}$~yr does not affect our results.
}
\label{fig:Fig7}
\end{figure}

\subsubsection{Motivation for the fiducial choice $\Delta t = 10^5$~yr}
\label{sec:2.7.2}

Figure\,\ref{fig:Fig7} shows the cosmic volume-normalized SN~Ia rate ($R_{\rm vol}$; the number of SNe~Ia per unit comoving volume per unit time) as a function of redshift for different observational time windows.
The upper three panels demonstrate the strong dependence of stochastic variance on $\Delta t$.
The $\Delta t =$ 1 and 3~yr mock campaigns contain relatively few explosions and therefore exhibit substantial run-to-run fluctuations.
For $\Delta t = 10$~yr, this sampling noise is already greatly suppressed.
Because our analysis divides the sample into many bins by progenitor type, host-galaxy property, and redshift, we require a substantially larger number of SNe~Ia and hosts to avoid sparse statistics that obscure the underlying physical trends.
We therefore adopt a long observational time window, $\Delta t = 10^{5}$~yr, and count all events exploding within this interval to maximize statistical power.
Operationally, a single realization with $\Delta t = 10^{5}$~yr is equivalent to $>$\,20,000 repeated independent realizations with the $\sim$3--5~yr baselines of modern time-domain SN~Ia surveys\footnote{E.g., SDSS-II \citep{Frieman2008SDSS2SN} obtained repeated imaging during three Sep--Nov seasons from 2005 to 2007; SNLS \citep{Astier2006SNLS} was a five-year rolling survey operating from 2003 to 2008; PS1 \citep{Rest2014PS1,Chambers2016PS1Surveys} formally began science operations in May 2010, with the first 1.5 years of science imaging spanning roughly 2010 Feb to 2011 Jun, and the Medium Deep Survey continuing until early 2014; DES-SN \citep{Smith2020DESSN} ran for five Aug--Feb seasons from 2013 to 2018; and ZTF \citep{Bellm2019ZTF,Dhawan2022ZTFSNIa} began public operations in Mar 2018, surveying the visible northern sky in an untargeted rolling mode with a typical three-night cadence during Phase~I (from 2018 Mar to 2020 Nov).}, while achieving statistically stable SN~Ia and host counts at far lower computational cost.
The bottom panel presents this fiducial $\Delta t=10^5$~yr model, which exhibits a statistically near-ideal trend.
Because the $10^5$~yr model consists, by construction, of only a single realization, any residual variation should be interpreted as model-driven structure (systematic uncertainty) rather than sampling noise.

Although $\Delta t = 10^{5}$~yr is much longer than the few-year baselines of current SN surveys, it remains negligible relative to the timescales of stellar, galactic, and cosmic evolution, and thus effectively samples an instantaneous snapshot feature of SNe~Ia and their host galaxies.
A possible concern is that binary orbital evolution can proceed on timescales of tens to thousands of years (Fig.\,\ref{fig:Fig4}), such that integrating over $10^{5}$~yr could bias inferred progenitor properties and hence the SN~Ia statistics.
We test this explicitly in the bottom panel of Fig.\,\ref{fig:Fig7} by splitting the $10^{5}$~yr window into two equal halves and recomputing the SN~Ia rates separately for the first and last $0.5\times10^{5}$~yr.
The two estimates are essentially indistinguishable, indicating that the long-window choice introduces no measurable bias while substantially improving statistical power.

\subsubsection{SN host fractions and per-host SN~Ia counts as functions of $\Delta t$}
\label{sec:2.7.3}

\begin{figure}
\includegraphics[width=0.93\columnwidth]{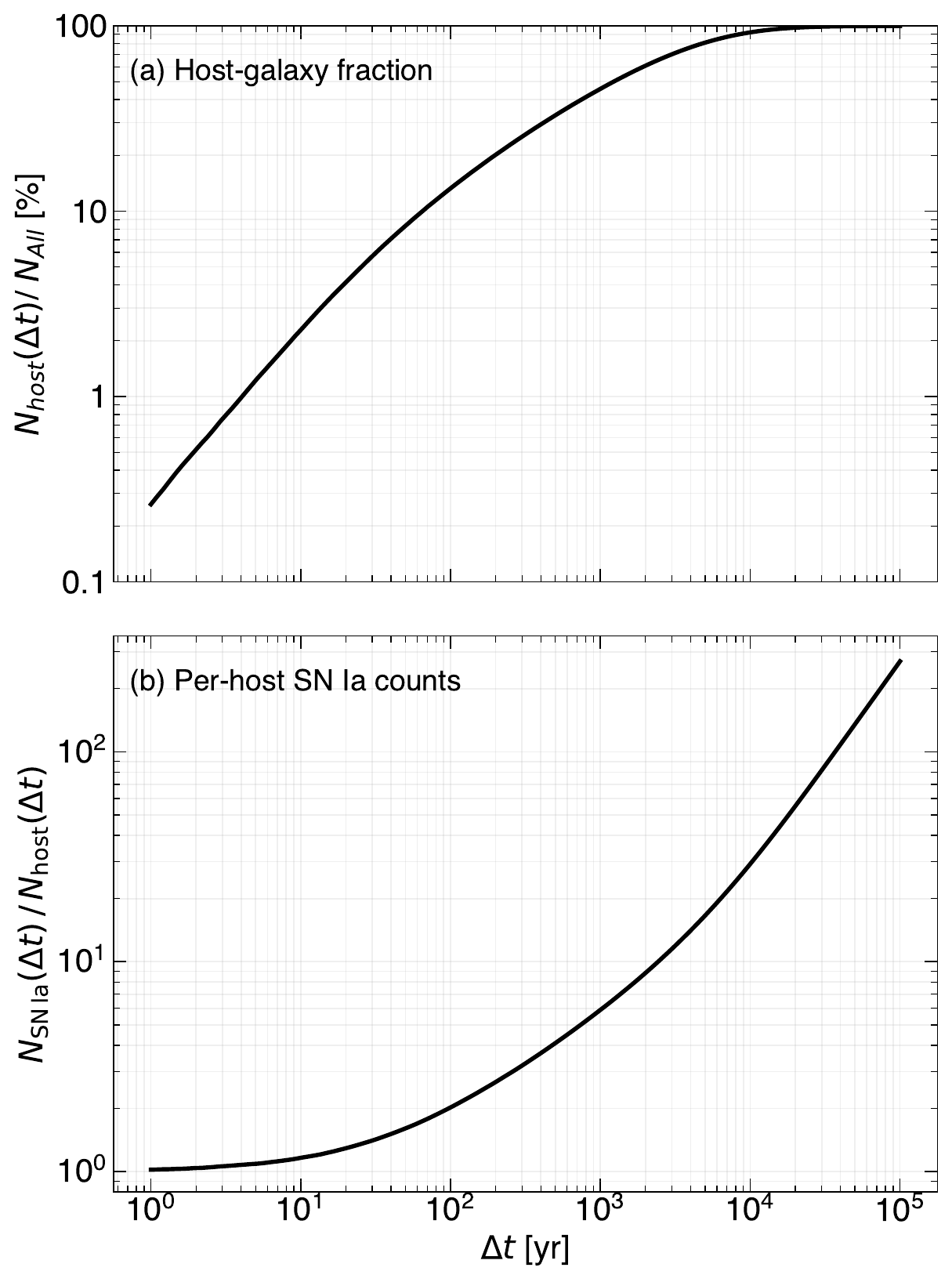}
\caption{
Host-galaxy fraction and per-host SN~Ia counts as functions of the observational time window, $\Delta t$, for all galaxies with $M_* \geq 10^{8.0}~\Msun$ over $0 \leq z \leq 5$.
\textbf{\textit{(a)}} Host fraction, $N_{\rm host}(\Delta t)/N_{\rm All}~[\%]$, versus $\Delta t$.
For $\Delta t = 10^5$~yr, essentially every galaxy (99.74\,\%) hosts at least one SN~Ia, whereas for $\Delta t = 3$~yr---comparable to the baselines of modern time-domain surveys---the fraction drops to $\sim$\,0.8\,\%.
\textbf{\textit{(b)}} Mean number of SNe~Ia per host, $N_{\rm SNIa}(\Delta t)/N_{\rm host}(\Delta t)$, versus $\Delta t$.
Hosts contain $\sim$\,270 events on average for $\Delta t = 10^5$~yr, but approach the single-event-per-host regime at $\Delta t = 3$~yr.
}
\label{fig:Hostfraction}
\end{figure}

Figure\,\ref{fig:Hostfraction} provides basic intuition for the role of $\Delta t$ by showing the SN~Ia host fraction and per-host SN~Ia counts for the full galaxy population over $0 \leq z \leq 5$.
Panel ($a$) shows the host fraction as a function of $\Delta t$.
Although the exact values vary modestly with the adopted galaxy low-mass cut, the statistics for our fiducial selection, $\logM > 8.0$, are as follows.
For $\Delta t = 10^5$~yr, essentially every galaxy (99.74\,\%) hosts at least one SN~Ia.
Reducing the window to $\Delta t =$ 100 yr and 10 yr lowers the host fraction to $\sim$\,13.3\,\% and $\sim$\,2.3\,\%, respectively, and for $\Delta t = 3$~yr---comparable to the baselines of modern surveys---the fraction further declines to $\sim$\,0.8\,\%.
Panel ($b$) shows the mean number of SNe~Ia per host as a function of $\Delta t$.
For $\Delta t = 10^5$~yr, hosts contain $\sim$\,270 SNe~Ia on average; for $\Delta t = 100$~yr, the mean falls to $\sim$\,2; and for $\Delta t = 3$ yr, it approaches unity, i.e., the single-event-per-host regime.

It is worth emphasizing that the host-demographic statistics considered here are defined in an SN~Ia event-weighted sense.
Within the observational time window of $10^{5}$\,$\mathrm{yr}$, we identify every SN~Ia and its host, recording each SN--host pair as a separate catalogue entry.
A galaxy that produces $N$ SNe over this interval therefore appears $N$ times, so the resulting host-demography is explicitly {\it event-weighted}; repeated appearances of the same host simply reflect its higher intrinsic SN~Ia rate.
Observed ``host demographics'' are likewise constructed on a one-host-per-SN basis, although only a tiny fraction of systems produce multiple events.
In the idealized limit of a long, unbiased survey, the empirical host distribution therefore converges to the same SN-rate--weighted distribution defined by our $10^{5}$\,$\mathrm{yr}$ construction, differing only by Poisson noise.
Our approach can thus be viewed as predicting the asymptotic host-galaxy demography that would be recovered by a survey with effectively unlimited temporal coverage.

For additional context, we quantify the sibling-SN statistics for a real-survey-like window of $\Delta t = 3$~yr.
A host fraction of $\sim$\,0.8\,\% corresponds to $\sim$\,$3.6\times10^8$ hosts, of which 94.7\,\% host one SN, 4.7\,\% host two sibling SNe, 0.5\,\% host three, and 0.2\,\% host four or more sibling SNe.
For a more direct comparison with observations, the $z \leq 0.3$ sample (matched to the ZTF redshift range) yields 97.6\,\% single-event hosts, 2.4\,\% two-event hosts, 0.1\,\% three-event hosts, and no hosts with four or more sibling SNe.
Although SN~Ia multiplicity is survey-dependent---varying with duration, cadence, targeting strategy, magnitude limit, redshift coverage, and host-matching quality---the empirical distribution\footnote{For the Pantheon+ subset with released host-galaxy names, we count 846 unique hosts, with multiplicities of: 1-event hosts: 827 (97.8\,\%), 2-event hosts: 16 (1.9\,\%), 3-event hosts: 3 (0.4\,\%), and 4-or-more-event hosts: 0.
For the ZTF subset with reported siblings, we count 3628 unique hosts in the total ($z \leq 0.3$) sample, with multiplicities of: 1-event hosts: 3603 (99.3\,\%), 2-event hosts: 25 (0.7\,\%), and 3-or-more-event hosts: 0.
We also count 945 unique hosts in the volume-limited ($z \leq 0.06$) sample, with multiplicities of: 1-event hosts: 939 (99.4\,\%), 2-event hosts: 6 (0.6\,\%), and 3-or-more-event hosts: 0.} is approximately: 1-event hosts comprise $\sim$\,98--99\,\%, 2-event hosts $\sim$\,0.5--2\,\%, 3-event hosts $\sim$\,0--0.4\,\%, and 4-or-more-event hosts 0\,\%.
In this sense, our prediction is broadly compatible with the observed sibling-SN incidence.

\subsection{Simulation--Survey Consistency in the SN~Ia Rate Definition}
\label{sec:2.8}

SN~Ia birth-rate calculations involve two distinct timescales: the cosmic-scale time axis used to define DTDs and the finite observational baseline of real and mock surveys.
We keep this distinction explicit by expressing DTDs in units of [Gyr$^{-1}$] and survey-duration-normalized rates in units of [yr$^{-1}$] throughout.
On the one hand, DTDs are defined as the SN~Ia rate as a function of delay time (in units of Gyrs) following a single burst of $1~\Msun$ SF. 
Hence, the DTD, as a fundamental response function, is SN~Ia counts per delay time on cosmic scales, i.e., SN~Ia rate [${\rm M}_{\odot}^{-1}~{\rm Gyr}^{-1}$].
We note that we apply this Gyr unit only to SN birth rates; for SFRs we follow standard usage, i.e., SFR\,[$\Msun~ {\rm yr}^{-1}$] and sSFR\,[${\rm yr}^{-1}$].

On the other hand, observational SN~Ia rates are defined by event counts normalized by the survey duration (in units of years).
Accordingly, dividing our simulated event counts by $10^{5}$~yr yields {\it intrinsic} rates that are directly comparable to bias-corrected observational measurements.
We use three conventional forms of intrinsic SN~Ia rates that differ only in how the same underlying event catalogue is normalized.
First, the \textit{galaxy--time} normalization defines the \textit{galaxy}--normalized rate, $R_{\rm gal}$ [galaxy$^{-1}$\,yr$^{-1}$]: the total number of SNe~Ia divided by the number of host galaxies and by the survey duration (or $\Delta t = 10^{5}$~yr in our model).
Second, the \textit{host mass--time} normalization defines the \textit{mass}--normalized rate, $R_{\rm mass}$ [$\Msun^{-1}$\,yr$^{-1}$]: the total number of SNe~Ia divided by host stellar mass and by the survey duration (or $\Delta t = 10^{5}$~yr in our model).
This $R_{\rm mass}$ is what is usually termed the ``specific SN~Ia rate'' and ``SN~Ia Unit per Mass (SNuM; in units of SNe per 100~yr per $10^{10}\,\Msun$).
Lastly, the \textit{volume--time} normalization defines the \textit{volume}--normalized rate, $R_{\rm vol}$ [cMpc$^{-3}$\,yr$^{-1}$]: the total number of SNe~Ia divided by the comoving volume and by the survey duration (or $\Delta t = 10^{5}$~yr in our model).
This $R_{\rm vol}$ is what is usually termed the ``cosmic SN~Ia rate'' (as used in \S\S\,\ref{sec:2.7} and \S\,\ref{sec:8}--\ref{sec:9}).

\begin{table*}
\centering
\caption{Full catalogue of our simulated SNe~Ia (see Appendix\,\ref{appendix:D} (Figs.\,\ref{fig:corner_all_z0to5}\,--\,\ref{fig:corner_host_z0}) and Appendix\,\ref{appendix:E} (Fig.\,\ref{fig:corner_host_observerframe})). 
The table includes one representative example for each progenitor channel considered in this study: SD (WD+MS, WD+GB, WD+He, and sub-$M_{\rm Ch}$ SD) and DD (WD+WD and sub-$M_{\rm Ch}$ DD) channels. 
}
\label{tab:snpm_catalogue}
\footnotesize
\setlength{\tabcolsep}{3.78pt}
\begin{tabular}{lccccccccccc}
\midrule
\midrule
\multicolumn{2}{l}{\textbf{IDs \& Properties of SNe~Ia \& their Hosts}} &   &   &   &   &   &   &   &   &   &  \\
\midrule
SN~Ia ID & 12533746 & 7521145 & 1920724 & 5143970 & 1055982 & 8797649 & \hspace*{0.8em}$\boldsymbol{\cdots}$$\boldsymbol{\cdots}$ \\
\midrule
Progenitor binary configuration &      &      &      &      &      &      &   \\
($M_1$\,[$\Msun$], $M_2$\,[$\Msun$], $P_{\rm orb}$\,[day]) & \scriptsize (2.83, 2.24, $10^{2.6}$) & \scriptsize (9.05, 5.08, $10^{1.4}$) & \scriptsize (8.77, 5.25, $10^{0.8}$) & \scriptsize (4.71, 2.12, $10^{4.7}$) & \scriptsize (5.47, 3.24, $10^{0.6}$) & \scriptsize (3.43, 1.40, $10^{3.1}$) & \hspace*{0.8em}$\boldsymbol{\cdots}$$\boldsymbol{\cdots}$ \\
Progenitor explosion channel & WD+MS & WD+GB & WD+He & Sub-$M_{\rm ch}$ SD & WD+WD & Sub-$M_{\rm ch}$ DD & \hspace*{0.8em}$\boldsymbol{\cdots}$$\boldsymbol{\cdots}$ \\
Progenitor age [Gyr] & 0.703 & 0.125 & 0.095 & 1.415 & 6.451 & 2.866 & \hspace*{0.8em}$\boldsymbol{\cdots}$$\boldsymbol{\cdots}$ \\
Progenitor metallicity ($\log Z[Z_{\odot}]$) & -0.66 & 0.23 & 0.31 & -0.29 & 0.48 & -0.27 & \hspace*{0.8em}$\boldsymbol{\cdots}$$\boldsymbol{\cdots}$ \\
Progenitor $t_{\rm explosion}$ [Gyr] & 4.248 & 2.138 & 10.740 & 3.224 & 13.640 & 7.774 & \hspace*{0.8em}$\boldsymbol{\cdots}$$\boldsymbol{\cdots}$ \\
Progenitor $\Delta t_{\rm lb}$ [Gyr] & 9.550 & 11.659 & 3.057 & 10.573 & 0.157 & 6.023 & \hspace*{0.8em}$\boldsymbol{\cdots}$$\boldsymbol{\cdots}$ \\
$t_{\rm 0}-t_{\rm arrival}^{\dagger}$\,[yr] within $\Delta t=10^5$\,yr & 17711 & 94127 & 4417 & 4044 & 31938 & 90590 & \hspace*{0.8em}$\boldsymbol{\cdots}$$\boldsymbol{\cdots}$ \\
\midrule
Host ID & \scriptsize 40000555225 & \scriptsize 25000156961 & \scriptsize 80000259669 & \scriptsize 33000106818 & \scriptsize 98000258416 & \scriptsize 62000565632 & \hspace*{0.8em}$\boldsymbol{\cdots}$$\boldsymbol{\cdots}$ \\
Host redshift & 1.512 & 3.013 & 0.252 & 2.038 & 0.011 & 0.622 & \hspace*{0.8em}$\boldsymbol{\cdots}$$\boldsymbol{\cdots}$ \\
Host age [Gyr] & 1.49 & 0.38 & 3.23 & 0.85 & 9.05 & 3.60 & \hspace*{0.8em}$\boldsymbol{\cdots}$$\boldsymbol{\cdots}$ \\
Host metallicity ($\log Z_*[Z_{\odot}]$) & -0.49 & -0.02 & 0.29 & 0.05 & 0.23 & -0.19 & \hspace*{0.8em}$\boldsymbol{\cdots}$$\boldsymbol{\cdots}$ \\
Host mass ($\log M_*[{\rm M}_{\odot}]$) & 8.34 & 9.42 & 10.37 & 10.03 & 11.24 & 8.62 & \hspace*{0.8em}$\boldsymbol{\cdots}$$\boldsymbol{\cdots}$ \\
Host sSFR ($\log{\rm sSFR}[{\rm yr^{-1}}]$) & -9.31 & -8.80 & -9.59 & -9.38 & -12.43 & -9.80 & \hspace*{0.8em}$\boldsymbol{\cdots}$$\boldsymbol{\cdots}$ \\
Distance from host center$^{\ddagger}$  [Kpc] & 1.60 & 0.50 & 0.49 & 2.07 & 29.58 & 1.05 & \hspace*{0.8em}$\boldsymbol{\cdots}$$\boldsymbol{\cdots}$ \\
\midrule
TNG100 snapshot No. & 40 & 25 & 80 & 33 & 98 & 62 & \hspace*{0.8em}$\boldsymbol{\cdots}$$\boldsymbol{\cdots}$ \\
TNG100 tile index & 65 & 94 & 19 & 79 & 1 & 37 & \hspace*{0.8em}$\boldsymbol{\cdots}$$\boldsymbol{\cdots}$ \\
TNG100 tile redshift range $(z_0, z_1)$ & \scriptsize (1.454, 1.513) & \scriptsize (2.954, 3.068) & \scriptsize (0.250, 0.265) & \scriptsize (2.024, 2.049) & \scriptsize (0.005, 0.015) & \scriptsize (0.610, 0.630) & \hspace*{0.8em}$\boldsymbol{\cdots}$$\boldsymbol{\cdots}$ \\
Shell-volume weight ($w_{i,t}$)$^{\ddagger\dagger}$ & 20279.19 & 43582.77 & 615.81 & 9747.96 & 0.86 & 3154.36 & \hspace*{0.8em}$\boldsymbol{\cdots}$$\boldsymbol{\cdots}$ \\
\midrule
\end{tabular}
\begin{minipage}{\textwidth}
\footnotesize
\textit{Note.}
$^{\dagger}$ $t_{\rm arrival} = t_{\rm explosion} + \Delta t_{\rm lb}$
$^{\ddagger}$ Galactocentric distance of each SN~Ia-hosting star particle (see \S\,\ref{sec:3}).
$^{\ddagger\dagger}$ The value $w_{i,t}$ denotes the shell-volume weight of tile $t$ in snapshot $i$ (see \S\S\,\ref{sec:2.6} and Eq.~\ref{eq:tile_weight}).
(This table is available in its entirety in machine-readable form in the electronic edition.)
\end{minipage}
\end{table*}

\subsection{Observation-driven Redshift Baselines for Global-volume and Local-volume Samples}
\label{sec:2.9}

Our full catalogue spans $0 \leq z \leq 5$, within which we construct samples for (i) all galaxies and (ii) SNe~Ia and their host galaxies.
The interval $0 \leq z \leq 5$ has both practical and symbolic significance: it effectively includes nearly the full galaxy population relevant to this work, except for a limited number of the earliest systems at the highest redshifts \citep[e.g.,][]{Madau2014,Behroozi2019,Forster-Schreiber2020,Robertson2022}.
Appendix\,\ref{appendix:D} presents the corresponding all-galaxy, SN~Ia, and SN-host-galaxy distributions over $0 \leq z \leq 5$ (Figures\,\ref{fig:corner_all_z0to5} and \ref{fig:corner_host_z0to5}).
For the fiducial analyses in the main text, however, we focus on two explicitly defined subsets: the global-volume and local-volume galaxy samples (hereafter ``global'' and ``local'' samples), spanning $0 \leq z \leq 3$ and $0 \leq z \leq 0.1$, respectively (see \S\S\,\ref{sec:2.9}).
For SN~Ia cosmology, the ``global-volume'' subset is taken to represent the full SN~Ia and host-galaxy population, while the ``local-volume'' subset provides the nearby reference frame for interpreting redshift evolution in the global-volume sample (see \S\S\,\ref{sec:2.9}).

We first define the global cosmological sample as all galaxies, SNe~Ia, and their hosts within $0 \leq z \leq 3$.
Although any finite redshift boundary is somewhat arbitrary, $0 \leq z \leq 3$ effectively captures the range over which SNe~Ia are observed in substantial numbers and are expected to dominate the cosmic SN~Ia budget.
Observationally, deep \textit{HST} surveys (CLASH and CANDELS) have measured SN~Ia rates to $z\simeq 2.4$--2.5, and \textit{JWST} has begun to identify normal SNe~Ia at $z\simeq 2.9$ \citep[e.g.,][]{Graur2014, Rodney2014,Vinko2025}.
Theoretically, our $0 \leq z \leq 5$ results in Appendix\,\ref{appendix:D} further support this fiducial choice: in the redshift--sSFR and mass--sSFR planes of host galaxies (Figure\,\ref{fig:corner_all_z0to5}), $z \simeq 3$ marks the onset of SN~Ia hosting in massive ($M_* \simeq 10^{11}~\Msun$; the most massive group at $z \simeq 3$), intermediate-SF galaxies (${\rm sSFR} \simeq 10^{-10}~{\rm yr}^{-1}$, about 1.5 dex below the star-forming main sequence at $z \simeq 3$), which are likely (ancestors of) early-type galaxies.
In other words, $z \simeq 3$ is the epoch at which massive galaxies first begin to host SNe~Ia in cosmic history.
Because high-redshift SN surveys are subject to selection biases that preferentially detect massive galaxies, restricting the host-galaxy sample to $z \leq 3$ is sufficient for our purposes.
Indeed, models that convolve empirical DTDs with the cosmic SFH predict that the volumetric SN~Ia rate at $z\gtrsim 2$ lies in a low-probability tail \citep[e.g.,][]{Dahlen2008, MaozMannucci2012}.
Taken together, these considerations make the $0 \leq z \leq 3$ regime a natural, observation-driven definition of the \emph{entire} SN~Ia and host populations.

We further define the local sample as all galaxies, SNe~Ia, and their hosts within $0 \leq z \leq 0.1$.
In SN~Ia cosmology, this interval anchors the low-redshift end of the Hubble diagram and sets the benchmark volumetric SN~Ia rate in the nearby Universe \citep[e.g.,][]{Leibundgut2001,Frohmaier2019}.
At $z \lesssim 0.1$, the combination of high angular resolution and high signal-to-noise enables unusually detailed constraints on host demographics and, for many events, on the immediate SN~Ia birthplace environment, including local color (e.g., $U-V$) and sSFR \citep[e.g.,][]{Rigault_2013,Roman2018,Kim2018,Kim2019,Rigault2020}.
Accordingly, the $z \lesssim 0.1$ SNe~Ia and their host populations serve as the natural observational reference point for interpreting redshift evolution in the global sample.

\begin{table*}
\centering
\caption{Summary statistics of the simulated dataset.
All statistics are evaluated over the full-sky comoving volume within the specified redshift range, except for the `TNG100' sample, whose values are taken directly from the native snapshot.
The fiducial model assumes subhalos with $M_*\ge10^{8}~\Msun$ and an observational time window of $\Delta t = 10^5$~yr, except for the ``3-yr Mock Survey for Full Catalogue'' sample, for which $\Delta t = 3$~yr is adopted.}
\label{tab:snpm_stats}
\footnotesize
\setlength{\tabcolsep}{14pt}
\begin{tabular}{lccccc}
\midrule
\midrule
\textbf{Quantity}   & \textbf{TNG100}     & \textbf{Full Catalogue}         & \textbf{Global Sample}        & \textbf{Local Sample}             & \textbf{3-yr Mock Survey} \\
                    & \textbf{($0 \leq z \leq 5$)} & \textbf{($0 \leq z \leq 5$)}  & \textbf{($0 \leq z \leq 3$)}  & \textbf{($0 \leq z \leq 0.1$)}    & \textbf{for Full Catalogue} \\
\midrule
$N$ of Subhalos  & $3 \times 10^{6}$ & $4.7 \times 10^{10}$ & $3.4 \times 10^{10}$ & $8.6 \times 10^{6}$ & $4.7 \times 10^{10}$ \\
$N$ of Star particles & $2 \times 10^{10}$ & $1.5 \times 10^{14}$ & $1.4 \times 10^{14}$ & $8.6 \times 10^{10}$ & $1.5 \times 10^{14}$ \\
\midrule
$N$ of SNe~Ia  & $6.9 \times 10^{8}$ & $1.3 \times 10^{13}$ & $9.7 \times 10^{12}$ & $1.2 \times 10^{9}$ & $3.8 \times 10^{8}$ \\
\midrule
$N$ of SDs & $4.3 \times 10^{8}$ & $1.0 \times 10^{13}$ & $7.4 \times 10^{12}$ & $4.7 \times 10^{8}$ & $3.0 \times 10^{8}$ \\
\ \ $-$ WD+MS & $1.8 \times 10^{8}$ & $4.7 \times 10^{12}$ & $3.2 \times 10^{12}$ & $1.9 \times 10^{8}$ & $1.4 \times 10^{8}$ \\
\ \ $-$ WD+GB & $1.2 \times 10^{8}$ & $2.6 \times 10^{12}$ & $2.0 \times 10^{12}$ & $1.5 \times 10^{8}$ & $8.2 \times 10^{7}$ \\
\ \ $-$ WD+He & $2.0 \times 10^{7}$ & $5.9 \times 10^{11}$ & $3.6 \times 10^{11}$ & $1.9 \times 10^{7}$ & $1.8 \times 10^{7}$ \\
\ \ $-$ Sub-$M_{\rm ch}$ (WD+He) & $1.0 \times 10^{8}$ & $2.3 \times 10^{12}$ & $1.8 \times 10^{12}$ & $1.1 \times 10^{8}$ & $6.6 \times 10^{7}$ \\
$N$ of DDs & $2.6 \times 10^{8}$ & $2.5 \times 10^{12}$ & $2.3 \times 10^{12}$ & $7.3 \times 10^{8}$ & $7.6 \times 10^{7}$ \\
\ \ $-$ CO+CO WDs & $1.9 \times 10^{8}$ & $2.0 \times 10^{12}$ & $1.8 \times 10^{12}$ & $5.2 \times 10^{8}$ & $5.9 \times 10^{7}$ \\
\ \ $-$ Sub-$M_{\rm ch}$ (CO+He WDs) & $7.1 \times 10^{7}$ & $5.7 \times 10^{11}$ & $5.7 \times 10^{11}$ & $2.1 \times 10^{8}$ & $1.7 \times 10^{7}$ \\
\midrule
\end{tabular}
\end{table*}

\subsection{Final Data Products: Catalogue and Codes}
\label{sec:2.10}

The \texttt{SN Ia Population Machine}, developed to synthesize galactic and cosmic SN~Ia populations by coupling cosmological hydrodynamic simulations with binary population synthesis, is publicly available at \url{https://doi.org/10.5281/zenodo.18603625}.
In the Zenodo repository, we provide the catalogue of SNe~Ia and their hosts generated in this work, together with the scripts and documentation required to reproduce the catalogue-level analyses.
Table~\ref{tab:snpm_catalogue} presents the full catalogue of our simulated SNe~Ia (see also Appendix\,\ref{appendix:D} (Figs.\,\ref{fig:corner_all_z0to5}\,--\,\ref{fig:corner_host_z0}) and Appendix\,\ref{appendix:E} (Fig.\,\ref{fig:corner_host_observerframe})), listing the SN~Ia identifier, progenitor properties (e.g., SD/DD and near-$M_{\rm Ch}$/sub-$M_{\rm Ch}$ classification and age), host-galaxy properties (e.g., redshift, age, metallicity, mass, and sSFR), and comoving shell-volume quantities (TNG100 snapshot No. and shell-volume weight).
The table includes one representative example for each progenitor channel considered in this study---SD (WD+MS, WD+GB, WD+He, and sub-$M_{\rm Ch}$ SD) and DD (WD+WD and sub-$M_{\rm Ch}$ DD) channels---as a guide to its structure and content.
In addition, Table~\ref{tab:snpm_stats} summarizes key statistics of the simulated dataset, including the total numbers of galaxies, star particles, SN~Ia--host pairs, and SD/DD events.

\section{Anatomy of SN I\MakeLowercase{a} Populations in Individual Galaxies}
\label{sec:3}

Before turning to the host-galaxy-level statistics of SNe~Ia, we first zoom in on individual galaxies to develop intuition for how SNe~Ia are distributed within their hosts.
To better characterize SN~Ia locations on galactic scales and to resolve the detailed distributions of SD and DD events, this section relies on the higher-resolution TNG50-1 simulation, whose spatial and mass resolution are substantially better than those of TNG100.
We begin by examining SNe~Ia in representative galaxies spanning different morphologies and redshifts (\S\S\,\ref{sec:3.1}).
We then turn to SNe~Ia in Milky Way-like galaxies at $z=0$ (\S\S\,\ref{sec:3.2}).

\subsection{SNe Ia in Representative Galaxies of Different Morphology and Redshift}
\label{sec:3.1}

\begin{figure*}
\includegraphics[width=0.95\textwidth]{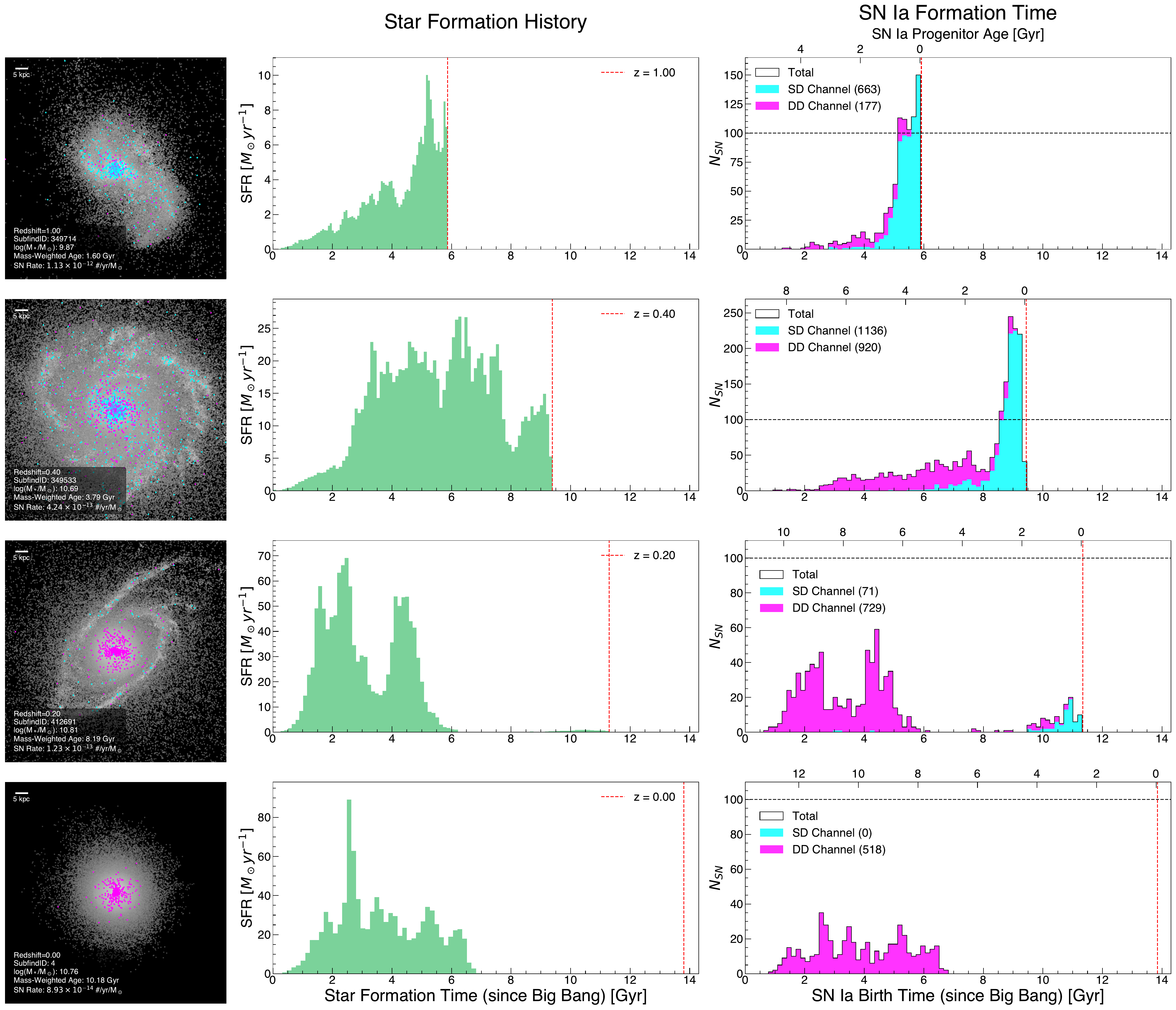}
\caption{
Representative SN~Ia host galaxies from TNG50-1 and their SFHs, together with their SN~Ia formation-time distributions observed in a mock survey with an observational time window of $\Delta t = 10^5$~yr.
\textbf{\textit{(Left column)}} Host-galaxy stellar maps with key properties in the legend.
White dots indicate star particles, shown with mass-weighted brightness, while cyan and magenta dots mark star particles that host SNe~Ia produced within the last $\Delta t = 10^5$~yr from the SD and DD channels, respectively.
\textbf{\textit{(Middle)}} SFH of each host as a function of cosmic time since the Big Bang.
The vertical dotted line marks the cosmic time corresponding to each host redshift.
\textbf{\textit{(Right)}} Formation-time distributions of SNe~Ia occurred over $\Delta t = 10^5$~yr, separated into SD (cyan) and DD (magenta) contributions, with event counts listed in the legend.
Top axes show progenitor age (delay time between SF and explosion).
Horizontal dotted lines at $N_{\rm SN}=100$ aid visual comparison across hosts with different total SN~Ia counts.
With an observational time window of $\Delta t = 10^{5}$~yr, a single galaxy hosts multiple SNe~Ia.
In the single-event-per-host regime (e.g., adopting $\Delta t = 3$ yr), these multiple-event birth times (i.e., progenitor age) and types (i.e., SD/DD) define a host-specific underlying probability distribution, from which the observed SN (including its birth time and type) is a single random draw.
}
\label{fig:single_gal_diffz}
\end{figure*}

Figure\,\ref{fig:single_gal_diffz} illustrates a selection of SN~Ia host galaxies from TNG50-1 and their SFHs, together with their SN-formation-time distributions observed in a mock survey with an observational time window of $\Delta t = 10^5$~yr.
We select a small set of representative galaxies that spans morphology (disk, elliptical, and merging) and redshift ($z=0-1$).
In the first column, we show the projected star-particle density map and overlay the locations of SNe~Ia formed within the past $10^{5}$\,yr; cyan and magenta star symbols indicate the star particles hosting SD and DD events, respectively.
SD SNe~Ia closely trace sites of recent SF, highlighting their strong association with young progenitors.
By contrast, DD SNe~Ia are preferentially associated with older stellar components, following the overall stellar-mass distribution rather than the star-forming substructure.
We emphasize that, with $\Delta t = 10^{5}$~yr, a single galaxy can host numerous SNe~Ia, often hundreds to thousands.
A galaxy that produces $N$ SNe~Ia contributes $N$ times as distinct SN--host pairs to all host-demography statistics that are explicitly {\it event-weighted}.
Relatedly, in the single-event-per-host regime (e.g., adopting $\Delta t = 3$ yr; see \S\S\,\ref{sec:2.7}), these multiple-event birth times (i.e., progenitor ages) and types (i.e., SD/DD) serve as a host-specific underlying \emph{probability distribution}, from which one SN (including its birth time and type) is randomly selected (see Fig.\,\ref{fig:Host_Prog_age}).

The second column presents the SFH of each galaxy, and the third column shows the corresponding SN~Ia formation history, with the SD and DD contributions separated. 
Together, these examples illustrate that a galaxy's SFH primarily governs its SN~Ia population. 
The top and second rows show a merging galaxy at $z=1.0$ and a spiral galaxy at $z=0.4$, respectively. 
Both systems exhibit ongoing SF at the observed epoch (vertical red dotted lines) and prominent prompt ($<$\,1\,Gyr) SNe~Ia, with the SD channel providing the dominant contribution, while only a smaller number of tardy SNe~Ia arise from older stellar populations. 
In contrast, the third and bottom rows show a quenched galaxy at $z=0.2$ and a massive elliptical galaxy at $z=0.0$, respectively. 
At their observed epochs, both galaxies are dominated by old stellar populations and exhibit predominantly old ($>$\,1\,Gyr) SNe~Ia, with the DD channel contributing the majority of events. 
The quenched galaxy at $z=0.2$ (third row) has an SN~Ia rate that broadly tracks its past SFH, while residual late-time SF near the observed epoch supplies a modest population of young events. 
The massive elliptical galaxy at $z=0.0$ (bottom row), which ceased SF nearly 6~Gyr ago, has an SN~Ia rate that is almost entirely sustained by the delayed ($>$\,1\,Gyr) DD channel, tracing the ancient stellar population. 

Beyond the individual examples above, our results emphasize a general picture in which SNe~Ia occur across both actively star-forming and long-quenched galaxies.
Although SNe~Ia are often associated with young star-forming galaxies, they are also frequently observed in massive old elliptical galaxies with little or no ongoing SF \citep[e.g.,][]{Kang_2020}, implying that a significant fraction of SNe~Ia originates from old stellar populations.
Indeed, our results show that, except in galaxies with strong ongoing SF, the DD channel contributes a larger share of the total SN~Ia production than the SD channel, especially at $z\lesssim0.5$ (see \S\,\ref{sec:9}).

\begin{figure*}
\includegraphics[width=0.95\textwidth]{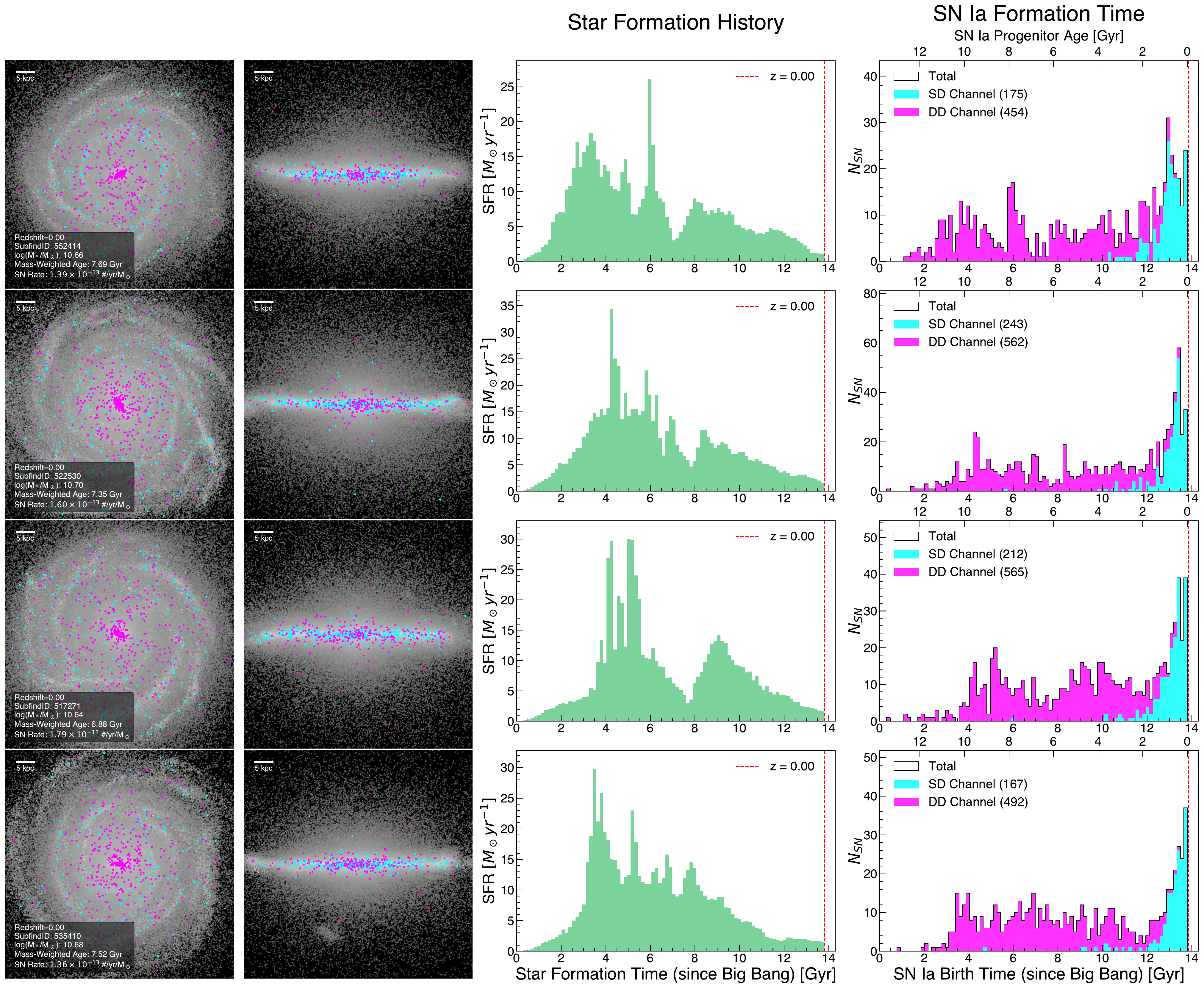}
\caption{
Same as Fig.\,\ref{fig:single_gal_diffz}, but for Milky Way-like galaxies located at $z=0$, with an additional second column displaying their edge-on view. 
These host galaxies have stellar masses in the range $4.4\text{--}5.0 \times 10^{10}~\Msun$ and sizes ($R_{90}$) in the range $12.5\text{--}16.7$~Kpc, comparable to the present-day Milky Way.
With an observational time window of $\Delta t = 10^{5}$~yr, a single galaxy hosts multiple SNe~Ia.
In the single-event-per-host regime (e.g., adopting $\Delta t = 3$ yr), these multiple-event birth times (i.e., progenitor age) and types (i.e., SD/DD) define a host-specific underlying probability distribution, from which the observed SN (including its birth time and type) is a single random draw.
}
\label{fig:single_gal_MWs}
\end{figure*}

\subsection{SNe Ia in Milky Way-like Galaxies at \texorpdfstring{{\boldmath $z=0$}}{z=0}}
\label{sec:3.2}

Figure\,\ref{fig:single_gal_MWs} is analogous to Fig.\,\ref{fig:single_gal_diffz}, but now focuses on Milky Way-like mock galaxies at $z = 0$. 
These hosts have stellar masses in the range $4.4$--$5.0 \times 10^{10}~\Msun$ and sizes ($R_{90}$) of $12.8$--$17.0$~Kpc, comparable to the present-day Milky Way. 
Over our adopted observational time window of $\Delta t = 10^{5}$~yr, the four Milky Way-like galaxies at $z = 0$ produce on average $717.50$ SNe~Ia per galaxy. 
This corresponds to a mean SN~Ia rate of $R_{\rm gal} \simeq 0.72 \times 10^{-2}~[{\rm galaxy}^{-1}\,{\rm yr^{-1}]}$, in good agreement with empirical estimates of local-universe SN~Ia rates for $L_\ast$ spirals: $(0.72\pm0.23)\times10^{-2}~[{\rm galaxy}^{-1}\,{\rm yr^{-1}]}$ \citep[e.g.,][and references therein]{Maoz2014}. 
Among the $717.50$ SNe~Ia, SD and DD events contribute on average $199.25$ (27.8\,\%) and $518.25$ (72.2\,\%), respectively. 
Notably, even in disk galaxies typified by the Milky Way, the DD channel dominates ($N_{\rm SD}:N_{\rm DD}\simeq3:7$), in line with DD dominance at $z \lesssim 0.5$ (as we will see in \S\,\ref{sec:9}).

In the face-on views (first column), SD SNe~Ia closely trace spiral arms, indicating their association with young progenitors, whereas DD SNe~Ia preferentially inhabit spheroidal components and inter-arm regions.
The edge-on views (second column) further reveal that SD events are confined to the thin plane, while DD events show a broader vertical extent.
We decompose stellar structures using the automated kinematic framework\footnote{For related kinematics-based decomposition schemes including explicit treatment of additional subcomponents and bars, see \citet{Zana2022}.} of \citet{Du2019, Du2020}.
Using this decomposition, we find that, on average, the kinematically cold disk (observational thin disk) hosts $\sim$\,64.0\,\% of all SNe~Ia, of which $\sim$\,43.2\,\% is SD events. 
The warm disk, bulge, and halo host the remaining SNe~Ia, contributing $\sim$\,15.5\,\%, $\sim$\,12.5\,\%, and $\sim$\,8.0\,\% of the total, respectively.
Consistent with the progressively older underlying stellar populations, the characteristic delay time increases from the cold disk to the warm disk, bulge, and halo.
The SFHs of the galaxies without structure decomposition (third column) show a dominant peak at $\sim$\,4--5~Gyr, followed by weaker SF and continued low-level SF to $z=0$.
The corresponding SN~Ia birth histories with the SD and DD contributions plotted separately (fourth column) demonstrate that recent SF at the observed epoch is effectively amplified by the DTD mainly through the SD channel, while older populations contribute predominantly via the DD channel.

\section{Comparative Demographics of All Galaxies and SN I\MakeLowercase{a} Host Galaxies}
\label{sec:4}

We now move beyond illustrative individual cases to the collective demographics of the full galaxy population and SN~Ia host galaxies in TNG100. 
SN~Ia hosts are not a random draw from all galaxies; they are selected by the convolution of each galaxy’s SFH with the DTD.
Host demographics therefore identify which regions of host-property space (e.g., stellar mass, sSFR, and redshift) dominate SN~Ia production, disentangling genuine progenitor-driven behaviour from demographic inheritance of the underlying galaxy population.
This perspective is central for interpreting observed host-dependent SN~Ia correlations (e.g., mass- and sSFR-related standardization residuals) and for determining which galaxies dominate the volumetric SN~Ia rate across cosmic time.
It also provides direct tests of survey representativeness, distinguishing observational selection effects from intrinsic SN~Ia physics.
In this section, we compare the full galaxy population to SN~Ia host-galaxy populations in the global ($0 \leq z \leq 3$) sample (\S\S\,\ref{sec:4.1}) and the local ($0 \leq z \leq 0.1$) sample (\S\S\,\ref{sec:4.2}).
We focus on representative combinations of the primary galactic parameters in the main text, and Appendix\,\ref{appendix:D} provides various sets of combinations across all parameters for all- and host-galaxies in the global- and local-samples.

\begin{figure*}
\includegraphics[width=\textwidth]{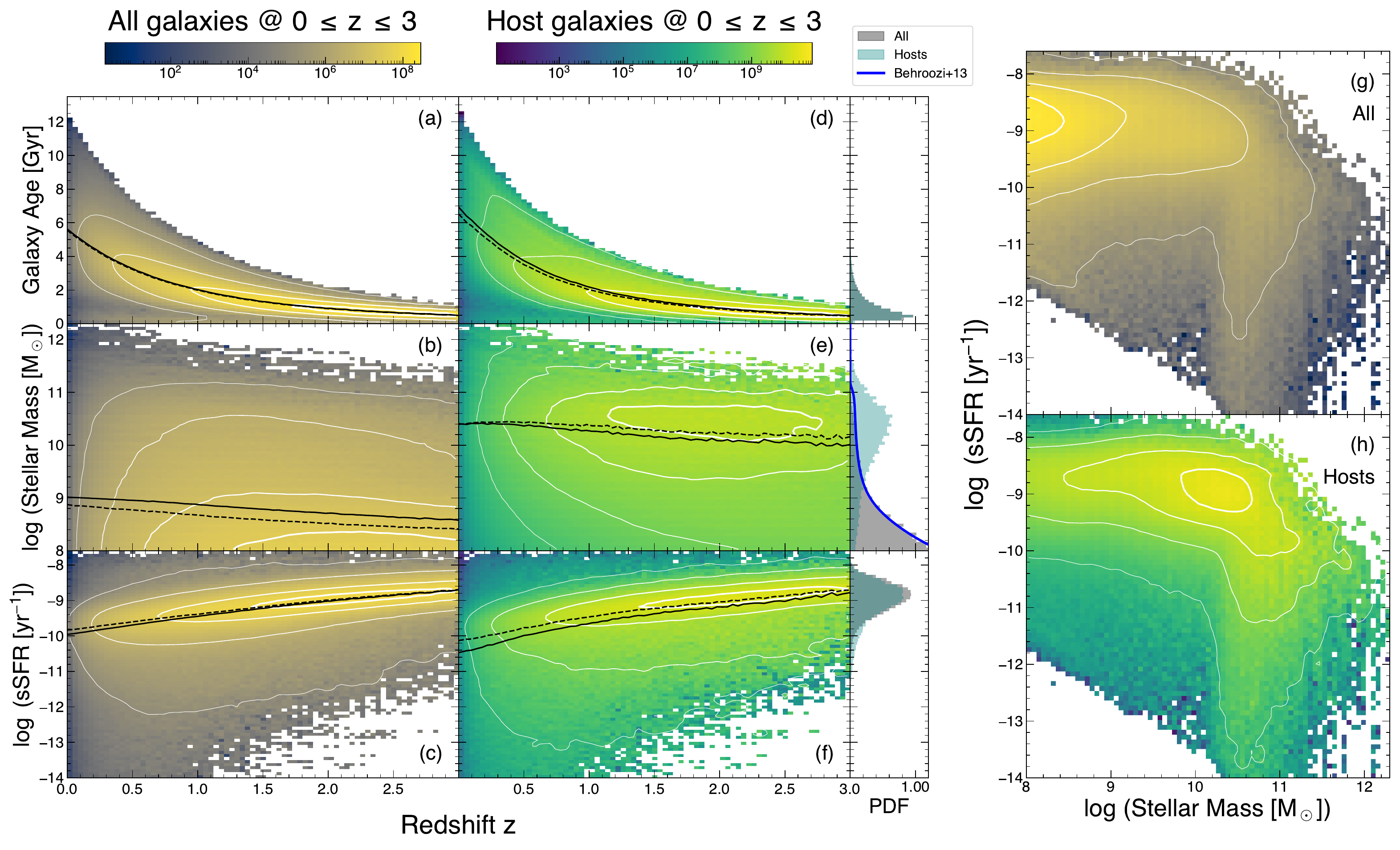}
\caption{
Basic demographics of \textit{all} galaxies and SN~Ia \textit{host} galaxies in TNG100 over $0 \leq z \leq 3$ (the global sample).
\textbf{\textit{(a--c)}} Redshift evolution of the star-particle-mass-weighted mean stellar age ($T_*$), total stellar mass ($M_*$; the sum of all constituent star-particle masses), and sSFR for the full galaxy population.
Colored 2D maps show the logarithmic galaxy number density (dark-blue-to-yellow `cividis' color scale), while white contours mark the 0.3, 1, 2, and 3\,$\sigma$ levels.
Solid and dashed black curves trace the mean and median trends, respectively.
These panels illustrate that the TNG100 global sample spans wide ranges in redshift, age, mass, and sSFR, yielding a well-conditioned parent population for modeling SN~Ia and host demographics across cosmic time.
\textbf{\textit{(d--f)}} The same redshift evolution for the SN~Ia host galaxies.
Because multiple SNe~Ia can occur in a single galaxy within our $\Delta t\,=\,10^{5}$~yr observational window, these host distributions are explicitly event-weighted: each host contributes once per SN~Ia event it produces.
Colored 2D maps are shown with a blue-green-yellow `viridis' color scale.
The most pronounced demographic shift is in stellar mass: compared to the full population, the host $M_*$ distribution is systematically displaced to higher masses.
The right-hand marginal panels show the corresponding 1D probability density functions (grey for all; pink for hosts); for reference, the empirical galaxy stellar mass function model of \citet{Behroozi2013} over $0 \leq z \leq 3$ (blue curve) is overplotted on the all-galaxy histogram.
\textbf{\textit{(g \& h)}} Demographics in the $M_*$--sSFR plane---the two principal host properties governing the SN~Ia populations---for all galaxies and SN~Ia hosts, respectively.
This projection compactly visualizes how SN~Ia hosts occupy the joint distribution of stellar mass and SF activity relative to the parent sample, highlighting that the dominant contributors to the SN~Ia-weighted host population lie near the overlap of high stellar mass and elevated SF rather than simply at the peak of the parent number density.
}
\label{fig:Demo_z3} 
\end{figure*}

\begin{figure*}
\includegraphics[width=\textwidth]{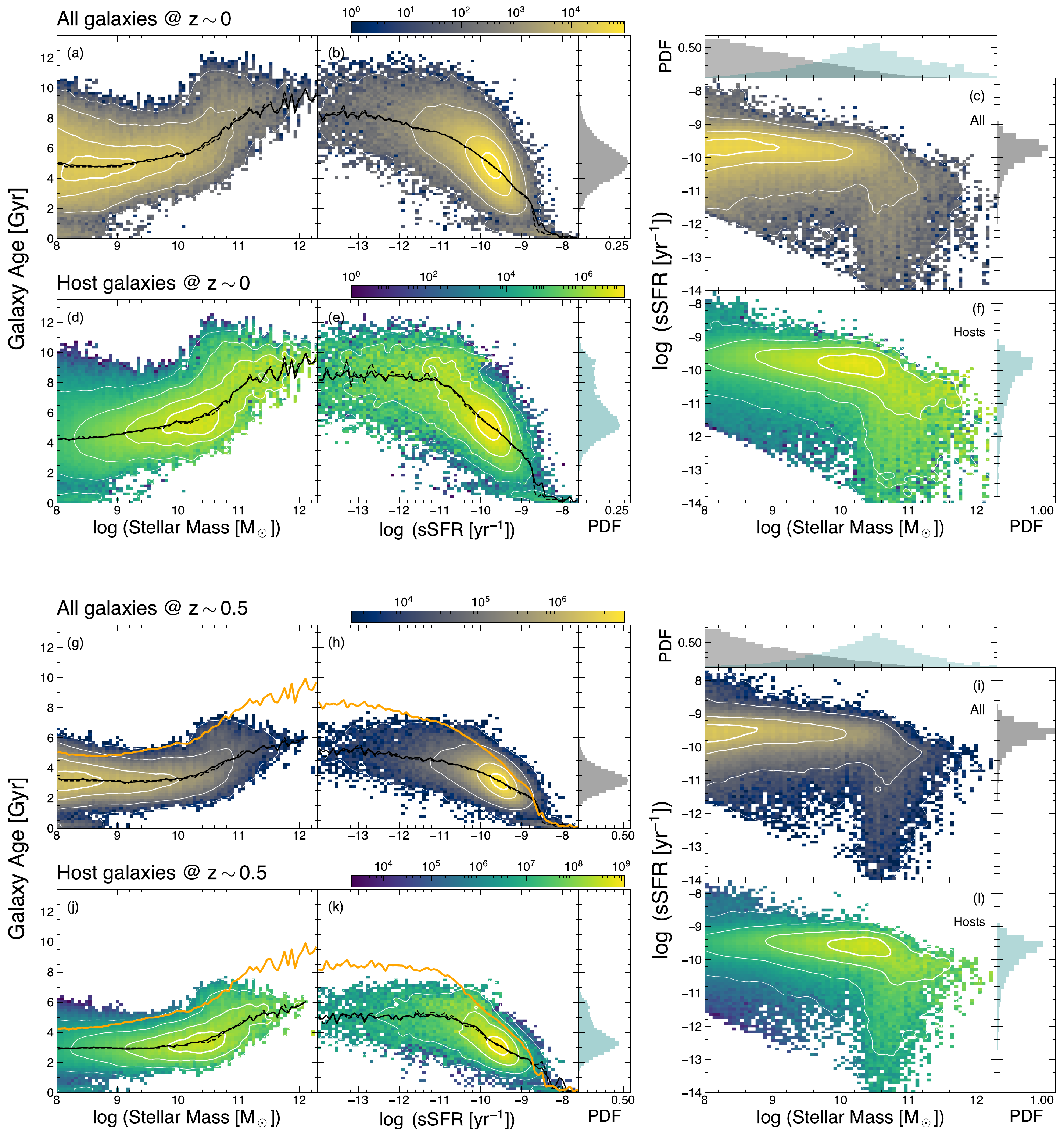}
\caption{
\textbf{\textit{(a--f)}} Local analogue of Figure\,\ref{fig:Demo_z3}, showing the demographics of \textit{all} galaxies ($a$--$c$) and SN~Ia \textit{host} galaxies ($d$--$f$) over $0 \leq z \leq 0.1$.
Panels ($a$ \& $d$) plot galaxy age versus stellar mass, while ($b$ \& $e$) plot galaxy age versus sSFR.
Colored 2D maps show the logarithmic galaxy number density (color bar), while white contours mark the 0.3, 1, 2, and 3\,$\sigma$ levels.
Solid and dashed black curves trace the mean and median trends, respectively.
Because multiple SNe~Ia can occur in a single galaxy within our $10^{5}$~yr observational window, the host distributions are explicitly event-weighted: each host contributes once per SN~Ia event it produces.
Panels ($c$ \& $f$) show the stellar mass--sSFR distribution for all galaxies and for hosts. 
The top and right-hand marginal panels compare the corresponding 1D probability density functions (grey for all; pink for hosts).
\textbf{\textit{(g--l)}} Same as ($a$--$f$), but for an intermediate-redshift slice at $z\simeq0.5$ ($0.45\le z\le 0.55$).
Orange curves in panels ($g$--$h$ \& $j$--$k$) indicate the corresponding mean trends at $z\simeq0$ from panels ($a$--$b$ \& $d$--$e$), respectively.
}
\label{fig:Demo_z0p1_and_z0p55}
\end{figure*}

A brief clarification helps interpret the demographic plots in this section (Figures\,\ref{fig:Demo_z3} and \ref{fig:Demo_z0p1_and_z0p55}).
As described in \S\S\,\ref{sec:2.7} in the context of host multiplicity, we identify all SNe~Ia and their associated hosts within an observational time window of $10^{5}$\,$\mathrm{yr}$ and record each SN--host pair as a separate entry in the host catalogue.
A galaxy that produces $N$ SNe~Ia within this interval therefore appears $N$ times, such that all host-demography statistics are explicitly {\it event-weighted}.
As shown in Fig.\,\ref{fig:Hostfraction}($a$), for the $\Delta t = 10^5$~yr simulation, effectively all parent galaxies eventually enter the host catalogue.
Consequently, the all-galaxy and host samples share the same overall distribution shape in the demographic plots; what changes is the relative weight assigned to each 2D bin, as reflected by the 0.3, 1, 2, and 3$\sigma$ contours.
Likewise, Fig.\,\ref{fig:Hostfraction}($b$) shows that galaxies produce, on average, $\sim$270 SNe~Ia over $\Delta t = 10^5$~yr.
Accordingly, as indicated by the color bars, the host sample (i.e., the SN~Ia--host pairs) contains correspondingly more entries than the full parent-galaxy sample.

\begin{figure*}
\includegraphics[width=\textwidth]{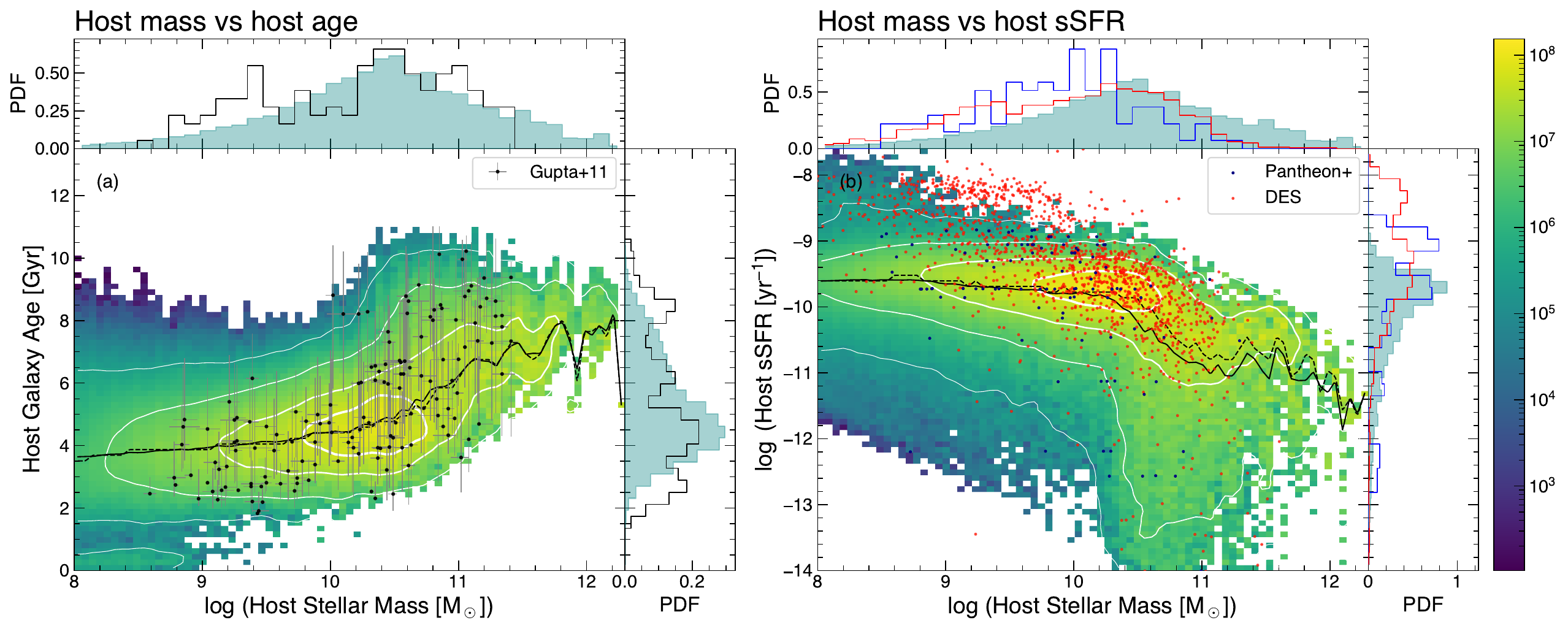}
\caption{
Analogous to Fig.\,\ref{fig:Demo_z0p1_and_z0p55}($d$ \& $f$), but for $0.1 \le z \le 0.3$ and including observational constraints.
\textbf{\textit{(a)}} Host mass versus host age diagram for the model density map (color bar) and observed 199 hosts.
Host masses and host ages are taken from \citet{Gupta_2011}; host ages are mass-weighted mean stellar ages derived from SED fitting to integrated multi-band photometry.
The top and right-hand marginal panels compare the corresponding 1D probability density functions (grey for model; black for observation).
The model predicts somewhat older hosts at the low-mass end, but it reproduces the key feature: a transition occurring near $\logM\simeq10.5$.
\textbf{\textit{(b)}} Host mass versus host sSFR diagram for the model density map (color bar) and observations.
The observed data are taken from the Pantheon+ sample (113 hosts) \citep{Pantheon2022} and the DES 5-yr sample (1447 hosts) \citep{Wiseman2020, Smith_2020}, both restricted to $0.1 \leq z \leq 0.3$.
The marginal panels compare the corresponding 1D probability density functions (grey: model; blue: Pantheon+; red: DES).
The model reproduces the main observed concentration near $\logM\simeq10.5$ and $\logsSFR\simeq-9.5$.
}
\label{fig:Demo_z0p1_0p3}
\end{figure*}

\subsection{Global-sample Demographics: All Galaxies versus SN~Ia Hosts}
\label{sec:4.1}

Figure\,\ref{fig:Demo_z3} summarizes the basic demographics of \textit{all} galaxies and SN~Ia \textit{host} galaxies in TNG100 over $0 \leq z \leq 3$ (the global sample).
Panels ($a$--$c$) show the redshift evolution of the mass-weighted mean stellar age ($T_*$), stellar mass ($M_*$), and sSFR for the full galaxy population.
In panel ($a$), $T_*$ increases steeply toward the present epoch.
The right-hand marginal histogram shows the $T_*$ distribution, with a pronounced peak at $\sim0.8$~Gyr and a long tail to old ages; the peak is dominated by galaxies at $z>2$, reflecting the prevalence of very young systems at early epochs.
By contrast, panels ($b$--$c$) show that stellar mass and sSFR evolve more gradually with decreasing $z$.
The mass histogram shows that the all-galaxy sample reaches the lower-mass threshold of $10^{8.0}\,\Msun$ and declines toward higher masses, broadly consistent with empirical stellar-mass function modeling of \citet{Behroozi2013}.
The sSFR histogram shows that the sSFR distribution peaks at $\logsSFR\simeq-9$, driven primarily by the abundant low-mass star-forming galaxies, again broadly consistent with observations \citep[e.g.,][]{Whitaker2014, Speagle2014, Schreiber2015, Ilbert2015}.
Taken together, these panels show that the TNG100 global sample spans broad ranges in redshift, age, mass, and sSFR, providing a well-conditioned baseline for modeling SNe~Ia and their host demographics across cosmic time.

Panels ($d$--$f$) show the redshift evolution for the SN~Ia host galaxies.
Because multiple SNe~Ia can occur in a single galaxy within our $10^{5}$~yr observational window, the host histograms are event-weighted (see \S\S\,\ref{sec:2.7}).
In panel ($d$), hosts broadly track the parent-galaxy population in $T_*$, but appear modestly older toward $z = 0$.
The marginal age histogram closely parallels that of all galaxies, peaking at $\sim$\,0.5~Gyr with only a weak old-age tail; the peak is driven primarily by $z>2$ hosts.
The strongest demographic shift is found in stellar mass: relative to all galaxies, the host $M_*$ distribution is displaced to substantially higher masses, with the mean rising by more than 1~dex, from $\logM \simeq 9$ (panel $b$) to $\gtrsim 10$.
This SN-weighted shift is consistent with SNe~Ia occurring preferentially in massive systems.
As expected from the right-hand marginal histogram, the mean $M_*$ of the all-galaxy sample depends sensitively on the imposed low-mass cut.
In contrast, the mean $M_*$ of the host sample is comparatively insensitive to the low-mass cut because low-mass galaxies with $\logM<8.0$ contribute only a very small fraction of the SN~Ia yield (see Fig.\,\ref{fig:Demo_z0p1_and_z0p55}).
In the sSFR panel, hosts follow the global redshift trend but are shifted slightly further downward toward $z=0$.
The extended lower envelope of the host distribution below the mean sSFR locus supports the interpretation that a significant fraction of events arise in passive, massive hosts (consistent with the elevated host masses), while the modest upper extension implies a non-negligible contribution from strongly star-forming, lower-mass systems.

Finally, panels ($g$--$h$) recast these demographics in the $M_*$--sSFR plane---the two principal host properties governing SN~Ia populations---for all galaxies and SN~Ia hosts, respectively.
This projection provides a compact view of how SN~Ia hosts populate the joint space of mass and SF activity, and how that distribution differs from the underlying parent-galaxy sample.
As implied in panels ($a$--$f$), the dominant contributors to the SN-weighted host population are not simply the most numerous galaxies in the parent sample, but systems with intermediate mass ($\logM$\,$\simeq$\,$10.5$) and high sSFR ($\logsSFR$\,$\simeq$\,$-9.0$).
Two branches are evident: a horizontal sequence at $\logM \lesssim 10.5$ and $\logsSFR \gtrsim -10.5$, and a vertical sequence at $\logM \gtrsim 10.5$ and $\logsSFR \lesssim -10.5$.
This behaviour fits naturally into the broader context of galaxy populations: galaxies are strongly bimodal in fundamental stellar-population properties, including stellar mass, stellar age, colour, spectral energy distribution (SED), and SFH \citep[e.g.,][]{Strateva2001, Kauffmann2003, Baldry2004, Bell2004, Mateus2006, Noeske2007, Peng2010}.
The galaxy bimodality is commonly described in the colour--magnitude diagram as a division between actively star-forming galaxies in the blue cloud and quiescent systems in the red sequence.
In this sense, the $M_*$--sSFR diagrams in panels ($g$--$h$) are closely analogous to the colour--magnitude diagram.

\subsection{Local and \texorpdfstring{{\boldmath $z\simeq0.5$}}{z~0.5} Sample Demographics: All Galaxies versus SN~Ia Hosts}
\label{sec:4.2}

Figure\,\ref{fig:Demo_z0p1_and_z0p55} is analogous to Fig.\,\ref{fig:Demo_z3}, showing the demographics of all galaxies and SN~Ia host galaxies over $0 \leq z \leq 0.1$ (the local sample; upper group of panels) and at $z\simeq0.5$ (lower group of panels).
Panels ($a$--$f$) summarize the demographics of the local sample.
For all galaxies, panels ($a$) and ($b$) map $T_*$ as a function of $M_*$ and sSFR; the right-hand marginal $T_*$ distribution peaks broadly at $\sim$\,$5$~Gyr and exhibits a weak tail toward old ages.
Panel ($c$) shows the corresponding $M_*$--sSFR plane, demonstrating that the local sample is shifted to higher stellar masses and lower sSFR relative to the global sample (Fig.\,\ref{fig:Demo_z3}($g$)).
For host galaxies, panels ($d$) and ($e$) show that the $T_*$ distribution is strongly bimodal, with a dominant peak at $\sim$\,5~Gyr and a secondary peak at $\sim$\,9~Gyr; the intervening dip near $\sim$\,7~Gyr is associated with hosts near $\logM=10.9$ and $\logsSFR=-10.4$.
Panel ($f$) shows the stellar mass--sSFR distribution for hosts.
Compared to the full population, SN~Ia hosts are preferentially drawn from older, more massive, and lower-sSFR systems.
The top marginal histogram indicates that hosts near $\logM=10.5$ dominate.
Likewise, the right-hand marginal histogram favors moderate-sSFR hosts, with the distribution peaking near $\logsSFR\simeq-9.7$.

Panels ($g$--$l$) provide an analogue of panels ($a$--$f$) for galaxies at $z \simeq 0.5$.
For all galaxies, panels ($g$) and ($h$) show $T_*$ as a function of $M_*$ and sSFR; the right-hand marginal distribution of $T_*$ exhibits a broad peak at $\sim$\,3~Gyr and only a short tail toward older ages.
Panel ($i$) shows the corresponding distribution in the $M_*$--sSFR plane, indicating that the $z \simeq 0.5$ population is shifted slightly toward lower stellar masses and higher sSFR relative to its $z \simeq 0$ counterpart.
For the host sample, panels ($j$) and ($k$) show that the $T_*$ distribution is bimodal, with a dominant peak at $\sim$\,3~Gyr and a secondary peak at $\sim$\,5~Gyr.
Panel ($l$) shows that, relative to the $z \simeq 0$ hosts, the $z \simeq 0.5$ hosts are depleted in the high-$M_*$ and low-sSFR regime; accordingly, their mean properties shift toward lower $M_*$ and higher sSFR, broadly following the evolution of the parent population.
Overall, the trends at $z \simeq 0.5$ remain qualitatively similar to those at $z \simeq 0$.

Having established the predicted host-galaxy demographics at low and intermediate redshift, we now compare the model with observations in Figure\,\ref{fig:Demo_z0p1_0p3}, in the $M_*$--$T_*$ plane (panel $a$) and the $M_*$--sSFR plane (panel $b$).
For panel ($a$), we use host age and mass measurements from \citet{Gupta_2011} and restrict both the model and the observational data to the common redshift interval $0.1 \le z \le 0.3$.
The model density map is analogous to those shown in Fig.\,\ref{fig:Demo_z0p1_and_z0p55}($d$) and ($j$), but it is plotted over $0.1 \le z \le 0.3$.
Although the model predicts somewhat older hosts at the low-mass end, by $\sim1$~Gyr for $\logM\lesssim9.5$, it successfully reproduces the key observational feature: a sharp age transition near $\logM\simeq10.5$.
For panel ($b$), we use host mass and sSFR measurements from the Pantheon+ sample \citep{Pantheon2022} and the DES 5-yr sample \citep{Wiseman2020, Smith_2020}, again restricting both the model and the observational data to $0.1 \le z \le 0.3$.
The corresponding model density map is analogous to those in Fig.\,\ref{fig:Demo_z0p1_and_z0p55}($f$) and ($l$), apart from the $0.1 \le z \le 0.3$ selection.
The model reproduces the main concentration of hosts near $\logM\simeq10.5$ and $\logsSFR\simeq-9.5$.

A closer inspection of panel ($b$) shows that the DES sample exhibits an excess near $\logM \simeq 9.3$ and $\logsSFR \simeq -8.4$, in a region that is almost unpopulated in our model.
This feature should be interpreted with caution, however, because SFR estimates from broadband photometric SED fitting are known to be highly uncertain and strongly dependent on the adopted modeling assumptions \citep{Smith_2020}.
At $z \lesssim 0.3$, star-forming main-sequence galaxies typically occupy $\logsSFR \simeq -10$, so even a $\sim$1~dex elevation would place them only near $\logsSFR \simeq -9$.
While a small fraction of galaxies at $\logsSFR \simeq -8.4$ could in principle be genuinely extreme systems, such as compact starbursts or very blue high-sSFR disks, the strength of the DES excess would imply an implausibly large population of strong main-sequence outliers.
A more likely explanation is systematic bias in the SED fitting.
Indeed, a detailed multi-wavelength study of DES SN~Ia hosts by \citet{Ramaiya2025} found that optical-only or optical+NIR-only SED fits can misidentify intrinsically passive, massive red galaxies as dusty, star-forming low-mass systems, thereby biasing sSFR high and stellar mass low.
They ascribed this effect to the well-known age--metallicity--dust degeneracy.
In this light, much of the apparent DES excess at $\logM \simeq 9.3$ and $\logsSFR \simeq -8.4$ is more naturally interpreted as an artifact of SED fitting than as evidence for a substantial population of genuinely extreme starburst SN~Ia hosts.

\section{SN I\MakeLowercase{a} Rate in Galactic Context: Host Dependence}
\label{sec:5}

The SN~Ia birth rate provides an integrative measure of how SN~Ia progenitor physics is embedded within galaxy formation and cosmic evolution \citep[e.g.,][]{Greggio05}.
In practice, it determines where and when real surveys harvest most events, and thus which host-galaxy populations at which epochs contribute most heavily to the observed SN~Ia sample.
This demographic weighting, further modulated by survey selection effects, can in turn influence cosmological inference if SN~Ia observables correlate with host properties.
In this section, we examine the SN~Ia rate in the galactic context by quantifying its variation across host-galaxy parameter space, identifying the regimes that dominate SN production, and isolating the physical drivers of these trends.
We begin with the mass-normalized SN~Ia rate ($R_{\rm mass}$; rate per unit stellar mass) in \S\S\,\ref{sec:5.1}, and then turn to the galaxy-normalized SN~Ia rate ($R_{\rm gal}$; rate per galaxy) in \S\S\,\ref{sec:5.2}.

We define $R_{\rm mass}$ and $R_{\rm gal}$ as the SN~Ia rates normalized by the total stellar mass and the total number of galaxies, respectively, in the full parent-galaxy population.
The parent sample includes all galaxies that satisfy the same survey-volume, galaxy-mass, and observational-time-window selection, regardless of whether they host an SN event.
Accordingly, these definitions measure the mean SN~Ia rate per unit stellar mass and per galaxy, rather than the SN-event-weighted distribution of hosts.
An event-weighted formulation would instead highlight where SNe preferentially occur, as in the host-demographics analysis of \S\,\ref{sec:4}.
In practice, we count all SNe~Ia over the observational time window, $\Delta t$, and define
\begin{equation}
R_{\rm mass} = \frac{N_{\rm SN}}{\sum_i M_{*,i}}\,\frac{1}{\Delta t}~~\&~~R_{\rm gal} = \frac{N_{\rm SN}}{N_{\rm gal}}\,\frac{1}{\Delta t}~,
\label{eq:R_massgal}
\end{equation}
where $N_{\rm SN}$ is the total number of SNe~Ia, ${\sum_i M_{*,i}}$ is the total stellar mass of all galaxies in the sample, including both hosts and non-hosts, and $N_{\rm gal}$ is the corresponding total number of galaxies.
For comparison, observational estimates are commonly expressed as
\begin{equation}
R_{\rm mass} \simeq \frac{N_{\rm SN}}{\sum_i M_{*,i}}\,\frac{V_{\rm gal}}{V_{\rm SN}}\,\frac{1}{\Delta t}~~\&~~R_{\rm gal} \simeq \frac{N_{\rm SN}}{N_{\rm gal}}\,\frac{V_{\rm gal}}{V_{\rm SN}}\,\frac{1}{\Delta t}~,
\label{eq:R_massgal_realsurvey}
\end{equation}
where all quantities are defined analogously to Eq.~\ref{eq:R_massgal}, and $V_{\rm SN}$ and $V_{\rm gal}$ denote the effective survey volumes of the SN and galaxy samples, respectively.
Our construction is the simulation-side analogue of these estimators, evaluated directly from the complete galaxy population for a given galaxy property (e.g., galaxy redshift, age, metallicity, mass, and sSFR).

\begin{figure*}
\includegraphics[width=0.90\textwidth]{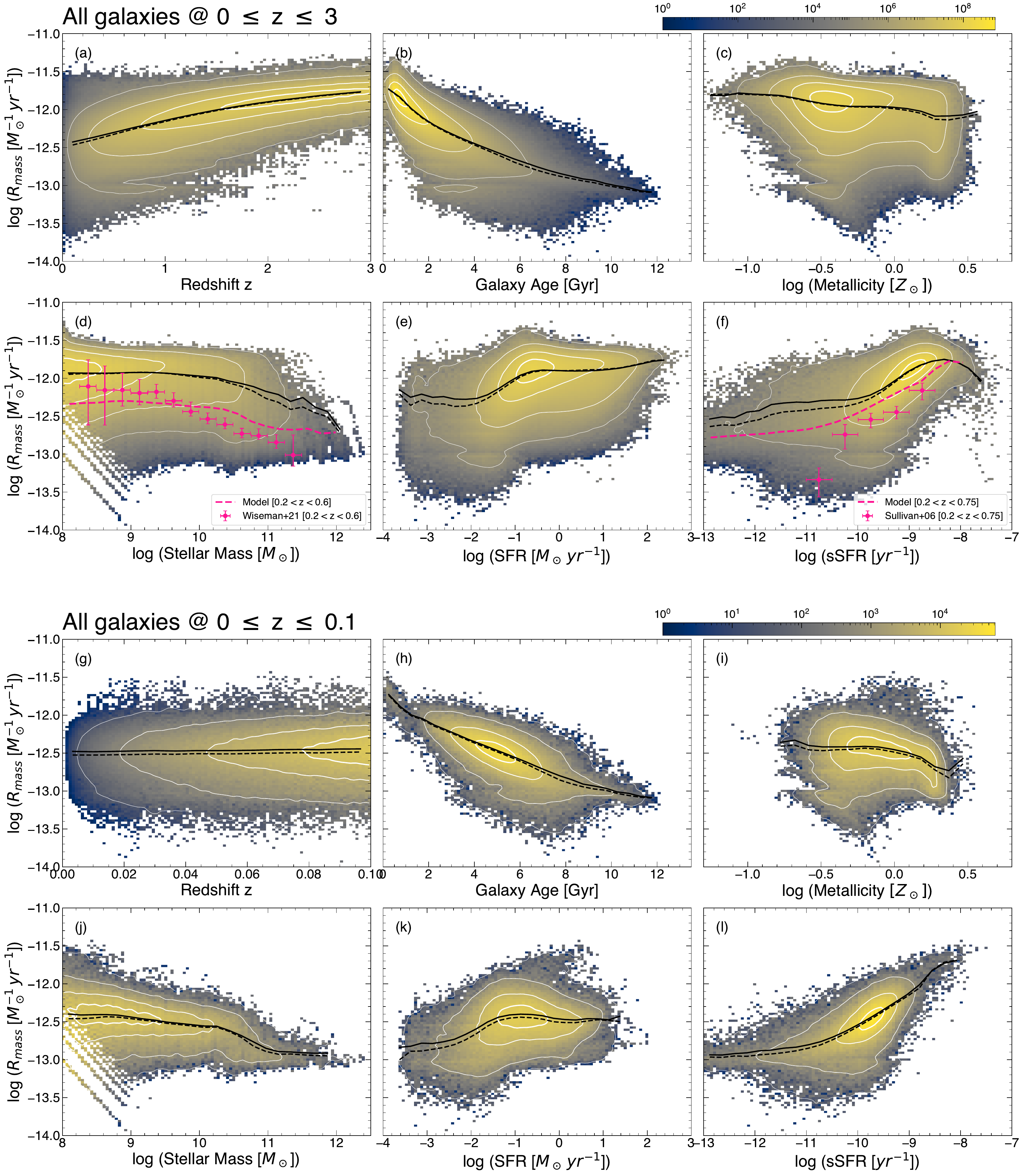}
\caption{
Mass-normalized SN~Ia rate ($R_{\rm mass}$) as a function of six key host properties.
\textbf{\textit{(a--f)}} $R_{\rm mass}$ versus ($a$) redshift, ($b$) mass-weighted mean stellar age, ($c$) mass-weighted mean stellar metallicity, ($d$) stellar mass, ($e$) SFR, and ($f$) specific SFR over the global sample ($0 \leq z \leq 3$).
Colored 2D maps encode the logarithmic galaxy number density (color bar), and white contours mark the 0.3, 1, 2, and 3\,$\sigma$ levels.
Solid and dashed black curves indicate the mean and median trends, respectively.
In panels ($d$ \& $f$), observational constraints are overplotted as pink dots with error bars (\citet{Wiseman2021} for the mass relation; \citet{Sullivan2006} for the sSFR relation).
To compare fairly to each dataset, the pink curves show the simulation (median lines) resampled to the observed redshift coverage by adopting a Gaussian distribution centered on the volume-weighted mean redshift of the corresponding observation, with $\sigma=0.15$.
\textbf{\textit{(g--l)}} Same as panels ($a$--$f$), but for the local sample ($0 \leq z \leq 0.1$), where redshift evolution is minimized and the dependence of $R_{\rm mass}$ on each parameter is more evident.
}
\label{fig:R_mass}
\end{figure*}

\begin{figure*}
\includegraphics[width=0.90\textwidth]{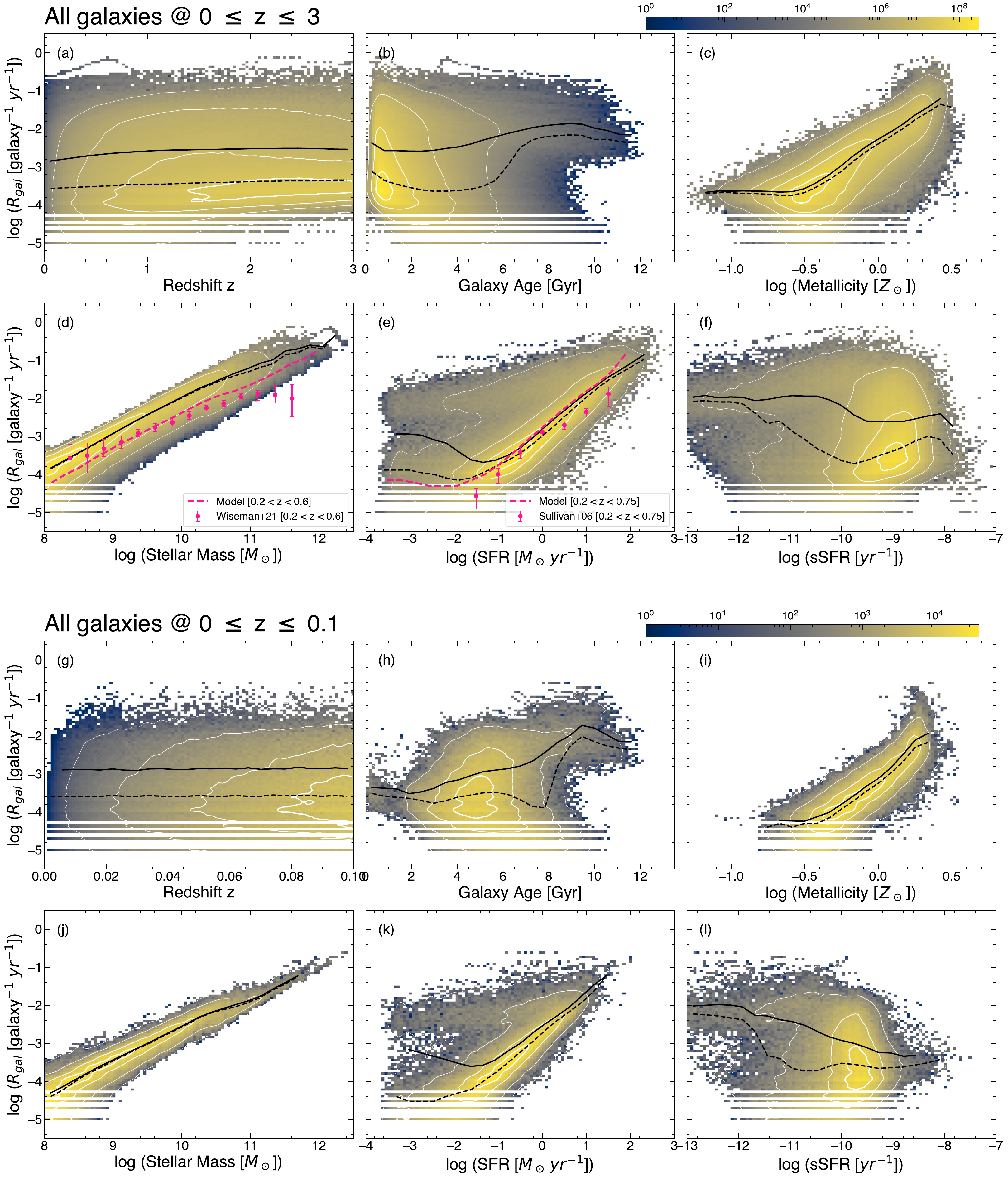}
\caption{
Same as Figure\,\ref{fig:R_mass}, but for galaxy-normalized SN~Ia rate ($R_{\rm gal}$). 
In panels ($d$ \& $e$), observational constraints are overplotted as pink dots with error bars (\citet{Wiseman2021} for the mass relation; \citet{Sullivan2006} for the SFR relation; $\mathrm{SFR}$ in \citet{Sullivan2006} is averaged over the past $\sim0.5$\,Gyr).
}
\label{fig:R_gal}
\end{figure*}

A brief clarification is useful for interpreting the SN~Ia rate plots in this section (Figures\,\ref{fig:R_mass} and \ref{fig:R_gal}).
    Under the definitions of $R_{\rm mass}$ and $R_{\rm gal}$ above, binning all parent galaxies along a given galaxy property assigns a single representative SN rate to each bin.
In the usual setting, this permits only a 1D trend (i.e., a rate--property relation), not a resolved 2D map on the SN-rate--galaxy-property plane.
In our model, however, we adopt a long observational time window, $\Delta t = 10^{5}\,\mathrm{yr}$, for which nearly the entire galaxy population ($99.74\,\%$) hosts at least one SN~Ia.
This exceptionally high host-occupancy of 2D cells strongly suppresses sparse-cell effects, so that the binned plane remains sufficiently well populated, making a resolved 2D density-map analysis feasible in our framework.

\subsection{Mass-normalized SN~Ia Rate and its Host Dependence} 
\label{sec:5.1}

Figure\,\ref{fig:R_mass}($a$--$f$) presents the mass-normalized SN~Ia rate, $R_{\rm mass}$, as a function of six key galaxy properties over $0 \leq z \leq 3$ (the global sample).
In each panel, the colored 2D density map shows the logarithmic number density of simulated SN~Ia galaxies.
Panel ($a$) shows $R_{\rm mass}$ as a function of redshift; toward higher $z$, younger galaxies in the younger universe tend to exhibit higher $R_{\rm mass}$.
Panel ($b$) shows $R_{\rm mass}$ versus the stellar age ($T_*$) with a peak at $T_* < 1$\,Gyr and $\log R_{\rm mass}\,[\Msun^{-1}\,{\rm yr}^{-1}] > -12$, reflecting the predominance of younger galaxies at higher $z$ in the underlying $0 \leq z \leq 3$ population.
The strong dependence of $R_{\rm mass}$ on $T_*$ is evident.
Panel ($c$) shows $R_{\rm mass}$ versus the stellar metallicity ($Z_*$).
Because of the well-known mass--metallicity relation, high-$Z_*$ hosts are generally more massive, so the apparent $R_{\rm mass}$--$Z_*$ distribution largely reflects a projection of the underlying $R_{\rm mass}$--$M_*$ relation (cf. panel $d$).

Panel ($d$) shows how $R_{\rm mass}$ varies with stellar mass ($M_*$).
At low masses, $R_{\rm mass}$ remains nearly constant, but begins to decline at $M_* \simeq 10^{10.5}\,\Msun$.
Our redshift-matched simulated hosts reproduce the observed $M_*$--$R_{\rm mass}$ relation \citep{Wiseman2021}, albeit with a $\sim$\,0.2--0.3\,dex offset at $\logM \gtrsim 10$.
Panel ($e$) shows $R_{\rm mass}$ as a function of SFR, revealing a step-like transition near $\logSFR \simeq -1.2$.
Panel ($f$) presents $R_{\rm mass}$ versus sSFR; because age and sSFR evolve in opposite directions, the strong sSFR dependence is consistent with the trend in the age panel ($b$).
Our redshift-matched simulated hosts lie $\lesssim$\,0.3~dex above the observed SFR--$R_{\rm mass}$ relation \citep{Sullivan2006}, while following a broadly similar overall trend, except at $\logsSFR \simeq -10.8$.

Figure\,\ref{fig:R_mass}($g$--$l$) is closely analogous to panels ($a$--$f$), but restricted to $0 \leq z \leq 0.1$ (the local sample), where redshift evolution is minimized and the influence of each parameter on $R_{\rm mass}$ can be seen more clearly.
Panel ($g$) shows that $R_{\rm mass}$ is nearly constant over $0 \leq z \leq 0.1$.
Panel ($h$) shows that younger hosts achieve a relatively higher $R_{\rm mass}$, with a peak at $T_* \simeq 1$\,Gyr and $\log R_{\rm mass}\,[\Msun^{-1}\,{\rm yr}^{-1}] \simeq -12.5$.
In panel ($i$), the $R_{\rm mass}$--$Z_*$ distribution is shifted rightward and downward relative to panel ($c$): the local sample is more metal-rich because of cosmic chemical evolution, and older at fixed $Z_*$, which lowers $R_{\rm mass}$.
Panel ($j$) highlights the strong dependence of $R_{\rm mass}$ on $M_*$, with a step-like feature near $M_* \simeq 10^{10.5}\,\Msun$.
In panel ($k$), the step-like feature is also present at $\log {\rm SFR} \simeq -1.5$.
The characteristic transitions in panels ($j$) and ($k$) reflect the well-known galaxy bimodality of star-forming and quenched galaxies.
Panel ($l$) shows that $R_{\rm mass}$ depends strongly on sSFR, consistent with the behavior in the age panel ($h$).
Overall, SN~Ia production within galaxies reflects an intricate dependence on redshift, stellar age, stellar metallicity, stellar mass, SFR, and sSFR.
By isolating each galaxy property in turn, we clarify how individual parameters regulate $R_{\rm mass}$ and how their combined effects give rise to the global trends.

Observational studies have long shown that the mass-normalized SN~Ia rate rises toward lower-mass galaxies rather than remaining constant, from nearby measurements based on host morphology and broad-band colour \citep{Mannucci2005,Li2011} to more recent untargeted surveys at low and intermediate redshift \citep{Smith2012,Brown2019,Wiseman2021}.
This trend is commonly parameterized as $R_{\rm mass} \propto M_*^{x}$, with reported slopes of roughly $x \simeq -0.3$ to $-0.5$ across different samples and host selections \citep{Sullivan2006,Smith2012,Brown2019,Wiseman2021}. 
Our results are broadly consistent with this behavior in panel ($j$), yielding $x \simeq -0.2$ and $-0.3$ over $\logM\simeq 9$ to 11 and 10 to 11, respectively.
Theoretically, the usual interpretation is that this dependence reflects stellar-mass-dependent galaxy SFHs \citep{Maoz2014,Graur2015}, and our model supports this picture: the dependence of $R_{\rm mass}$ arises primarily because more massive galaxies have lower sSFRs and therefore older stellar populations.
At the same time, recent studies have argued that SFH alone may not fully account for the observed trend, and that an additional metallicity-dependent enhancement in SN~Ia production efficiency may be required, particularly in dwarf galaxies, through an increased close-binary fraction at low $Z$ \citep{Kistler2013,Gandhi2022,Johnson2023}.
A further possibility is that the DTD itself depends on metallicity; in this regard, our model predicts a higher total SN~Ia yield at lower $Z_*$ (\S\S\,\ref{sec:6.2}, Fig.\,\ref{fig:DTDs_yield_alpha_z0_0p1}).

\subsection{Galaxy-normalized SN~Ia Rate and its Host Dependence}
\label{sec:5.2}

Figure\,\ref{fig:R_gal} is analogous to Fig.\,\ref{fig:R_mass}, but for the galaxy-normalized SN~Ia rate, $R_{\rm gal}$.
Panels ($a$--$f$) show $R_{\rm gal}$ as a function of galaxy properties over $0 \leq z \leq 3$ (the global sample).
Panel ($a$) presents $R_{\rm gal}$ versus redshift and shows that galaxies with low $R_{\rm gal}$ contribute most to the total SN~Ia budget.
Panel ($b$) shows $R_{\rm gal}$ versus stellar age ($T_*$), with galaxies concentrated at young ages ($<2$~Gyr); this reflects the predominance of younger galaxies at higher $z$ in the underlying $0 \leq z \leq 3$ population.
Panel ($c$) shows $R_{\rm gal}$ versus stellar metallicity ($Z_*$).
Because of the well-known mass--metallicity relation, high-$Z_*$ galaxies are generally more massive, so the apparent $R_{\rm gal}$--$Z_*$ trend largely represents a projection of the underlying $R_{\rm gal}$--$M_*$ relation (cf.\ panel $d$).
In addition, the metallicity-dependent DTD yield, which decreases mildly by $\sim$\,10\,\% from the lowest to highest $Z_*$ (\S\S\,\ref{sec:6.2}, Fig.\,\ref{fig:DTDs_yield_alpha_z0_0p1}), also contributes to making the $R_{\rm gal}$--$Z_*$ slope shallower.

Panel ($d$) shows that $R_{\rm gal}$ depends strongly on $M_*$.
At low $M_*$, galaxies are typically star-forming and can produce SNe~Ia at elevated rates per galaxy, thereby flattening the relation relative to a strict one-to-one scaling.
Quantitatively, increasing the stellar mass by 4.0~dex, from $\logM =$ 8 to 12, raises the rate by only $\sim$\,3.4~dex, from $\log R_{\rm gal} = -4.0$ to $-0.6$.
This behavior is consistent with empirical two-component (`A+B') rate models \citep{Scannapieco2005,Mannucci2006,Sullivan2006,Childress2014}.
Our redshift-matched simulated galaxies also reproduce the observed $M_*$--$R_{\rm gal}$ relation \citep{Wiseman2021}, except for a $\sim$\,0.4\,dex offset at $\logM \simeq 11.5$.
Panel ($e$) shows $R_{\rm gal}$ as a function of SFR and reveals a large vertical spread at low SFR, with a possible bifurcation emerging in this regime.
Our redshift-matched simulated galaxies lie $\lesssim$\,0.3\,dex above the observed SFR--$R_{\rm gal}$ relation \citep{Sullivan2006}, but follow a broadly similar trend.
Panel ($f$) shows the distribution of $R_{\rm gal}$ versus sSFR.
Taken in isolation, the preceding SFR (= sSFR\,$\times$\,mass) panel ($e$) could be misread as implying that sSFR directly regulates $R_{\rm gal}$, since high SFR may be intuitively associated with high \emph{specific} SFR.
In fact, $R_{\rm gal}$ depends only weakly on sSFR, and its increase is driven primarily by the underlying dependence on mass rather than on sSFR itself.
Massive galaxies often show weak or no recent SF, yet still maintain elevated $R_{\rm gal}$ because of their large stellar-mass reservoirs.
Accordingly, massive galaxies dominate the low-sSFR, high-$R_{\rm gal}$ sequence, whereas lower-mass galaxies populate the higher-sSFR, lower-$R_{\rm gal}$ cloud, with a broad peak around $\logsSFR \simeq -9$ (see also Fig.\,\ref{fig:Demo_z0p1_and_z0p55}).

Figure\,\ref{fig:R_gal}($g$--$l$) is analogous to panels ($a$--$f$), but restricted to $0 \leq z \leq 0.1$ (the local sample).
Panel ($g$) shows that $R_{\rm gal}$ is nearly constant over $0 \leq z \leq 0.1$.
Panel ($h$) shows that younger galaxies ($<\!7$\,Gyr) tend to have lower rates ($\log R_{\rm gal}\,[{\rm galaxy}^{-1}\,{\rm yr}^{-1}]\,<\,-2.8$), whereas older galaxies ($>\!7$\,Gyr), although largely quenched, still sustain relatively high rates ($\log R_{\rm gal}\,[{\rm galaxy}^{-1}\,{\rm yr}^{-1}]\,>\,-2.8$) because of their larger stellar masses.
In panel ($i$), the $R_{\rm gal}$--$Z_*$ distribution shifts downward relative to panel ($c$) because the local sample excludes high-redshift SF galaxies, and shifts rightward because the $z\simeq 0$ universe is, on average, more metal-rich.
In panel ($j$), the strong dependence of $R_{\rm gal}$ on $M_*$ remains evident in the local universe, with the density peak shifted slightly toward higher-mass galaxies.
In panel ($k$), the bifurcation into two sequences is clearer than in panel ($e$): a high-$R_{\rm gal}$ branch dominated by massive galaxies, and a low-$R_{\rm gal}$ branch populated by lower-mass galaxies, with an apparent deficit between them.
Panel ($l$) likewise shows the two branches as in panel ($f$): a high-$R_{\rm gal}$, nearly horizontal sequence at $\logsSFR \lesssim -10.5$, and a low-$R_{\rm gal}$, nearly vertical group at $\logsSFR \gtrsim -10.5$.
Overall, SN~Ia production depends on redshift as well as on galaxy age, metallicity, mass, SFR, and sSFR, whose combined influence shapes $R_{\rm gal}$.

\section{Delay-Time Distributions of SN\MakeLowercase{e}~I\MakeLowercase{a}}
\label{sec:6}

In SN~Ia studies, the delay time and the progenitor age provide two complementary descriptions of the same underlying clock.
The \emph{delay time} ($\tau$) is the fundamental timescale: the elapsed interval between progenitor formation in a SF episode and the eventual explosion.
Because $\tau$ is a universal time lag, its distribution---the DTD---is often modeled as a universal response function to SF events, independent of host environment (but see \S\S\,\ref{sec:2.4} for progenitor-metallicity dependence).
The \emph{progenitor age} is the age at explosion, and its distribution at a given epoch is obtained by convolving the DTD with the host SFH.
The DTD and the SN-progenitor-age distribution (SPAD) thus encode the same information viewed in different frames: the former isolates the intrinsic timescales of binary evolution and explosion physics, whereas the latter describes how those timescales are realized across evolving galaxy populations and host properties.
Together, they provide a unified language for interpreting host demographics, SN~Ia rates, and their cosmic evolution.
With this framing, Section~\ref{sec:6} and Section~\ref{sec:7} report the SN~Ia population-machine results in the delay-time and progenitor-age views, respectively.
In this section, we present our recovered DTDs decomposed by progenitor channels (\S\S\,\ref{sec:6.1}) and discuss the non-universality of DTDs (\S\S\,\ref{sec:6.2} and \S\S\,\ref{sec:6.3}) inspired by the channel- and metallicity-dependent star-particle-level DTDs (as shown in \S\S\,\ref{sec:2.4}).

\begin{figure*}
\includegraphics[width=0.97\textwidth]{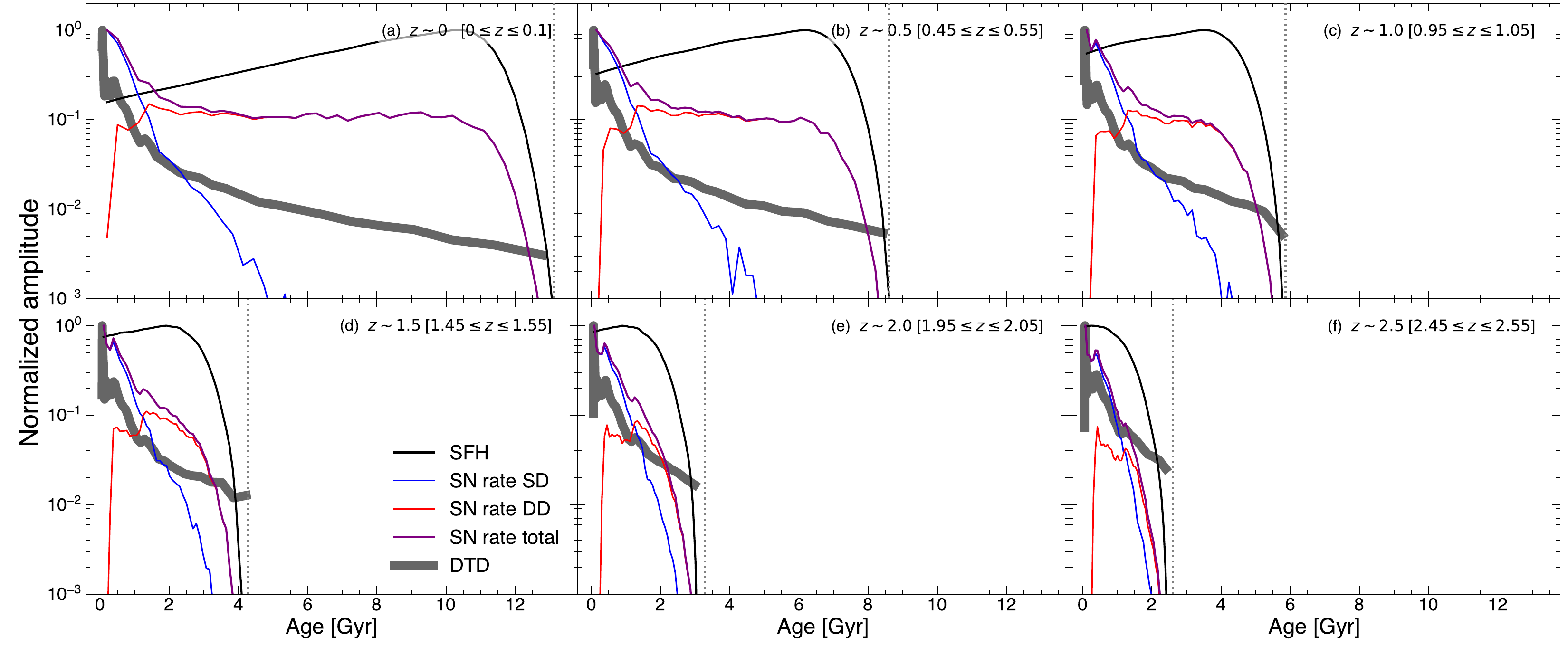}
\caption{
Recovery of the DTDs for SNe~Ia: ($a$) the local sample ($0 \leq z \leq 0.1$); ($b$--$f$) higher-redshift slices centered at $z\simeq 0.5$, 1.0, 1.5, 2.0, and 2.5.
Thin black lines show the SFH of all galaxies, while thin blue, red, and purple lines show the corresponding SN~Ia rates from the SD and DD channels and their sum, respectively.
Deconvolution of the SN~Ia rate with the SFH yields the recovered DTD, shown by the thick grey line.
All curves are rescaled to unity at their maxima to enable a direct shape comparison.
Panels ($b$--$f$) isolate the ``redshift-windowing'' effect: at higher $z$, the younger cosmic age (vertical dotted lines) progressively shortens the available SFH baseline and the associated SN~Ia rate history, producing increasingly truncated DTD reconstructions.
Because the truncation removes the oldest SFH tail first (i.e., typical DD progenitors), the DD contribution is preferentially suppressed relative to SD, leading to a cosmic transition in which the dominant progenitor channel shifts from SD to DD systems with time (see \S\,\ref{sec:9}).
}
\label{fig:DTDs_SFH_SPAD}
\end{figure*}

\subsection{Recovered DTDs and Channel-ordered Contributions}
\label{sec:6.1}

Figure\,\ref{fig:DTDs_SFH_SPAD} demonstrates the recovery of the DTDs for SNe~Ia in the $0 \leq z \leq 0.1$ local sample and in higher-$z$ slices at $z\simeq$ 0.5, 1.0, 1.5, 2.0, and 2.5.
Panel ($a$) shows the stacked SFH of the full parent-galaxy population\footnote{The stacked SFH used for the DTD recovery is constructed from the full parent-galaxy population---namely, all galaxies that were eligible to contribute SN events under the same volume, mass, and observational-time-window definitions used for the SN~Ia rate measurement.
In this stack, each galaxy enters once, irrespective of how many SNe~Ia it produces over $\Delta t$; this galaxy-weighted SFH therefore represents the total stellar population available to produce SNe~Ia, providing the reference stellar population against which the event counts are normalized.} at $0 \leq z \leq 0.1$ (black thin line) together with the corresponding SN~Ia rates from the SD and DD channels and their sum (blue, red, purple thin lines). 
The de-convolution of the SN~Ia rate with the stacked SFH\footnote{Compared to the observed cosmic SFH (\citealt{Madau1996} and subsequent work; e.g., \citealt{Madau2014}), our redshift-binned SFHs are older by $\lesssim$\,1~Gyr at fixed $z$ because we restrict the sample to galaxies with $M_*>10^{8.0}~\Msun$, which excludes low-mass, young systems; the impact is negligible because such low-mass galaxies contribute only a very small fraction of the SN~Ia rate (see Fig.\,\ref{fig:Demo_z0p1_and_z0p55}).} yields the recovered DTD (grey thick line).
Panels ($b$--$f$) highlight the redshift-windowing effect: at higher $z$, the younger cosmic age progressively shortens the available SFH baseline and thus the associated SN~Ia rates, yielding increasingly truncated DTD reconstructions.
This truncation removes the oldest SFH tail first (i.e., typical oldest DD progenitors), preferentially suppressing the recovered DD contribution relative to SD.
Equivalently, at higher $z$, many DD events that would occur at later cosmic times have simply not yet occurred.\footnote{
For example, at $z = 0.5$ ($t_{\rm lb}$\,$\simeq$\,$5.2$\,Gyr), events with $\tau$\,$>$\,8.6\,Gyr---which would explode over the subsequent 5.2\,Gyr, composing the DD tail---are necessarily absent.}
As a result, the DD contribution cannot rise with redshift as rapidly as the SD contribution, making the SD/DD ratio appear to ``evolve'' with redshift even for an identical physical DTD kernel (i.e., without invoking explicit kernel non-universality such as metallicity dependence).
The net result is an SN~Ia demographic transition in which the dominant progenitor channel shifts from SD to DD systems with cosmic time, as will be discussed in \S\,\ref{sec:9}.

\begin{figure*}
\includegraphics[width=0.9\textwidth]{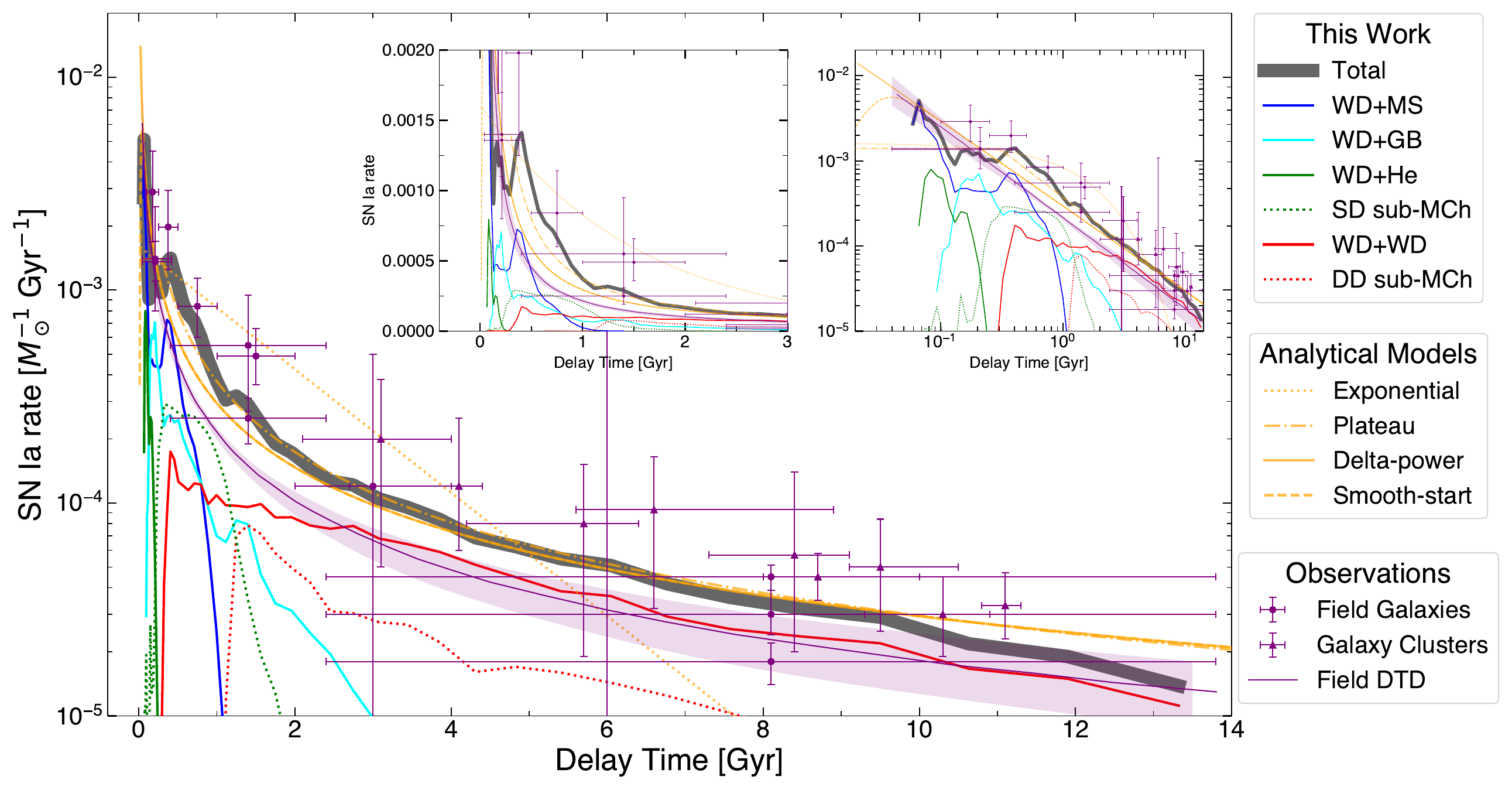}
\caption{
Recovered SN~Ia DTD for $0 \leq z \leq 3$ (the global sample).
The main panel uses a linear $x$-axis and logarithmic $y$-axis to emphasize the DTD morphology; the insets show the data in linear--linear scaling (left; zoomed at short delays) and in log--log scaling (right).
Blue, cyan, and green solid curves show the SD near-$M_{\rm Ch}$ channels (WD+MS, WD+GB, and WD+He), and red solid curve shows the DD near-$M_{\rm Ch}$ channel (WD+WD). 
Green and red dotted curves denote sub-$M_{\rm Ch}$ progenitors for SD and DD channels, respectively.
Thick grey curve is the all-channel sum.
The recovered DTD exhibits a steep early rise followed by a long tail, reflecting the time-ordered emergence of distinct progenitor channels (see also Fig.\,\ref{fig:Fig5}).
Orange curves show commonly adopted analytic models: exponential, plateau, delta-power, and smooth--start \citep{Greggio05, Schronrich09,Childress2014, Weinberg17, Wiseman2021}.
Purple points with error bars represent observational constraints from field galaxies and galaxy clusters at $0<z<1.45$ \citep{Totani2008, Maoz2010, Maoz2011, Maoz2012, Graur2013}, while the purple solid line and shaded band indicate the field DTD inferred from volumetric rates at $0<z<2.25$ \citep{Maoz2017}.
In absolute (unnormalized) units, our DTD tracks the observed trend and reproduces the late-time behaviour out to the last observed point at $\tau\simeq11$~Gyr.
The observational reconstructions exhibit a relative normalization offset, and our DTD lies between them, indicating overall consistency in both normalization and slope.
The field-galaxy and galaxy-cluster DTDs yield $\chi^2_{\nu}=1.71$ and Pearson $r=0.86$, whereas the field DTD yields $r=0.97$.
}
\label{fig:DTD}
\end{figure*}

Figure\,\ref{fig:DTD} presents the recovered DTD of SNe~Ia over $0 \leq z \leq 3$ (the global sample).
The main panels employ a linear $x$-axis and logarithmic $y$-axis to highlight the DTD morphology, while the insets display the same relationship in linear--linear and log--log scalings.
The DTD is characterized by a steep early rise followed by a long tail, reflecting the time-ordered emergence of distinct progenitor channels as demonstrated in \S\S\,\ref{sec:2.4} (Fig.\,\ref{fig:Fig5}).
We overplot the commonly adopted analytic models \citep{Greggio05, Schronrich09, Childress2014, Weinberg17, Wiseman2021} and observational measurements \citep{Totani2008, Maoz2010, Maoz2011, Maoz2012, Graur2013, Maoz2017}, showing that our DTD is consistent, in overall form, with models and observations.
The analytic curves (normalized for visual comparison) include an exponential form and several power-law prescriptions that differ primarily at early delays but converge toward $\alpha\simeq-1$ at late times (e.g., the plateau, delta-power, and smooth-start models).
Among these, the delta-power model most closely matches our recovered DTD, capturing both the rapid early rise and the subsequent power-law tail.
In the observational comparison, our DTD closely follows the measured trend in absolute (unnormalized) units, and reproduces the late-time behaviour well\footnote{Although the measurements span $0<z<1.45$ for DTDs derived from field galaxies and galaxy clusters and $0<z<2.25$ for a field DTD derived from volumetric rates, a direct comparison to our global sample is appropriate. 
This is because, although the star-particle-level DTD varies systematically with progenitor metallicity (\S\S\,\ref{sec:2.4}), the DTDs are, to first order, only mildly altered between our global and redshift-binned samples.} out to the last observed point at $\tau\simeq11$~Gyr.
The observational reconstructions show a relative offset in normalization, and our DTD lies between them, indicating overall consistency in the normalization and slope.
Quantitatively, the DTDs derived from field galaxies and galaxy clusters yield $\chi^2_{\nu}=1.71$ and Spearman $\rho=0.86$, whereas the field DTD derived from volumetric rates yields $\rho=0.97$.

The approximately $\tau^{-1}$ form and early-time rise are broadly consistent with observational constraints on the onset and peak time, $\tau_{\rm peak}$, of the SN~Ia DTD.
Over the past two decades, such constraints have moved the inferred $\tau_{\rm peak}$ from a delay of several Gyr to a substantially earlier onset.
Early high-redshift rate studies argued for a relatively long characteristic delay of $\sim$\,2--3\,Gyr \citep{Dahlen2008}, whereas host-galaxy analyses soon provided evidence for a substantial short-delay component, including signals at $\lesssim\,180$~Myr \citep{Aubourg2008}.
Subsequent DTD reconstructions increasingly favored a continuous, approximately power-law form rather than a narrow peak, with measurements consistent with ${\rm DTD}\propto \tau^{-1}$ over $\sim$\,0.1\,--\,10\,Gyr \citep{Totani2008,Maoz2012}.
More recent volumetric- and host-based studies have reinforced this picture, showing that the data are broadly consistent with a $\tau^{-1}$-like DTD beginning at very short delays, typically parameterized as $\tau_{\rm peak} \simeq 40$\,Myr \citep{Rodney2014}, while direct recent constraints that leave the onset time free find values consistent with $\tau_{\rm peak} \simeq 50$\,Myr, albeit with substantial uncertainty \citep{Castrillo2021,Wiseman2021}.
However, the commonly quoted range of $\sim$\,0.03\,--\,0.3\,Gyr largely reflects differences in time resolution, binning, and parameterization, rather than evidence for fundamentally different intrinsic peak times; accordingly, one should be cautious about treating either $\tau_{\rm peak}$\,$\simeq$\,0.03 or 0.3\,Gyr as a uniquely established value.

\begin{figure*}
\includegraphics[width=0.9\textwidth]{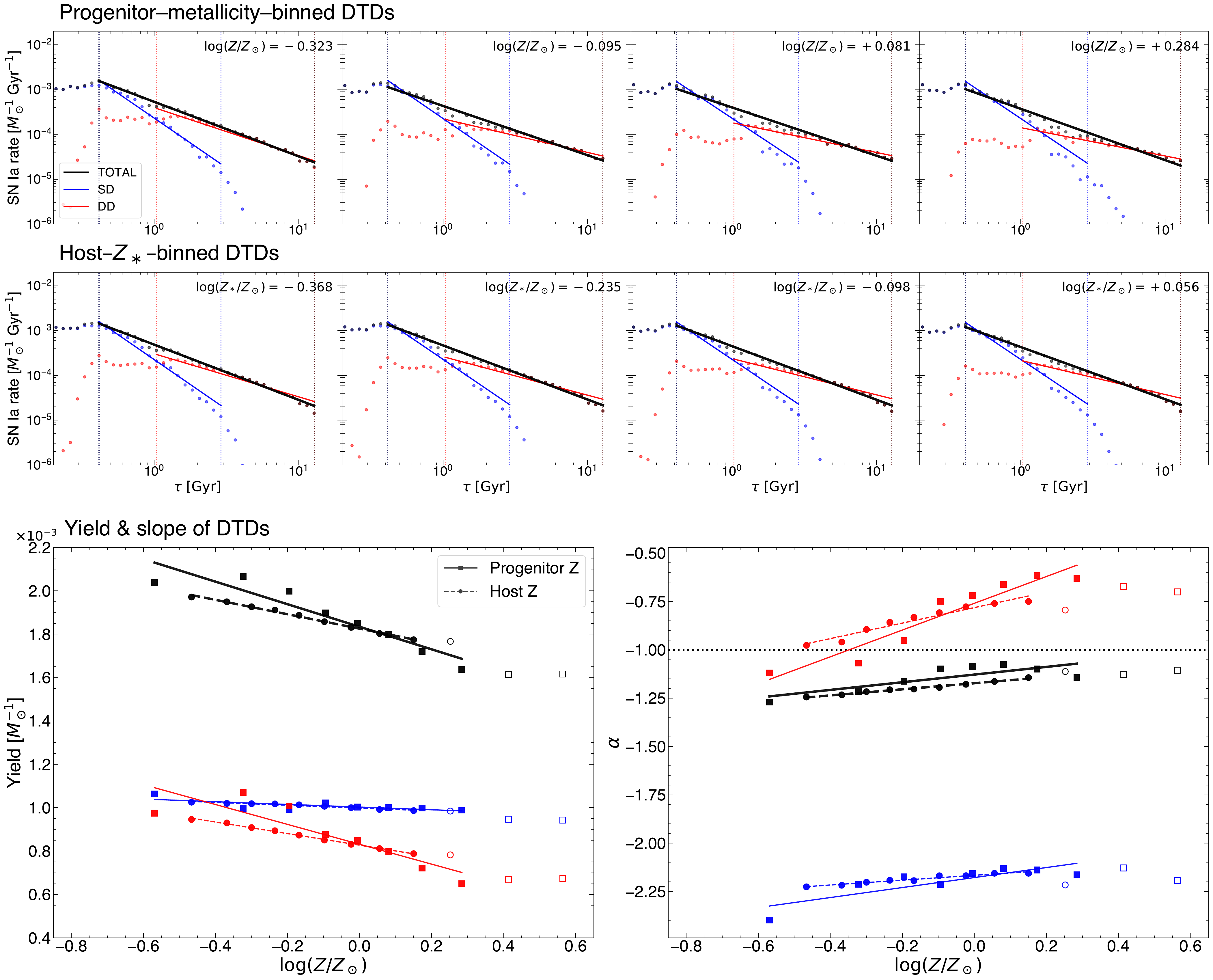}
\caption{
Metallicity-dependent SN~Ia DTDs in the local sample ($0\leq z \leq 0.1$).
\textbf{\textit{(Top row)}} DTDs binned by star-particle (progenitor) metallicity, $Z_{{\rm pro},i}$, into 10 equal-number intervals, of which four representative bins are shown.
Blue and red points show the SD and DD DTDs, respectively, while black points denote their pointwise sum.
Solid lines indicate the corresponding power-law fits for the SD, DD, and total channels, with the total channel highlighted by thicker lines.
The SD DTD exhibits a complex structure out to $\sim$0.4~Gyr; beyond this timescale, it transitions into a smooth, approximately power-law monotonic decline, followed by a more rapid drop after 3~Gyr (vertical blue dotted lines at 0.4 and 3~Gyr).
By contrast, the DD DTD remains approximately stable from the end of its early oscillatory phase at 1~Gyr out to a Hubble time (vertical red dotted lines at 1.0 and 13.8~Gyr).
We adopt these fitting ranges to maximize the constraining power for each channel, and measure $\alpha_{\rm SD}$ and $\alpha_{\rm DD}$ from the SD and DD fits, respectively; the total slope, $\alpha_{\rm tot}$, is fit over $\tau=0.4$--$13.8~{\rm Gyr}$.
The yields ($Y_{\rm SD}$, $Y_{\rm DD}$, and $Y_{\rm tot}$) are computed by integrating the pointwise DTDs over the entire range ($\tau=0.0-13.8$~Gyr) and are therefore independent of the slope-fitting choices.
\textbf{\textit{(Middle row)}} Same as the top row, but binned by host-galaxy stellar metallicity, $Z_*$.
\textbf{\textit{(Bottom row)}} Derived yields (left) and slopes (right) versus $Z_{{\rm pro},i}$ (filled squares; solid lines) and $Z_*$ (filled circles; dashed lines).
Open symbols mark bins excluded from the slope fits because $\geq 20\%$ of SN-hosting particles lie outside the \texttt{COMPAS} metallicity grid, $Z\in[0.001,0.03]$.
The yields are $Y_{\rm SD} \sim 1.0 \times 10^{-3}~{\rm M_\odot^{-1}}$ and $Y_{\rm DD} \sim 0.9 \times 10^{-3}~{\rm M_\odot^{-1}}$, with $Y_{\rm tot}=Y_{\rm SD}+Y_{\rm DD} \sim 1.9 \times 10^{-3}~{\rm M_\odot^{-1}}$.
$Y_{\rm SD}$ is largely insensitive to $Z$, whereas $Y_{\rm DD}$ decreases toward higher $Z$, driving a corresponding decline in $Y_{\rm tot}$.
The inferred slopes span a wide range (e.g., $\alpha_{\rm SD}\simeq-2.2$ to $\alpha_{\rm DD}\simeq-0.8$), while $\alpha_{\rm tot}$ remains $\sim-1.15$ that is close to the canonical $\alpha\simeq-1$.
$\alpha_{\rm SD}$ varies only weakly with $Z$, whereas $\alpha_{\rm DD}$ becomes shallower with increasing $Z$; $\alpha_{\rm tot}$ shows little metallicity dependence.
}
\label{fig:DTDs_yield_alpha_z0_0p1}
\end{figure*}

\begin{figure*}
\includegraphics[width=0.9\textwidth]{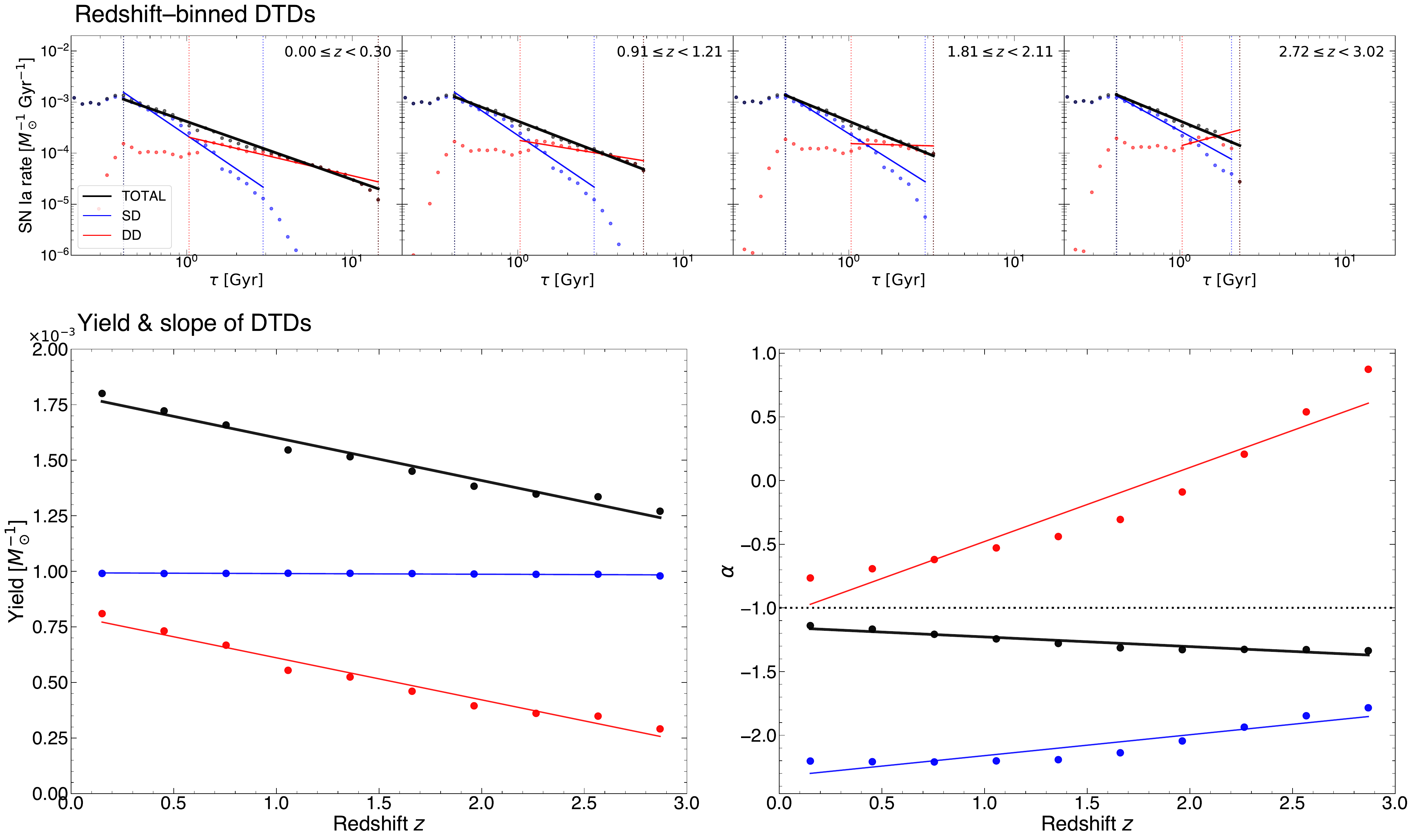}
\caption{
Analogous to Fig.\,\ref{fig:DTDs_yield_alpha_z0_0p1}, but for redshift-dependent SN~Ia DTDs.
\textbf{\textit{(Upper row)}} DTDs binned into 10 equal SN~Ia-number redshift slices; four representative bins are shown ($z \simeq$ 0, 1, 2, and 3).
We adopt identical fit ranges for all bins, measuring $\alpha_{\rm SD}$ and $\alpha_{\rm DD}$ from the SD and DD fits, respectively.
The dominant systematic is look-back-time truncation: at higher redshift, the available $\tau$ baseline is progressively shortened from the old-age end; the late-time DD events that support the tail of the total DTD become increasingly absent, and the inferred $\alpha_{\rm tot}$ correspondingly steepens.
\textbf{\textit{(Lower row)}} Derived yields (left) and slopes (right) versus redshift.
In all redshift bins, the fraction of SN-hosting particles that lie outside the \texttt{COMPAS} metallicity grid, $Z\in[0.001,0.03]$, remains $<20\%$.
In the left panel, $Y_{\rm SD}$ is nearly constant at $\sim 1.0 \times 10^{-3}~{\rm M_\odot^{-1}}$, while $Y_{\rm DD}$ increases with cosmic time; as a result, $Y_{\rm tot}$ ($=Y_{\rm SD}+Y_{\rm DD}$) rises from $\sim 1.25 \times 10^{-3}~{\rm M_\odot^{-1}}$ to $\sim 1.80 \times 10^{-3}~{\rm M_\odot^{-1}}$ with cosmic time.
In the right panel, the inferred channel slopes span a wide range, from $\alpha_{\rm SD}\simeq-2.0$ to $\alpha_{\rm DD}\simeq0.0$.
$\alpha_{\rm SD}$ varies only weakly with redshift, whereas $\alpha_{\rm DD}$ steepens toward $z=0$.
Accordingly, $\alpha_{\rm tot}$ evolves from $\alpha\simeq-1.4$ to $-1.1$ with cosmic time, approaching the conventional $\alpha\simeq-1$ at $z=0$.
This redshift evolution of the DTD, together with progenitor-channel and progenitor-metallicity dependence (\S\S\,\ref{sec:6.2}), propagates directly into the cosmic-time evolution of SD and DD demographics (\S\,\ref{sec:9}).
}
\label{fig:DTDs_yield_alpha_z_evol}
\end{figure*}

\subsection{Non-universal DTDs: Progenitor-channel and Metallicity Dependence}
\label{sec:6.2}

In \S\S\,\ref{sec:2.4}, we show that the star-particle-level SN~Ia DTD varies systematically across progenitor channels and with progenitor metallicities; even within a single channel, the DTD is not a universal kernel, but an environment-conditioned response function, ${\rm DTD}(\tau;Z)$, whose normalization and slope are both shaped by progenitor metallicity.
Progenitor metallicity is host-dependent, so it naturally translates into host-specific signatures in SN~Ia populations.
Since inter-galaxy metallicity variations can span a factor of $\sim$\,30 (see Appendix\,\ref{appendix:D}), the resulting changes in DTD shape are likely to be both observationally measurable and physically consequential.
In this section, we examine the host dependence of DTDs in detail and use this trend to motivate the non-universal DTD hypothesis.

Figure\,\ref{fig:DTDs_yield_alpha_z0_0p1} shows metallicity-dependent SN~Ia DTDs in the local sample ($0\leq z \leq 0.1$) for different channels (SD, DD, and total) on log--log axes.
In the first row, we bin events by star-particle metallicity (progenitor $Z$; $Z_{{\rm pro},i}$) into 10 intervals and display four representative bins.
Blue and red dots give the SD and DD DTDs, and black dots show their pointwise sum; solid lines are the corresponding power-law fits for SD, DD, and total.
The second row repeats the analysis using host stellar metallicity ($Z_*$).
In both rows, the SD DTD exhibits a complex structure out to $\sim$0.4~Gyr, reflecting the superposed imprints of stellar and CE evolution.
Beyond this timescale, it transitions into a smooth, approximately power-law monotonic decline, followed by a more rapid drop after 3\,Gyr, reaching an SN rate of $\sim$\,$10^{-6}$\,[$\Msun^{-1}\,{\rm Gyr}^{-1}$] by 5\,Gyr.
By contrast, the DD DTD remains approximately stable from the end of its early oscillatory phase at 1~Gyr out to a Hubble time.
We adopt these fitting ranges ($\tau=0.4$--$3.0~{\rm Gyr}$ for SD; $1.0$--$13.8~{\rm Gyr}$ for DD) as our default to maximize the constraining power for each channel, and measure the power-law slopes, $\alpha_{\rm SD}$ and $\alpha_{\rm DD}$, from the SD and DD fits, respectively; the total slope, $\alpha_{\rm tot}$, is fit over $\tau=0.4$--$13.8~{\rm Gyr}$.
Meanwhile, we compute yields ($Y_{\rm SD}$, $Y_{\rm DD}$, and $Y_{\rm tot}$) by integrating the DTDs over the entire $\tau$ range ($\tau=0.0$--$13.8~{\rm Gyr}$); we use the original pointwise data, so the specific slope-fitting ranges do not affect the yield values.

The third row summarizes the resulting SN~Ia DTD yield (left) and slope (right) versus $Z_{{\rm pro},i}$ (filled squares, solid lines) and $Z_*$ (filled circles, dashed lines).
Open symbols mark bins excluded from the fits because $\geq 20\,\%$ of the SN-hosting particles have metallicities outside the COMPAS grid $Z\in[0.001,0.03]$.\footnote{In constructing these bins, we do not extrapolate in $Z$, but assign $Z<0.001$ to $Z = 0.001$ and $Z>0.03$ to $Z = 0.03$ (i.e., piling up at the boundaries).
The open-symbol bins nonetheless largely follow the fitted trends, and including them in the fits yields little change overall, except that the $\alpha_{\rm DD}$ dependences on $Z_*$ and $Z_{{\rm pro},i}$ shift from ${\Delta }\alpha/\Delta\log Z$ = 0.40 to 0.31 and 0.69 to 0.46, respectively.}
In the left panel, $Y_{\rm SD}$ and $Y_{\rm DD}$ are at $\sim$\,$1.0 \times 10^{-3}~{\rm M_\odot^{-1}}$ and $\sim$\,$0.9 \times 10^{-3}~{\rm M_\odot^{-1}}$, respectively, and $Y_{\rm tot}$ ($=Y_{\rm SD}+Y_{\rm DD}$) at $\sim$\,$1.9 \times 10^{-3}~{\rm M_\odot^{-1}}$.
Notably, the SD- and DD-DTDs in our model produce comparable integrated production (i.e., yield) without any additional normalization adjustment, although this agreement does not appear to be physically enforced.
$Y_{\rm SD}$ shows little dependence on $Z$, while $Y_{\rm DD}$ decreases toward higher $Z$, causing a corresponding decline in $Y_{\rm tot}$; the trend is stronger for $Z_{{\rm pro},i}$ than for $Z_*$, indicating that $Z_{{\rm pro},i}$ underlies the yield--metallicity relation.
In the right panel, the inferred channel slopes differ markedly, ranging from a steeper $\alpha_{\rm SD}\simeq -2.2$ to a shallower $\alpha_{\rm DD}\simeq -0.8$.
Interestingly, despite this pronounced channel-to-channel difference, their combination around $\tau \simeq 1.5$\,Gyr yields a total DTD that remains broadly consistent with the conventional $\alpha_{\rm tot}\simeq -1$, again without any explicit normalization tuning.
Consistent with the yield behaviour, $\alpha_{\rm SD}$ varies only weakly with $Z$, whereas $\alpha_{\rm DD}$ increases with increasing $Z$; $\alpha_{\rm tot}$ shows weak metallicity dependence.
The metallicity response is systematically larger for $Z_{{\rm pro},i}$ than for $Z_*$: the $\alpha_{\rm DD}$ response is larger by a factor of 1.7 in $Z_{{\rm pro},i}$ binning than in $Z_*$ binning, reinforcing that $Z_{{\rm pro},i}$ drives the apparent trends.
Overall, the SD and DD slopes differ by 2.8 times, and both yield and slope are more sensitive to $Z_{{\rm pro},i}$ than to $Z_*$, as expected because $Z_*$ traces only the mean of the broad within-galaxy $Z_{{\rm pro},i}$ distribution (Appendix\,\ref{appendix:D}).

Empirical reconstructions have suggested that the SN~Ia DTD is broadly described by a power law, ${\rm DTD}(\tau)\propto \tau^{\alpha}$ with $\alpha\,\simeq\,-1$ across a range of environments, motivating its widespread use as an approximately universal response kernel for convolution with galaxy SFHs \citep[e.g.,][]{Greggio05,Maoz2012,Maoz2014,Maoz2017}.
However, strict universality is not necessarily expected on physical grounds, because the mapping from initial binary parameters to exploding systems may depend on metallicity through winds, core growth, and compact-object masses.
For instance, \citet{Meng2009} showed with BPS calculations that the SD DTD depends on progenitor metallicity, with lower-$Z$ populations tending to yield more delayed responses.
From a volumetric SN~Ia rate observation, \citet{Strolger2010} inferred a DTD weighted toward longer delays and suggested that environmental factors, including metallicity, contribute to the apparent discrepancies with more prompt-weighted inferences.
In this context, our forward modeling of SN~Ia populations suggests that the DTD is intrinsically non-universal, and that the commonly adopted single DTD is more appropriately interpreted as a population-averaged approximation.
Within each progenitor channel, both the normalization and the slope vary systematically with progenitor metallicity and therefore with host-galaxy properties such as stellar mass and mean metallicity.

\subsection{Non-universal DTDs: Redshift Evolution}
\label{sec:6.3}

The star-particle-level DTD shows a systematic dependence on progenitor metallicity (\S\S\,\ref{sec:2.4} and \S\S\,\ref{sec:6.2}).
The implication is immediate: the SN~Ia DTD is not a single universal kernel, but an environment-dependent response function with metallicity-dependent normalization and effective slope.
Because progenitor metallicity itself evolves with redshift, this metallicity dependence is expected to map directly onto redshift-dependent trends in SN~Ia populations.
The effect should not be dramatic---the mean metallicity of the overall galaxy population shifts by only $\sim$\,0.3\,dex over $0 \leq z \leq 3$ (Appendix\,\ref{appendix:D})---but it should be measurable.
In this section, we quantify the redshift dependence of the DTD and use it to motivate the non-universal DTD picture.

Figure\,\ref{fig:DTDs_yield_alpha_z_evol} is analogous to Fig.\,\ref{fig:DTDs_yield_alpha_z0_0p1}, but for redshift-dependent SN~Ia DTDs.
The upper row repeats the Fig.\,\ref{fig:DTDs_yield_alpha_z0_0p1} analysis after binning events into 10 equal-number redshift intervals, and shows four representative bins ($z \simeq$ 0, 1, 2, and 3).
A key effect is the look-back-time truncation: toward higher redshift, the $\tau$ range available for fitting the DD DTD slope is progressively cut off from the old-age end, and the DD DTD slope becomes shallower.
More importantly, the late-time DD events that support the tail of the total DTD are increasingly absent at high $z$, and the inferred $\alpha_{\rm tot}$ becomes correspondingly steeper.
The lower row summarizes the resulting yields (left) and slopes (right) as functions of redshift.
In the left panel, $Y_{\rm SD}$ remains nearly constant at $\sim$\,$1.0 \times 10^{-3}~{\rm M_\odot^{-1}}$, whereas $Y_{\rm DD}$ increases with cosmic time; consequently, $Y_{\rm tot}$ ($=Y_{\rm SD}+Y_{\rm DD}$) rises from $\sim$\,$1.25 \times 10^{-3}~{\rm M_\odot^{-1}}$ to $\sim$\,$1.8 \times 10^{-3}~{\rm M_\odot^{-1}}$ with cosmic time. 
In the right panel, the inferred channel slopes differ markedly, spanning from $\alpha_{\rm SD}\simeq -2.0$ (steeply declining) to $\alpha_{\rm DD}\simeq -1.0$ to $0.6$ depending on $z$ (shallower, flattening, and even rising).
The $\alpha_{\rm SD}$ value shows only weak redshift dependence, whereas $\alpha_{\rm DD}$ steepens toward $z=0$.
Notably, $\alpha_{\rm tot}$ evolves from $\alpha\simeq -1.4$ to $-1.1$ with cosmic time, reaching the conventional $\alpha\simeq -1$ at $z = 0$.
Taken together, (i) the strong contrast between $Y_{\rm SD}$ and $Y_{\rm DD}$, (ii) the redshift evolution of the total yield ($Y_{\rm tot}$), (iii) the distinct behaviours of $\alpha_{\rm SD}$ and $\alpha_{\rm DD}$, and (iv) the redshift evolution of the combined slope ($\alpha_{\rm tot}$)---in addition to the metallicity dependence established in the previous section---make a compelling case for a non-universal SN~Ia DTD.
This redshift evolution of the DTD, combined with progenitor-channel and progenitor-metallicity dependence, propagates directly into the cosmic-time evolution of SD and DD demographics, as we show in \S\,\ref{sec:9}.

\section{Progenitor Ages of SN\MakeLowercase{e}~I\MakeLowercase{a}}
\label{sec:7}

We now shift from the delay-time description to the progenitor-age frame, which is the most direct bridge between intrinsic explosion timescales and the observed host demographics of cosmology samples.
The progenitor age is the age of the SN-producing system at explosion; its distribution (i.e., SPAD) at a given epoch is obtained by convolving the DTD with each host’s SFH.
Unlike the DTD, which isolates intrinsic binary-evolution timescales, the SPAD explicitly shows how these timescales are realized across heterogeneous galaxy populations and thus how they can correlate with host properties.
In this section, we model progenitor ages across the host-galaxy population and characterize their dependence on stellar mass and sSFR.

We organize this section into two complementary representations of the progenitor age predicted for each individual SN event in our model: the \emph{host-level mean} progenitor age and the \emph{event-level} progenitor age.
First, following the philosophy of \S\,\ref{sec:4}, we construct a host-aggregated quantity by averaging the progenitor ages of all individual SN events occurring within a given host galaxy, and we use this host-level mean to expose the physical connections to host properties.
Second, reflecting the observational situation in which $\sim$\,99\,\% of host galaxies contribute only a single SN~Ia (\S\S\S\,\ref{sec:2.7.3}), we analyze progenitor age as an event-level quantity, linking each SN~Ia directly to its host properties.
The two representations are complementary, differing in the physical unit of interpretation (galaxy versus event) and in the observational mapping from the model to survey data.
Accordingly, the host-level summary provides an efficient description of galaxy-scale behaviour, particularly steps/transitions, while the event-level treatment is appropriate for survey comparison and for assessing cosmology bias in standardization, because the bias may arise from individual SNe entering the Hubble diagram.
\S\S\,\ref{sec:7.1} examines the host-level mean progenitor age, whereas \S\S\,\ref{sec:7.2} and \ref{sec:7.3} focus on the event-level progenitor ages of individual SNe~Ia.

\begin{figure*}
\includegraphics[width=0.95\textwidth]{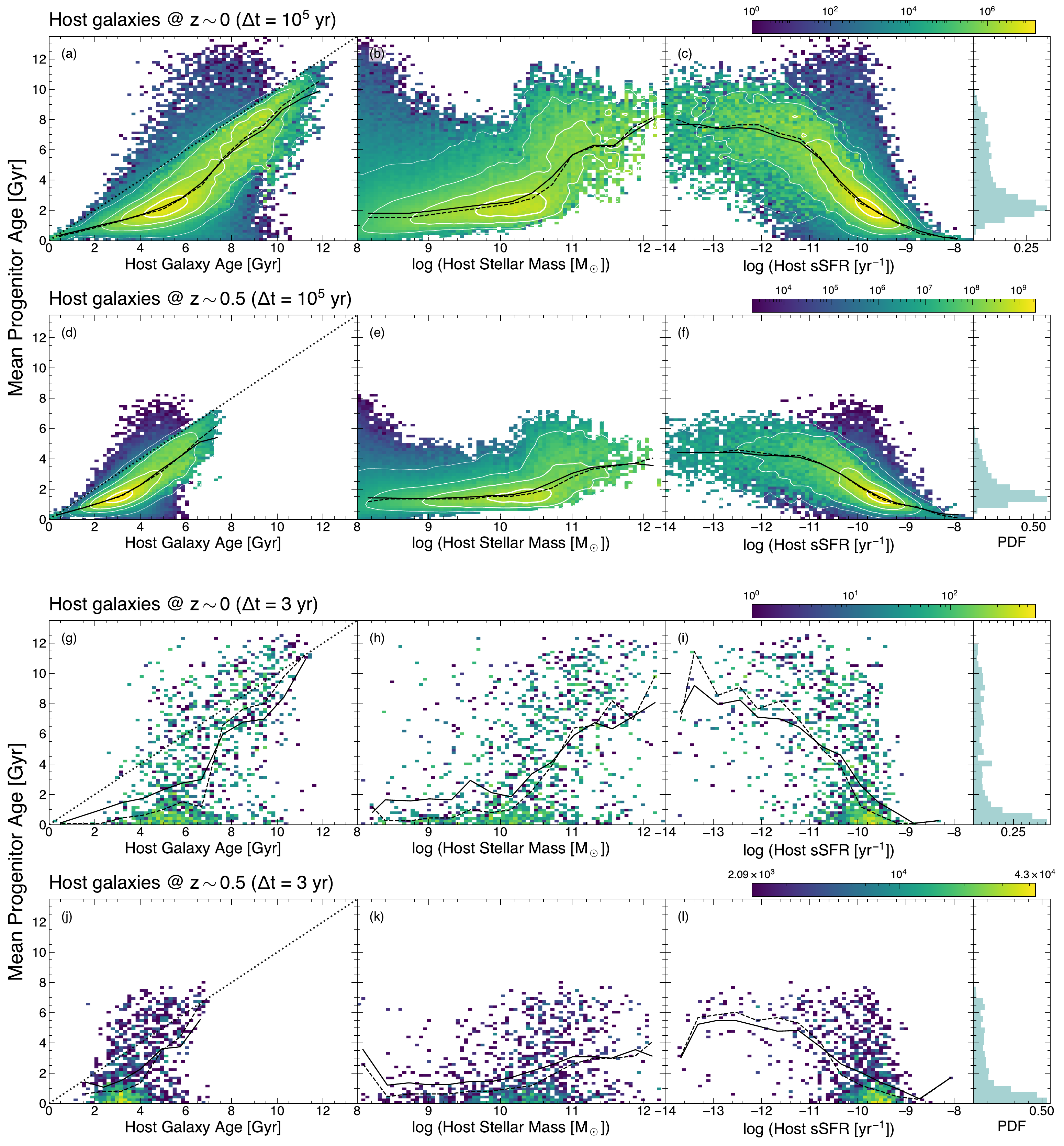}
\caption{
Host mean SN~Ia progenitor age ($T_{\rm pro}$) as a function of host-galaxy properties for two redshift slices and two observational time-window choices.
\textbf{\textit{(a--f)}} Two redshift slices---$z \simeq 0$ ($0.0\le z\le 0.1$; $a$--$c$) and $z \simeq 0.5$ ($0.45\le z\le 0.55$; $d$--$f$)---for $\Delta t = 10^{5}\,\mathrm{yr}$.
Panels ($a$ \& $d$) show $T_{\rm pro}$ versus host age ($T_*$), with the dotted line indicating equality, and panels ($b$ \& $e$) and ($c$ \& $f$) show $T_{\rm pro}$ versus host stellar mass ($M_*$) and sSFR, respectively.
Colored 2D maps indicate the logarithmic galaxy number density (color bar); white contours mark the 0.3, 1, 2, and 3\,$\sigma$ levels.
Solid and dashed curves trace the mean and median trends, respectively.
We identify SNe~Ia and their associated hosts within $\Delta t = 10^{5}$\,$\mathrm{yr}$ and record each SN--host pair as a separate entry in the host catalogue; accordingly, a galaxy producing $N$ SNe~Ia over this interval appears $N$ times, and all host-demography statistics are explicitly {\it event-weighted}.
At $z\simeq0$, $T_*$ captures the overall trend of $T_{\rm pro}$, motivating $T_*$ as a practical proxy.
The relations of $T_{\rm pro}$ with both $M_*$ and sSFR are nonlinear and S-shaped, each exhibiting a rapid transition; these transitions are sharper than the trends in the $M_*$--$T_*$ and sSFR--$T_*$ planes (Fig.\,\ref{fig:Demo_z0p1_and_z0p55}).
The right-hand marginal panels show the corresponding 1D probability density functions of $T_{\rm pro}$.
At $z\simeq0.5$, the same qualitative structure persists but is shifted toward younger ages, reflecting the younger Universe.
\textbf{\textit{(g--l)}} Same as ($a$--$f$), but using $\Delta t = 3~\mathrm{yr}$, enforcing the single-event-per-host limit.
In this regime, the plots are effectively the \emph{event-level} representation of progenitor age discussed in \S\S\,\ref{sec:6.2} (Fig.\,\ref{fig:Prog_age_event}).
Despite the substantial counting noise, the same qualitative dependencies remain visible, indicating that $T_*$, $M_*$, and sSFR still retain useful information about $T_{\rm pro}$ (equivalently, progenitor age in the single-event regime).
}
\label{fig:Host_Prog_age}
\end{figure*}

\subsection{Host-level Mean Progenitor Ages: Host-property and Redshift Dependence}
\label{sec:7.1}

In this section, we adopt a \emph{host-level} mean progenitor age ($T_{\rm pro}$), defined as the average of progenitor ages ($T_{{\rm pro},i}$) for the SN~Ia events in a given host galaxy.
The $T_{\rm pro}$ value is implicitly weighted by the event multiplicity: more massive star particles contribute more progenitors, and the contribution is also modulated by star-particle age through the age-dependent production efficiency encoded in the DTD.
Thus, $T_{\rm pro}$ is best viewed as an event-weighted (rather than purely mass- or age-weighted) summary statistic of the progenitor population within each host.
The use of the host-level mean is advantageous in two respects:
(i) it reduces sampling variance, sharpening the central tendency of underlying host-property--$T_{{\rm pro},i}$ signals;
and (ii) it offers a compact summary statistic tied to the host properties, enabling a clean description of galaxy-scale phenomenology (e.g., $T_{\rm pro}$ as a function of $T_*$, $M_*$, or sSFR).

Figure\,\ref{fig:Host_Prog_age} presents how the host-level mean progenitor age ($T_{\rm pro}$) depends on host stellar age ($T_*$), stellar mass ($M_*$), and sSFR. 
The upper two rows compare $0 \leq z \leq 0.1$ (panels $a$--$c$) and $z\simeq0.5$ (panels $d$--$f$). 
We identify SNe~Ia and their hosts within $\Delta t=10^5$~yr and record each SN--host pair as a separate entry in the host catalogue; accordingly, all host-demography statistics are {\it event-weighted}.
At $z\simeq0$, panel ($a$) shows that $T_*$ captures the overall trend of $T_{\rm pro}$, motivating $T_*$ as a practical proxy: this is important because progenitor ages cannot be measured directly.
Panels ($b$) and ($c$) show that $T_{\rm pro}$ exhibits nonlinear, S-shaped relations with both $M_*$ and sSFR, each characterized by a rapid transition that we term the ``progenitor-age step''; these transitions are sharper than the corresponding trends in the $M_*$--$T_*$ and sSFR--$T_*$ planes (Fig.\,\ref{fig:Demo_z0p1_and_z0p55}).
At $z\simeq0.5$, the same qualitative structure is retained but shifted to younger ages in the younger Universe.
The $T_*$--$T_{\rm pro}$ relation remains similar, whereas both hosts and progenitors are younger.
The S-shaped nonlinearity persists in $M_*$--$T_{\rm pro}$ and sSFR--$T_{\rm pro}$ relations, but the step amplitudes are reduced relative to $z\simeq0$, whereas the characteristic host mass and sSFR at which the transitions occur remain broadly unchanged.
We discuss the implications of these redshift trends further in \S\S\,\ref{sec:7.3}.

\begin{figure*}
\includegraphics[width=0.95\textwidth]{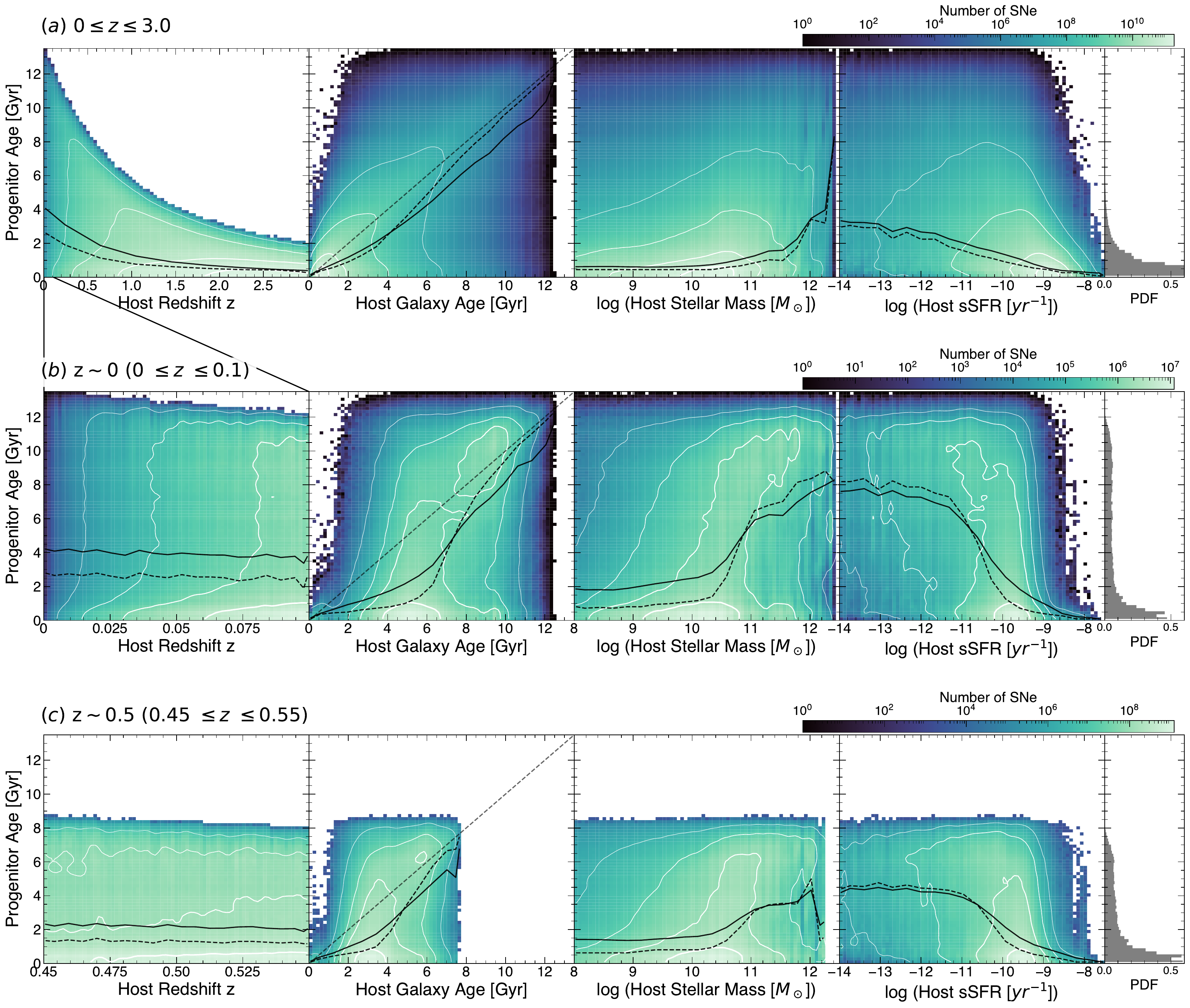}
\caption{
Progenitor ages of individual SNe~Ia as a function of host-galaxy properties: ($a$) $0 \leq z \leq 3$ (global sample), ($b$) $0 \leq z \leq 0.1$ (local sample), and ($c$) $0.45 \leq z \leq 0.55$ (the $z$\,$\simeq$\,0.5 sample).
From left to right, panels show progenitor age versus host redshift, host mass-weighted mean stellar age, host stellar mass, and host sSFR.
Colored 2D maps show the logarithmic number density of SNe~Ia (dark-blue-to-teal-to-pale-cyan ``mako'' color scale), with white contours indicating the 0.3, 1, 2, and 3\,$\sigma$ levels.
Solid and dashed black curves denote the mean and median relations, respectively, and the right-hand marginal panels show the 1D PDFs of progenitor ages.
\textbf{\textit{(a)}} The second panel shows a tight correspondence ($r=0.99$) between host mass-weighted mean stellar age and the mean SN~Ia progenitor age.
The third and fourth panels show the distributions with host stellar mass and sSFR, respectively.
\textbf{\textit{(b \& c)}} Same as ($a$), but for the local and $z$\,$\simeq$\,0.5 samples.
The $z$\,$\simeq$\,0.5 sample is limited to progenitor ages $\lesssim$\,8.5~Gyr and host ages $\lesssim$\,7.5~Gyr; otherwise, the two samples exhibit broadly similar structure.
The third and fourth panels exhibit strong nonlinearities, reflecting a sharp transition from young to old progenitors.
If HRs depend on progenitor age, the observed host-mass and host-sSFR steps are most naturally interpreted as a ``progenitor-age step.''
}
\label{fig:Prog_age_event}
\end{figure*}

Our model allows each host to contribute multiple SN~Ia events by construction.
Averaging over multiple events suppresses shot noise (i.e., Poisson/counting noise from finite event statistics) and can therefore overstate the apparent tightness of the host-property--$T_{{\rm pro},i}$ mapping.
Accordingly, the relations in ($a$--$f$) should be interpreted as an upper limit on the achievable precision and as mean trends that likely underestimate the true per-event scatter.
To connect directly to the observational regime in which most hosts contribute only one SN, the lower two rows of Fig.\,\ref{fig:Host_Prog_age} adopt the observational time window $\Delta t = 3~\mathrm{yr}$ comparable to the baselines of actual surveys, enforcing the single-event-per-host limit.
In this regime, the figure is essentially identical to the event-level representation of progenitor age discussed in the next section (\S\S\,\ref{sec:7.2}, Fig.\,\ref{fig:Prog_age_event}), because each host is assigned the progenitor age of its individual SN~Ia rather than being reduced to a single mean value over multiple events.
For $\Delta t = 3$ yr, we have a host fraction of $\sim$\,0.8\,\% that corresponds to $\sim$\,$3.6\times10^8$ hosts, of which a vast majority of hosts ($\sim$\,95\,\%) contain one SN~Ia.
Despite the substantial counting noise, the qualitative dependencies remain evident, indicating that $T_*$, $M_*$, and sSFR still encode useful information about host-level $T_{\rm pro}$ (equivalently, the progenitor age of the observed SN~Ia) even in the single-event regime.

\begin{figure*}
\includegraphics[width=0.8\textwidth]{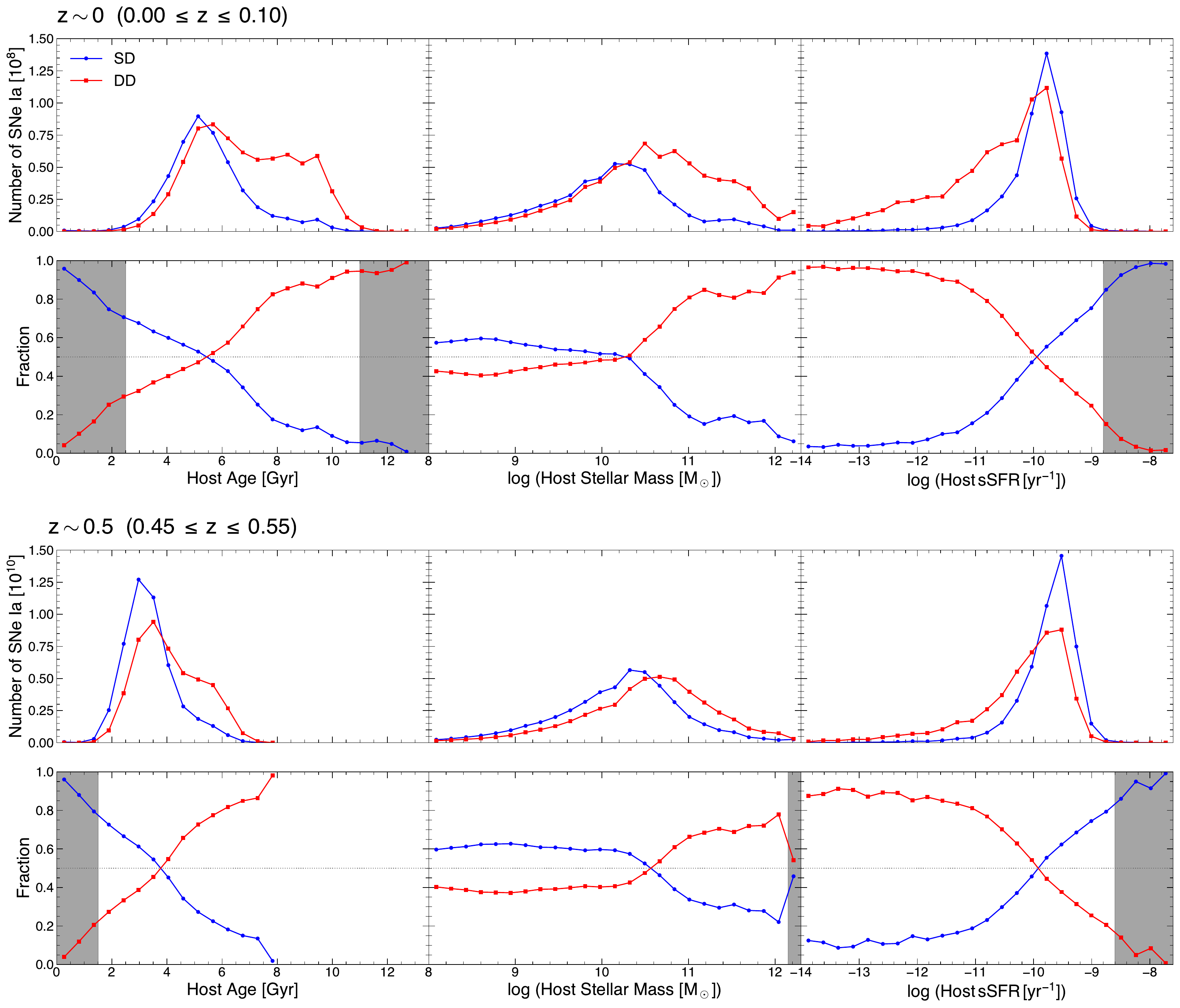}
\caption{
The SD (blue) and DD (red) event counts and their fractional contributions as functions of host-galaxy properties in two redshift slices: $0 \leq z \leq 0.1$ (upper two rows) and $0.45 \leq z \leq 0.55$ (lower two rows).
\textbf{\textit{(Upper two rows)}} For the local sample, each column shows host age, host mass, and host sSFR, matching the 2nd--4th panels of Fig.\,\ref{fig:Prog_age_event}($b$--$c$).
The first row gives the binned SN~Ia counts, and the second row shows the corresponding SD/DD fractions (dotted line: 50\,\%; grey shading: bins with negligible contributions from both channels).
In the local sample, young ($<5.5$~Gyr), low-mass ($<10^{10.4}~{\rm M}_{\odot}$), and high-sSFR ($>10^{-10.0}~{\rm yr}^{-1}$) hosts yield comparable SD and DD counts, whereas old ($>5.5$~Gyr), massive ($>10^{10.4}~{\rm M}_{\odot}$), and low-sSFR ($<10^{-10.0}~{\rm yr}^{-1}$) hosts are DD-dominated.
\textbf{\textit{(Lower two rows)}} For the $z\simeq0.5$ sample, the same qualitative dependencies on host age, mass, and sSFR persist, but the host-age distribution is truncated at $\lesssim 8$~Gyr by the younger cosmic age.
The mass- and sSFR-dependent trends remain similar, while the overall DD fraction is reduced relative to the local sample.
}
\label{fig:SDDD_abs_frac}
\end{figure*}

\subsection{Event-level Individual Progenitor Ages: Progenitor-age Step as an Origin of the Mass/sSFR Steps in HR}
\label{sec:7.2}

In this section, we adopt an event-level representation of progenitor age, in which each host is assigned the progenitor age of its individual SN~Ia rather than being reduced to a single mean value over multiple events.
This choice is advantageous in three respects:
(i) it is motivated by, and well matched to, the observational reality that most galaxies contribute only one SN over a finite survey baseline, making the data intrinsically event-sampled;
(ii) it preserves the full shape of the progenitor-age distribution---its width, tails, and channel mixtures (e.g., SD/DD mixtures)---whereas host averaging would erase the internal structure that can modulate cosmology-relevant biases;
and (iii) it connects most directly to standardization systematics: if HRs depend on progenitor age, the operative quantity is the age of each SN (i.e., which events are intrinsically brighter or fainter), not the host-averaged age.

Figure\,\ref{fig:Prog_age_event} presents how progenitor ages of individual SNe~Ia ($T_{{\rm pro},i}$) depend on their host-galaxy properties for ($a$) $0 \leq z \leq 3$ (the global sample), ($b$) $0 \leq z \leq 0.1$ (the local sample), and ($c$) $0.45 \leq z \leq 0.55$ (the $z$\,$\simeq$\,0.5 sample).
In the top row ($a$), the first panel shows the redshift versus $T_{{\rm pro},i}$ distribution.
At high redshift, progenitor ages are tightly concentrated at young values, whereas toward low redshift, the distribution broadens markedly as cosmic aging allows a much wider range of progenitor ages to be realized within galaxies.
By late times, the emergence of massive, quiescent hosts leads to the production of tardy SN~Ia from old stellar populations, progressively shifting the low-$z$ SN~Ia population toward older progenitors.
The second panel shows the $T_*$ versus $T_{{\rm pro},i}$ distribution, revealing a tight mapping between $T_*$ and $T_{{\rm pro},i}$ ($r=0.99$).
The $M_*$ (third) and sSFR (fourth) panels show the scarcity of very young SNe~Ia in high-mass and low-sSFR galaxies.
This behaviour is expected because such galaxies are typically quiescent, lack recent SF, and therefore rarely host the young stellar populations required to produce prompt SNe~Ia \citep[e.g.,][]{Childress2014}.
Overall, for the global sample, the progenitor age correlates with all four host properties, but the mass-weighted mean stellar age is the most direct and physically informative proxy.

In the middle and bottom rows ($b$ and $c$) of Fig.\,\ref{fig:Prog_age_event}, we repeat the analysis over $0 \leq z \leq 0.1$ and $0.45 \leq z \leq 0.55$ to illustrate the local and $z$\,$\simeq$\,0.5 relations, respectively.
Aside from the fact that the $z$\,$\simeq$\,0.5 sample is confined to progenitor ages $\lesssim$\,8.5~Gyr and $T_*$ $\lesssim$\,7.5~Gyr, the overall structure of the two samples remains similar.
In the third panels, $M_*$ serves only as an indirect proxy for the mean progenitor age because the relation is strongly nonlinear, yet it offers valuable insight into observed SN~Ia behaviour: the ``mass step'' in the host-mass versus HR plane \citep{Kelly2010,Sullivan2010,Lampeitl2010,Childress2013,Rigault2020, Wiseman2022, Chung2023} can be interpreted as an observational projection of a sharply nonlinear $M_*$--progenitor-age relation.
In our model, the nonlinearity emerges because $T_{\rm pro}$ rises rapidly around $M_* \simeq 10^{10.5}\,\Msun$.
This characteristic transition reflects the well-known bimodality between star-forming and quenched galaxies.
The host-mass step therefore likely arises not from a direct causal dependence of SN luminosity on stellar mass itself, but from a transition in the underlying SN~Ia progenitor population.
In the fourth panels, a closely analogous interpretation applies to sSFR: the nonlinear sSFR--progenitor-age relation mirrors the $M_*$ trend.
Here again, the transition occurs around ${\rm sSFR}\simeq10^{-10.5}~\mathrm{yr}^{-1}$, where the mean progenitor age rises rapidly.
This progenitor-age step may therefore underlie the reported correlations between host sSFR and HRs \citep[e.g.,][]{Sullivan2010, Rigault2020}.
Overall, if HRs depend on progenitor age, then the observed host-mass and host-sSFR steps are most naturally interpreted as manifestations of a ``sharp prompt-to-tardy transition'' \citep{Chung2023} or, equivalently, a ``progenitor-age step''.

To provide insight into the SD/DD mixture, Fig.\,\ref{fig:SDDD_abs_frac} presents the SD and DD event counts and their fractional contributions as functions of host-galaxy age, stellar mass, and sSFR, matching the second through fourth panels of Fig.\,\ref{fig:Prog_age_event}($b$--$c$), for the local and $z\simeq0.5$ samples.
In the first row, for the local sample, young ($<5.5$~Gyr), low-mass ($<10^{10.4}~{\rm M}_{\odot}$), and high-sSFR hosts ($>10^{-10.0}~{\rm yr}^{-1}$) produce comparable numbers of SD and DD events, whereas old ($>5.5$~Gyr), massive ($>10^{10.4}\,{\rm M}_{\odot}$), and low-sSFR hosts ($<10^{-10.0}\,{\rm yr}^{-1}$) are DD-dominated.
The second row shows the corresponding channel fractions and makes the SD-to-DD turnover along the host sequences explicit.
Importantly, this turnover occurs at the same loci where the mean progenitor age rises sharply, namely near the quasi-inflection points of the nonlinear relations, around $M_*\,\simeq\,10^{10.5}\,{\rm M}_{\odot}$ and ${\rm sSFR}\simeq10^{-10.5}\,\rm{yr}^{-1}$ (Fig.\,\ref{fig:Prog_age_event}).
In the third row, the $z\simeq0.5$ sample exhibits qualitatively similar dependence on host age, mass, and sSFR, although the host-age distribution is truncated at $\lesssim 8$~Gyr because of the younger cosmic age.
The mass- and sSFR-dependent trends remain similar, although the overall DD contribution is lower than in the local sample.
The fourth row shows the corresponding SD and DD fractions for the same sample.

\subsection{Cosmological Implications of Host-dependent SN~Ia Standardization}
\label{sec:7.3}

In the previous section (\S\S\,\ref{sec:7.2}), we modeled progenitor ages across the host-galaxy population and quantified how they vary with host properties.
The variation can be compactly described as a ``progenitor-age step'' along the stellar-mass and sSFR sequences.
Under the hypothesis that standardized SN~Ia HRs carry a progenitor-age dependence, mapping this intrinsic demographic step into standardized luminosities naturally yields the host-mass and host-sSFR magnitude steps observed in HRs \citep[e.g.,][]{RoseGarnavichBerg2019,Rose2021}.
If borne out, the SN-progenitor-age step provides a unified physical origin for these two widely discussed host-dependent standardization trends: the host-mass and host-sSFR magnitude-step phenomenology.

\begin{figure*}
\includegraphics[width=\textwidth]{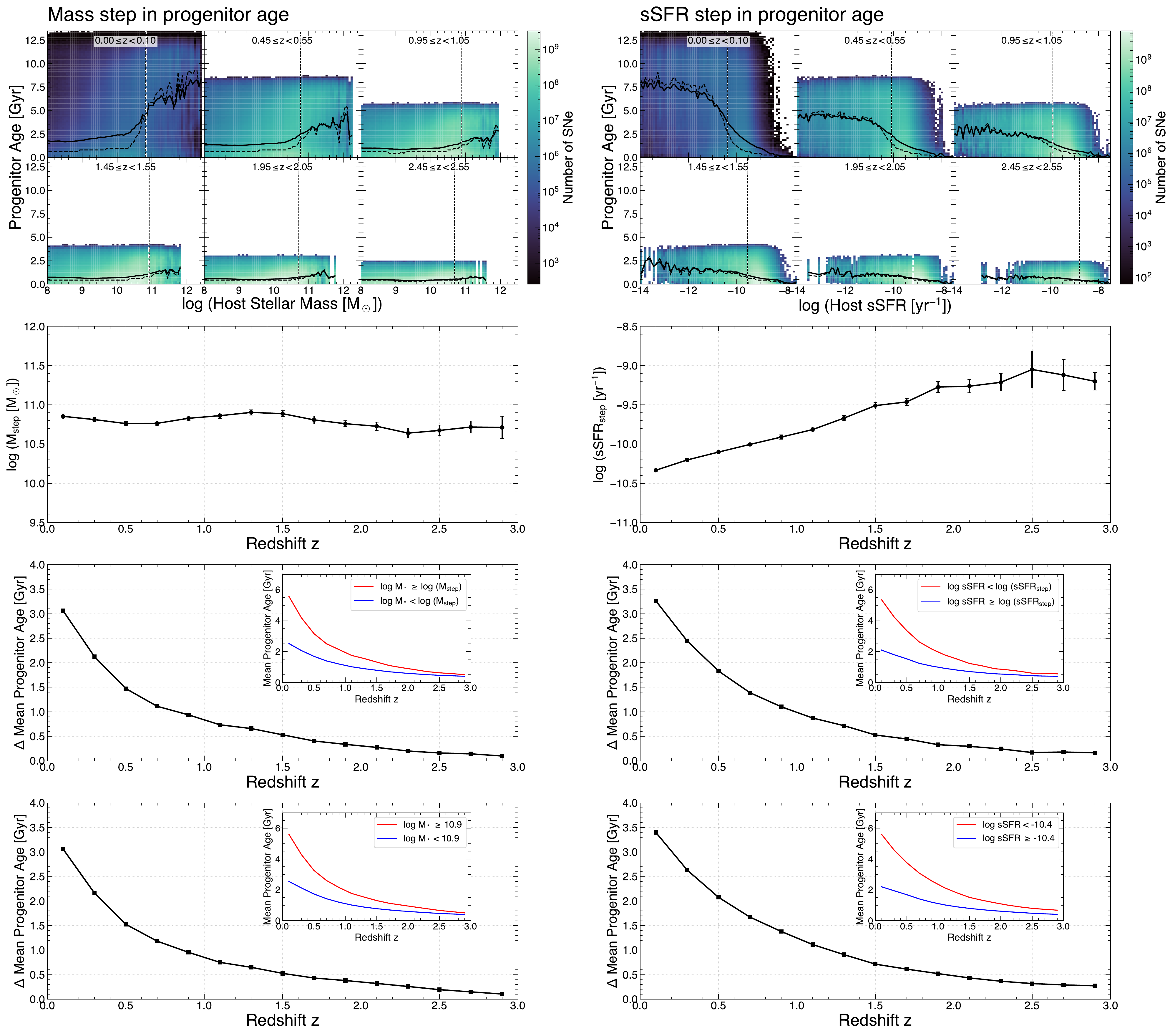}
\caption{
Cosmic evolution of the host-mass-driven and host-sSFR-driven ``steps'' in progenitor ages. 
\textbf{\textit{(Left column)}} Host-mass-driven progenitor-age step in the progenitor age versus host mass plane.
Top-row six panels show colored 2D maps of the logarithmic SNe~Ia number density (color bar) in six redshift bins, $z \simeq$ 0.0, 0.5, 1.0, 1.5, 2.0, and 2.5; solid and dashed black curves trace the mean and median relations, respectively. 
In each bin, $M_{\rm step}$ is defined as the mass threshold (vertical dashed lines) that maximizes the contrast in mean progenitor age between the two resulting subsamples (i.e., the progenitor-age ``step''). 
Rather than identifying the step location from a piecewise-constant fit, we fit a `sigmoid' function to the mean progenitor-age relation and adopt its center as the effective step location. 
This provides a more robust summary of the transition scale when the underlying relation is smooth and gradually varying. 
In particular, at high redshift the progenitor-age dependence on host sSFR becomes compressed and less sharply step-like, making a discontinuous-step model poorly suited. 
The quoted error bars in the second-row panels correspond to the uncertainty in the fitted sigmoid center.
The reduced step amplitude toward higher redshift is not sensitive to the precise choice of transition points; instead, cosmic-age (look-back time) compression of the progenitor-age distribution suppresses the contrast and dominates over the details of how the host split is defined.
Second-row panels show $\log M_{\rm step}[\Msun]$ versus redshift; over $0 \le z \le 3$, the transition mass is approximately constant.
For SN~Ia peak-luminosity standardization, the conventional mass-step correction (the $\gamma$ term) can be conducted using $M_{\rm step}$ at $z=0$. 
Third row shows the evolution of the mean progenitor-age difference between subsamples split by $M_{\rm step}(z)$ ($\Delta$ mean progenitor age); insets show the mean progenitor-age evolution of each subsample. 
The progenitor-age step grows steadily toward $z=0$ and exhibits strong redshift dependence.
Bottom row show the mean progenitor-age difference for a fixed split mass at $z=0$, $\log M_{\rm step}[\Msun]=10.9$.
\textbf{\textit{(Right column)}} Same as the left column, but for the host-sSFR-driven progenitor-age step.
In the top-row six panels, vertical dashed lines mark ${\rm sSFR}_{\rm step}$.
Second row shows $\log{\rm sSFR}_{\rm step}$ versus redshift; the transition sSFR increases by $\sim$1~dex toward higher $z$ over $0 \le z \le 3$.
For SN~Ia peak-luminosity standardization, the sSFR-based correction should be conducted within each separate redshift interval.
Third row shows the evolution of the mean progenitor-age difference between subsamples split by ${\rm sSFR}_{\rm step}(z)$. 
The progenitor-age step grows steadily toward $z=0$ and exhibits strong redshift dependence.
Bottom row shows the mean progenitor-age difference for a fixed split sSFR at $z=0$, $\log {\rm sSFR}_{\rm step}[{\rm yr}^{-1}]=-10.4$.
Applying host-sSFR step correction using the $z=0$ split points at other redshifts is unlikely to introduce a systematic bias.
}
\label{fig:Step_evolution_deltas}
\end{figure*}

To build intuition for the conventional host-mass-step correction (the $\gamma$ term) and its host-sSFR analogue in SN~Ia standardization, Figure\,\ref{fig:Step_evolution_deltas} shows how SN~Ia progenitor ages vary with host mass and host sSFR across redshift.
In the first row, we identify the sharp progenitor-age transition along each host-property sequence and quantify how its location drifts with redshift.
Within each redshift bin, we determine the transition point in the host property (mass or sSFR) by fitting a sigmoid function to the mean progenitor age, and define the ``step'' location as the center ($x_0$) of the best-fit sigmoid.
Because the cosmic age decreases toward higher redshift, the intrinsic progenitor-age distribution becomes progressively compressed, so the age contrast induced by any mass- or sSFR-based split correspondingly weakens with increasing $z$.
The second row shows that, over $0 \le z \le 3$, the transition mass remains nearly constant, reinforcing the view that stellar mass acts as an indirect proxy rather than as the direct physical driver.
In contrast, the transition sSFR shifts upward by as much as $\sim$1.3~dex toward higher $z$.
This indicates that sSFR responds more directly than mass to galaxy evolution, making it the more direct axis for capturing the redshift dependence of the transition.

Given the transition mass and sSFR at each redshift, the third row shows the mean progenitor ages on either side of the split and their difference across redshift.
For both, the progenitor-age step grows steadily toward $z=0$ and exhibits strong redshift dependence, implying that any HR mass step that is mediated by progenitor age should itself be redshift-dependent.
The fourth row demonstrates that these trends do not depend on allowing the split points to drift with redshift.
Adopting fixed $z=0$ thresholds, $\logM=10.9$ and $\logsSFR=-10.4$, produces nearly identical redshift evolution in the mean progenitor ages and in their difference.
Accordingly, applying host-mass and host-sSFR step corrections using the $z=0$ split points at other redshifts is unlikely to, by itself, introduce a systematic bias.

In a framework in which standardized SN~Ia luminosities remain directly or indirectly sensitive to progenitor ages \citep[e.g.,][]{Gallagher2008}, host-driven ``steps'' are best viewed as proxy corrections for an underlying progenitor-age step.
Our results imply that even an optimally chosen transition mass or sSFR does not, in general, yield a redshift-invariant step amplitude.
Hence, a single redshift-invariant step coefficient (e.g., a constant $\gamma$) should be regarded as an approximation whose adequacy should be tested by allowing $\gamma$ (and any analogous sSFR-step coefficient) to vary with redshift \citep[e.g.,][]{Scolnic2018,Vincenzi2024,Childress2014}.
Practically, this motivates performing a host-mass step correction or an sSFR-based step correction within each redshift interval (but not both simultaneously, since applying both risks over-correction), while using the $z=0$ split points (in mass or sSFR) at other redshifts is acceptable.

\section{SN I\MakeLowercase{a} Rates in Cosmic Contexts: Redshift Evolution}
\label{sec:8}

The cosmic SN~Ia rate evolution encodes information about single and binary stellar evolution within galaxies, galactic and cosmic chemical enrichment, and cosmological inference \citep[e.g.,][]{Matteucci2009, Dubay24, DES2024}.
Most previous studies have inferred this evolution by convolving an observationally constrained cosmic SFH with parameterized DTDs \citep[e.g.,][]{Childress2014,Wiseman2021,Wiseman2022,Palicio24}.
We instead obtain the cosmic SN~Ia rate via fully forward modeling: we start from individual star particles, evolve their binary populations, and directly count SN~Ia events.
In this section, we place these simulated SN~Ia event counts self-consistently in a cosmological context by presenting the cosmic volumetric SN~Ia rate as a function of redshift.

\begin{figure*}
\includegraphics[width=0.90\textwidth]{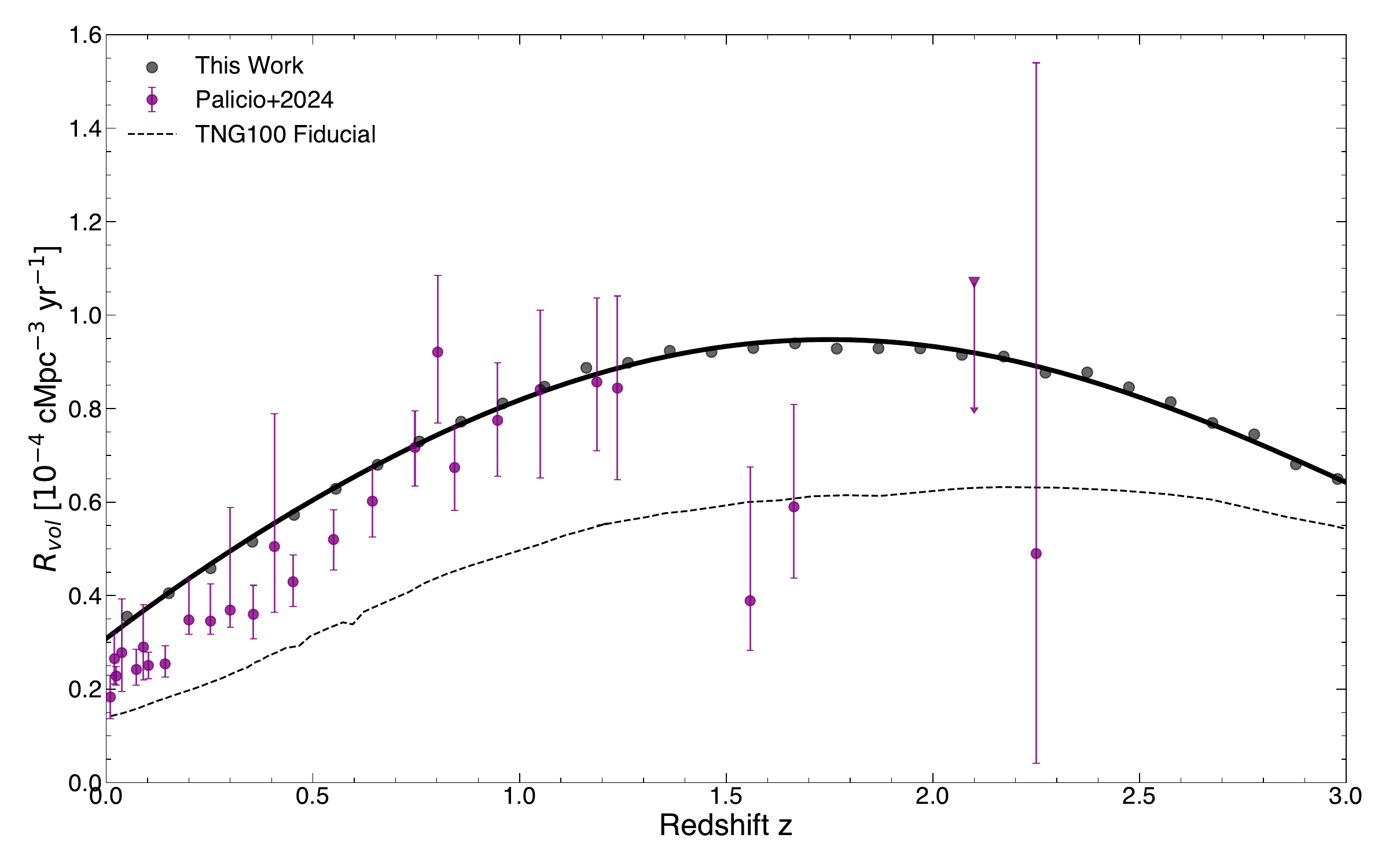}
\caption{
Cosmic evolution of the volume-normalized SN~Ia rate ($R_{\rm vol}$) over $0 \le z 
\le 3$.
Black points show our model predictions for $\Delta t = 10^5$ yr as a function of redshift; black solid curve represents a sixth-order polynomial fit, adopted for consistency with Fig.\,\ref{fig:Cosmic_crossover}.
Purple circles with error bars show observational cosmic $R_{\rm vol}$ measurements from \citet{Palicio24}.
Purple inverted triangle at $z=2.1$ is an upper limit and is excluded from the Spearman-$\rho$ and reduced-$\chi^2$ calculations.
Overall, the model reproduces the observed evolution ($\rho=0.85$) and yields 133.5\,\% of the observed normalization
The agreement at low $z$ is tight in both slope and amplitude: restricting to $z\lesssim1.5$, we obtain 125.7\,\% of the observed normalization with $\rho=0.95$. 
Black dashed curve shows the fiducial SN~Ia rates implemented in TNG100.
There is trend-level agreement with the TNG fiducial rate ($\rho=0.94$), although the TNG normalization is 56\,\% of ours.
}
\label{fig:R_vol}
\end{figure*}

Figure\,\ref{fig:R_vol} presents the volume-normalized SN~Ia rate, $R_{\rm vol}$, as a function of redshift.
Our model prediction increases with redshift and then turns over, peaking at $z\simeq1.7\!-\!1.8$.
For comparison, we overplot the cosmic $R_{\text{vol}}$ measurements from \citet{Palicio24}.
They presented $R_{\rm vol}(z)$ as a literature compilation of observed volumetric SN~Ia rates versus redshift, drawing on many surveys over the past $\sim$\,25 years and expressing the measurements, where possible, as redshift-binned weighted-average rates with uncertainties.
Compared with the observations, the model reproduces the overall evolution ($\rho=0.85$) and yields 124.3\,\% of the observed normalization.
The agreement is particularly strong at low $z$, where the model closely tracks both the redshift dependence and amplitude, while the discrepancy is primarily driven by the higher-$z$ data points.
Restricting to $z\lesssim1.5$, we obtain 115.9\,\% of the observed normalization with $\rho=0.95$.

For additional context, we compare our results with the fiducial SN~Ia rate implemented in TNG100 and find trend-level agreement ($\rho=0.94$).
The TNG fiducial rate, however, has a normalization that is 60.0\,\% of ours.
This offset is unsurprising: the TNG fiducial prescription does not explicitly follow individual binary evolution, but instead assigns SNe~Ia to each star particle treated as an SSP using an analytic, globally imposed DTD.
Systematic differences in both normalization and the detailed timing of events can therefore arise even when the qualitative DTD shapes are similar.
In this sense, our results provide an independent, physics-driven reassessment of the SN~Ia efficiency within TNG and highlight where current subgrid prescriptions may be refined.

\section{Cosmic Evolution of SN~I\MakeLowercase{a} Progenitor Channel Mixture}
\label{sec:9}

Although earlier observational and theoretical work has suggested that the relative importance of SD and DD channels may evolve with redshift \citep[e.g.,][]{Yungelson2000, Mannucci2006}, this evolution has not been directly quantified in a fully cosmological setting. 
Table~\ref{tab:sd_dd_cosmic} gives a summary of previous key studies related to the cosmic evolution of the SD--DD channel mixture and how this present work extends them.
Observational DTD reconstructions generally do not disentangle the SD and DD components, and empirical DTD models, by construction, do not yield an SD--DD transition without parameter tuning.
Meanwhile, traditional BPS calculations, though physically detailed, lack cosmological evolution, whereas cosmological simulations typically lack progenitor-channel decomposition.
By unifying these ingredients in a single forward-modeled framework, we provide a direct and self-consistent determination of SD and DD channel fractions across cosmic time.
We track how the relative contributions of progenitor channels (i.e., SD and DD) evolve across cosmic time (\S\S\,\ref{sec:9.1}), expound threefold drivers of the SD--DD demographic mixture evolution (\S\S\,\ref{sec:9.2}), and assess the cosmological implications of the resulting SN Ia demographic transition (\S\S\,\ref{sec:9.3}).

\begin{table*}
\centering
\footnotesize
\setlength{\tabcolsep}{3pt}
\renewcommand{\arraystretch}{1.06}
\caption{Summary of previous key works related to the cosmic evolution of SD--DD progenitor channels and how this study extends them.}
\label{tab:sd_dd_cosmic}
\begin{tabular}{>{\raggedright\arraybackslash}p{3.85cm}p{1.7cm}p{1.7cm}p{9.75cm}}
\midrule
\midrule
\textbf{Study} & \textbf{SD--DD} \newline \textbf{Separation} & \textbf{Cosmic} \newline \textbf{Evolution} & \textbf{Remarks} \\
\midrule
\citet{Kobayashi2009, Kobayashi2020} & No & \textbf{Yes} &
Builds galactic chemical evolution models with metallicity-dependent SD prescriptions and predicts SN~Ia rates/enrichment in galaxies and over cosmic time. \\
\midrule
\citet{Ruiter2009, Ruiter2011} & \textbf{Yes} & No &
Presents BPS predicting channel-resolved DTDs and rates (including SD and DD pathways) for comparison with observed DTDs/rates. \\
\midrule
\citet{Sharon2010} & No & Partial &
Measures the galaxy-cluster SN~Ia rate at $0.5<z<0.9$ with HST, providing rate constraints for predominantly old stellar populations in dense environments. \\
\midrule
\citet{Maoz2012}; \citet{Graur2013}; \citet{Graur2015} & No & No &
Uses SDSS data to infer an empirical DTD from SN rates and host SFHs, and measures SN rate trends with host stellar mass and SF activity at $z\simeq0.1$. \\
\midrule
\citet{Rodney2014} & No & \textbf{Yes} &
Focuses on CANDELS SN~Ia rates and classification methodologies (STARDUST), providing the companion high-$z$ dataset. \\
\midrule
\citet{Graur2014} & \textbf{Yes} & \textbf{Yes} &
Combines CLASH and literature rates to constrain BPS DTDs, showing that DD models provide a significantly better match to observations than SD models. \\
\midrule
\citet{Maoz2014} & Partial & No &
Reviews theoretical expectations for SN~Ia progenitors (with emphasis on DD pathways) and the qualitative shapes of expected DTDs. \\
\midrule
\citet{Childress2014} & No & \textbf{Yes} &
Convolves empirical galaxy mass-assembly histories with assumed DTDs to infer SN~Ia progenitor-age distributions as a function of cosmic epoch. \\
\midrule
\citet{Claeys_2014} & \textbf{Yes} & No &
Explores how uncertain binary-evolution ingredients impact predicted SN~Ia rates and DTDs for multiple progenitor channels. \\
\midrule
\citet{Scalzo2014a} & Partial & No &
Infers the distribution of SN~Ia ejecta masses from observations, constraining the prevalence of sub-Chandrasekhar-mass explosions. \\
\midrule
\citet{Schaye2015}; \newline \citet{Pillepich2018a} & No & \textbf{Yes} &
Runs cosmological simulations implementing time-delayed SN~Ia enrichment in subgrid models to follow SN~Ia contributions to galaxy chemical evolution across time. \\
\midrule
\citet{Heringer2017} & No & No &
Uses host-galaxy colours/luminosities with an assumed DTD to infer the field-galaxy SN~Ia DTD slope and normalization. \\
\midrule
\citet{Levanon2019} & Partial & No &
Models the early light-curve excess of SN~2018oh via ejecta interaction with disk-originated matter and compares against companion-interaction scenarios. \\
\midrule
\citet{Strolger2020} & No & \textbf{Yes} &
Infers the SN~Ia DTD by jointly fitting the cosmic SN~Ia rate history and matching SN~Ia yields to GOODS/CANDELS galaxy SFHs. \\
\midrule
\citet{Wiseman2021} & No & \textbf{Yes} &
Measures DES SN~Ia rates per galaxy over $0.2<z<0.6$ versus host stellar mass and constrains the DTD using galaxy assembly histories. \\
\midrule
\citet{Joshi_2024} & No & Partial &
Tests how reliably DTDs can be recovered from individual host SFHs using \texttt{IllustrisTNG} mock galaxies at multiple redshifts. \\
\midrule
\textbf{This work} & \textbf{Yes} & \textbf{Yes} &
Unifies cosmological galaxy-formation simulations and detailed BPS to produce galaxy-resolved SD/DD fractions and an astrophysically grounded cosmic SN~Ia rate with explicit channel decomposition. \\
\midrule
\midrule
\end{tabular}
\end{table*}

\subsection{SN~Ia Progenitor-channel Dominance Crossover with Cosmic Time}
\label{sec:9.1}

\begin{figure*}
\includegraphics[width=\textwidth]{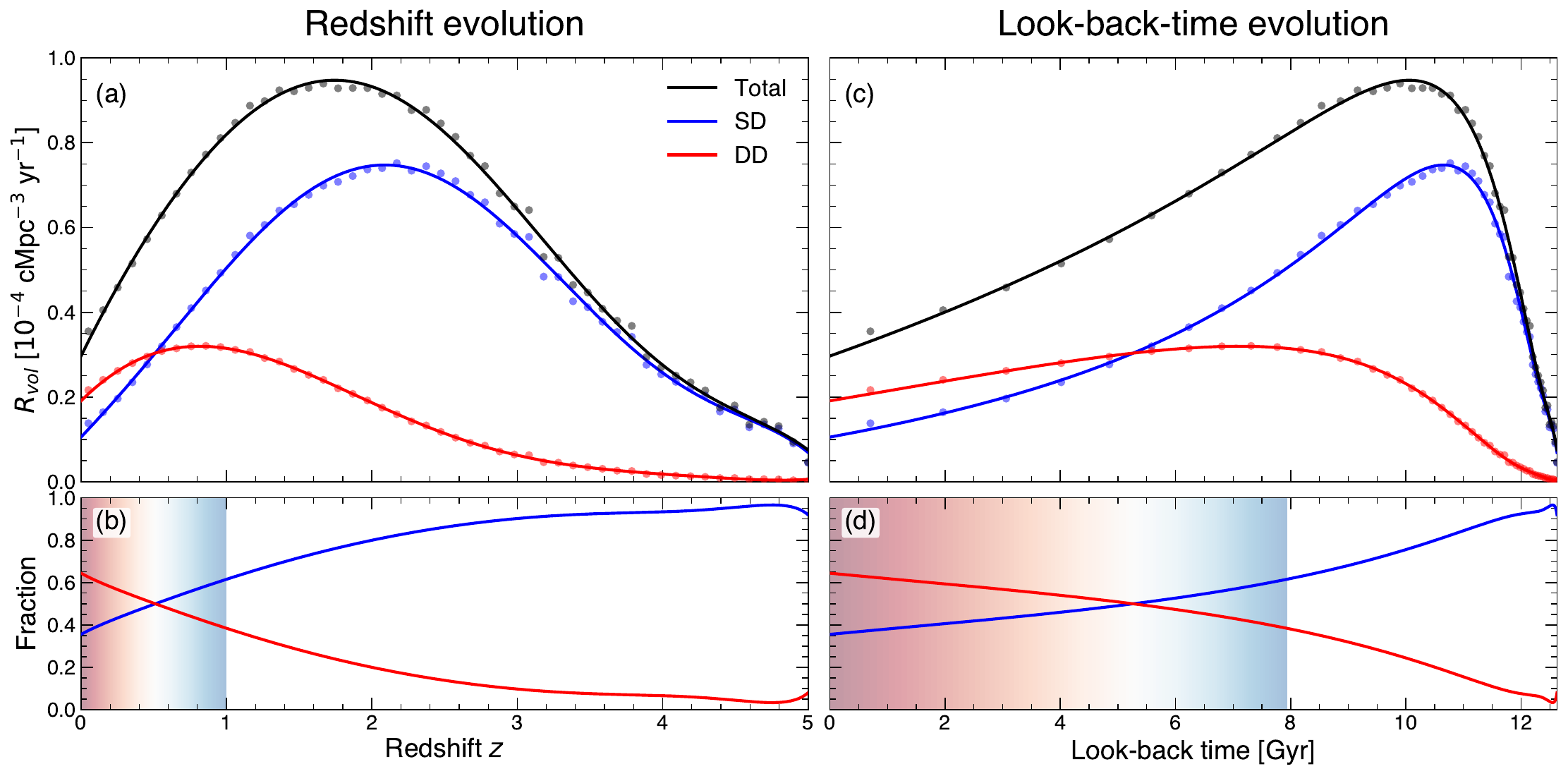}
\caption{
Analogous to Fig.\,\ref{fig:R_vol}, but decomposing the total SN~Ia rate by explosion channels and showing $R_{\rm vol}$ versus redshift ($0 \le z \le 5$; left) and look-back time ($0 \le \Delta t_{\rm lb} \le 12.6$~Gyr; right).
The upper panels plot $R_{\rm vol}$ for the SD (blue) and DD (red) channels, while the lower panels show the corresponding fractional contributions.
In panel ($a$), the black points and solid curves are identical to Fig.\,\ref{fig:R_vol}, and each channel curve is fit with a sixth-order polynomial in redshift; in panel ($c$), the same fits are mapped onto look-back time.
The weak end-point kink at $z>4.6$ and $\Delta t_{\rm lb}>12.5$~Gyr is a fitting artifact.
At high redshift, the SD pathway is favored by the prevalence of young, actively star-forming galaxies, whereas at later times the DD contribution rises and then mildly declines, reflecting a delayed and temporally broader response to the same cosmic SFH.
This difference in response functions (i.e., the SD- and DD-DTDs) drives strong redshift evolution in the relative channel contributions.
At higher redshift, look-back-time-driven truncation of SFHs further suppresses DD formation (Fig.\,\ref{fig:DTDs_SFH_SPAD}), while toward low redshift the metallicity dependence of the DD DTD weakly suppresses DD production in the increasingly metal-rich Universe, yielding a mild decrease toward $z=0$ ($\Delta t_{\rm lb}=0$).
In panels ($b$ \& $d$), the interplay of these effects drives an SD--DD dominance crossover in the cosmic SN~Ia budget at $z\simeq0.5$ and $\Delta t_{\rm lb}\simeq5.2$~Gyr.
We mark $0 \le z \le 1$ and $0 \le \Delta t_{\rm lb} \le 7.94$ with colored boxes to emphasize the regime in which (i) SN~Ia cosmology is most strongly constrained, including the ``sweet spot'' near $z \simeq 0.5$, and (ii) the SD/DD mixture evolves most rapidly.
}
\label{fig:Cosmic_crossover}
\end{figure*}

Figure\,\ref{fig:Cosmic_crossover} is analogous to Fig.\,\ref{fig:R_vol}, but here we explicitly decompose the volumetric SN~Ia rate into the SD and DD channels and present $R_{\rm vol}$ as a function of redshift (left) and look-back time (right).
Panel ($a$) shows that the SD channel tracks the cosmic SFH closely, indicating a prompt and temporally narrow response.
It reaches its maximum at $z \simeq 2.2$, slightly earlier than the cosmic SFH peak at $z \simeq 2$ \citep[e.g.,][]{Madau1996, Madau2014, Madau2017}, and then declines toward $z=0$.
By contrast, the DD channel rises more gradually, consistent with a more delayed and temporally broader response to the same cosmic SFH.
Its fractional contribution increases toward the present epoch, while its rate peaks at $z \simeq 0.8$.
These distinct response functions (i.e., SD- and DD-DTDs) naturally imprint strong redshift evolution on the relative channel contributions, producing a clear shift in progenitor-channel dominance.
Moreover, at higher redshift, the look-back-time-driven truncation of SFHs preferentially suppresses longer-delay DD formation more effectively (Fig.\,\ref{fig:DTDs_SFH_SPAD}).
In addition, the metallicity dependence of the DD DTD weakly suppresses DD production in the increasingly metal-rich low-redshift Universe, allowing a mild decrease in the DD rate toward $z=0$.
Together, these three effects yield an emergent SD--DD mixture drift with a channel dominance crossover at $z$\,$\simeq$\,0.5, in the cosmic SN~Ia budget (panel $b$).
A detailed discussion of the three factors driving the mixture evolution is deferred to \S\S\,\ref{sec:9.2}.

We note that the SD--DD crossover redshift ($z_{\rm crossover}$) is mildly model-dependent and, in particular, depends on the mock galaxy low-mass cut.
Raising the stellar-mass floor preferentially up-weights massive, more frequently quenched galaxies, suppressing the SD contribution and boosting the DD contribution in the ensemble average, which shifts $z_{\rm crossover}$ to higher redshift.
Since most observational surveys effectively probe $M_* \gtrsim  10^{8.0}~\Msun$ (see Fig.\,\ref{fig:Demo_z0p1_0p3}), we verify that $z_{\rm crossover}$ is insensitive to variations in the low-mass cut: increasing it by a factor of 30 from $M_* =10^{8.0}~\Msun$ changes $z_{\rm crossover}$ by only 0.07, from $z_{\rm crossover}=0.50$ for $M_* >10^{8.0}~\Msun$ to $z_{\rm crossover}=0.57$ for $M_* >10^{9.5}~\Msun$.\footnote{The crossover look-back time varies from 5.20\,Gyr for $M_* >10^{8.0}~\Msun$ to 5.68\,Gyr for $M_* >10^{9.5}~\Msun$.}
Moreover, galaxies with $M_* <10^{8.0}~\Msun$ contribute negligibly to SN~Ia production (see Figs.\,\ref{fig:Demo_z0p1_and_z0p55} and \ref{fig:Demo_z0p1_0p3}), so lowering the stellar-mass floor has little effect on $z_{\rm crossover}$.
Taken together, the test indicates that, within the stellar-mass regime relevant to current SN surveys, the predicted $z_{\rm crossover}$ is largely insensitive to the survey flux limit.

Panels ($c$--$d$) further sharpen the same narrative by presenting the same decomposition of the cosmic SN~Ia rate as a function of look-back time ($\Delta t_{\rm lb}$) instead of redshift.
Because $\Delta t_{\rm lb}$ maps monotonically to $z$, the time-domain representation preserves the same qualitative trends.
The SD contribution peaks at large look-back time ($\Delta t_{\rm lb} \simeq 11$~Gyr), even earlier than the cosmic SFH peak ($\Delta t_{\rm lb} \simeq 10.5$~Gyr at $z\simeq 2$), and then declines toward $\Delta t_{\rm lb}=0$~Gyr.
By contrast, the DD contribution is broader and more delayed: it rises more gradually, reaches a broad maximum around $\Delta t_{\rm lb}=7.5$~Gyr, and dominates at small look-back time.
In this time-domain view, the DD curve can be interpreted as a combined effect of a smoothed, time-shifted response of the DD DTD to the cosmic SFH, the look-back-time-driven loss of the SFH portion that contributes to DD at higher redshift, and an additional mild late-time suppression aided by the metallicity dependence of the DD DTD in the metal-rich present-day Universe (see \S\S\,\ref{sec:9.2}).
Together, these effects result in an SD--DD dominance crossover at $\Delta t_{\rm lb} \simeq 5.2$~Gyr in the cosmic SN~Ia budget.

\subsection{Threefold Drivers of the SD--DD Demographic Mixture Evolution with Cosmic Time}
\label{sec:9.2}

The redshift evolution of the SD--DD demographic mixture is shaped by a threefold set of effects.
Each effect is rooted directly or indirectly in the DTD formalism and is therefore structurally inevitable when SNe~Ia are produced via DTD-based convolution.

\smallskip \noindent \textit{(i) Channel-dependent DTDs as different response functions.} ---
In the DTD framework, the SN~Ia rate is a convolution of the SFH with the DTD response function, $R(t)=\int_0^{t} {\rm SFR}(t-\tau)\,{\rm DTD}(\tau)\,d\tau$ (Eq.~\Ref{eq:SN Ia rate}).
Because SD and DD have different ${\rm DTD}(\tau)$, they respond differently to a systematically evolving cosmic SFH.
A kernel concentrated at short delay times (a prompt SD-like response) yields $R(t)$ that closely tracks ${\rm SFR}(t)$, whereas a broad kernel extending to long delays (a tardy DD-like response) produces a delayed and temporally smoothed rate.
Thus, even for fixed SD- and DD-channel kernels, the relative SD/DD contributions can drift with $z$ as a direct consequence of convolution against an evolving SFH.

\smallskip \noindent \textit{(ii) Look-back-time-driven ``truncation of SFH.''} ---
In a redshift bin centered at $z$, the SNe~Ia observed at cosmic time $t(z)$ must originate from earlier SF and explode after a delay time $\tau$, so they are subject to the unavoidable causal bound $\tau \le t(z)$.
As the Universe becomes younger toward higher redshift, the maximum allowed delay time correspondingly decreases, truncating the long-delay tail of the population first.
Equivalently, long-delay explosions that would appear in lower-$z$ bins have simply not yet occurred at such higher $z$.
The first systems to be lost are therefore those with the largest $\tau$, namely the oldest DD progenitors.
These oldest SNe~Ia (mostly DD) differ most strongly from the young SNe~Ia (largely SD) in their properties, so the impact of mixture evolution becomes most evident in the contrast between low- and high-$z$ bins.
Notably, such ``demographic evolution'' can arise even without any kernel non-universality, simply from the look-back-time windowing inherent to $z$-binned observations (see \S\,\ref{sec:6}).

\smallskip \noindent \textit{(iii) Metallicity-dependent DTDs.} ---
Finally, the DTD is not a single universal kernel: its normalization and power-law slope can depend on progenitor metallicities, particularly for the DD channels (\S\S\,\ref{sec:2.4} and \S\,\ref{sec:6}).
Because the IMF, binary fraction, and binary-evolution physics can vary with metallicity, SD and DD exhibit distinct intrinsic metallicity-dependence in their DTD shapes, implying a host- and redshift-dependent mapping from SFH to SN~Ia rates.
As the mean stellar metallicity of galaxies and, in turn, the progenitor metallicity of individual SNe increase toward $z=0$, the relative DD contribution can decrease without any change in the SD contribution, providing an additional modulation of the SD--DD crossover.

\begin{table*}
\centering
\caption{Taxonomy of redshift-dependent evolution pathways relevant to SN~Ia cosmology.}
\label{tab:taxonomy}
\renewcommand{\arraystretch}{1.25}
\setlength{\tabcolsep}{8pt}
\begin{tabular}{>{\raggedright\arraybackslash}p{0.29\textwidth} >{\raggedright\arraybackslash}p{0.260\textwidth} >{\raggedright\arraybackslash}p{0.355\textwidth}}
\midrule
\midrule
\textbf{What Evolves with Redshift} & \textbf{How It May Bias SN~Ia Cosmology} & \textbf{Key References} \\
\midrule
\multicolumn{3}{c}{\textit{\textbf{(i) Within-component $z$-evolution: non-invariant standardization at fixed type}}}\\[0pt]
\midrule
Even within a single progenitor type, the standardized HR can show $z$ trends if standardized luminosity depends on progenitor conditions. &
Within-class redshift-dependent luminosity shift driven by progenitor ages and/or metallicities evolving with cosmic time. &
\citet{Drell2000,Dominguez2001,Timmes2003,Podsiadlowski2006,Riess2006,Kang_2020,Lee2020,Zhang2021,Lee2022,Wang2023,Chung2023,Chung2025,Son2025} \\
\midrule
\multicolumn{3}{c}{\textit{\textbf{(ii) Between-component $z$-evolution: mixture-prior drift}}}\\[0pt]
\midrule
Even if multiple components are each standardizable on their own, an apparent evolution can arise as the mixture drifts with $z$, shifting the population-averaged HR. &
Between-class redshift-dependent luminosity shift driven by the SD/DD and near-$M_{\rm Ch}$/sub-$M_{\rm Ch}$ mixture evolving with cosmic time. &
\citet{Dominguez2001,Timmes2003,Podsiadlowski2006,Fink2010,Greggio2010,Pakmor2010,vanKerkwijk2010,Maoz2014,LivioMazzali2018} \\
\midrule
\multicolumn{3}{c}{\textit{\textbf{(iii) Survey-level bias: $z$-dependent progenitor/host demography coupled to selections}}}\\[0pt]
\midrule
Even without (i) \& (ii), SN--host/environment correlations, coupled with $z$-dependent selection/measurement/analysis cuts, can re-weight host/progenitor subpopulations with $z$, yielding a $z$-dependent mean HR. &
Evolving progenitor/host demographics and dust/environment populations coupled to selection; $z$ evolution in $x_1$ and $c$ distributions interacting with survey thresholds. &
\citet{Mannucci2006,Howell2007,Kelly2010,Lampeitl2010,Sullivan2010,Childress2013,Roman2018,Kim2018,Rigault2020,BroutScolnic2021} \\
\midrule
\midrule
\end{tabular}
\end{table*}

\subsection{Cosmological Implications of SN~Ia Demographic Transition}
\label{sec:9.3}

Our results indicate that the dominant SN~Ia pathway evolves with cosmic time, with the balance shifting from SD to DD systems and, correspondingly, from prompt to tardy explosions.
The key implication of this demographic crossover is that it can undermine the assumptions behind empirical luminosity standardization.
Conventional distance estimation presumes that a compact set of observables---typically the peak apparent $B$-band magnitude $m_B$, a stretch parameter $x_1$, a color parameter $c$, and one or more host-galaxy proxies---can be mapped onto standardized luminosities using a \emph{single}, \emph{redshift-independent} set of nuisance parameters, schematically $\mu = m_B - M + \alpha x_1 - \beta c + \Delta_{\rm host} + \cdots$ \citep{Guy2007,Betoule2014,Scolnic2018,Brout_2022,popovic2026}.
This approach implicitly relies on two assumptions:
(i) SNe~Ia constitute an effectively homogeneous population describable by a single set of nuisance parameters, and (ii) the mixture (i.e., the relative fraction) of any physically distinct sub-populations does not evolve with redshift.
However, SNe~Ia arise from multiple progenitor channels (e.g., SD/DD and near/sub-$M_{\rm Ch}$ systems), which can differ in their light-curve, spectral, and yield distributions (e.g., ejecta mass and $^{56}{\rm Ni}$ mass), violating assumption (i), and their relative contributions are expected to evolve with cosmic time, violating assumption (ii).

As a result, different sub-populations can occupy distinct regions in the nuisance-parameter space and possess different standardized luminosity zero-points even at fixed parameters.
Thus, if there exists any channel--luminosity correlation, applying a single, globally trained standardization across a redshift-dependent population mix can systematically mis-correct subsets of events, imprinting a redshift-dependent distance bias, $\Delta\mu(z)$ (a drift in the mean HR), which can mimic ``extra dimming'' and may be partially degenerate with cosmological parameters.
In this sense, a ``single and universal'' standardization can absorb demographic drift as an apparent $\Delta\mu(z)$, biasing cosmological inference if left unmodeled.
In follow-up work, our cosmology--BPS framework will enable a quantitative test of this scenario by explicitly connecting the redshift evolution of progenitor age, metallicity, and channel mix to the implied standardization HRs in a fully forward-modeled, end-to-end manner.
\color{black}

This idea sits within a long-standing SN~Ia cosmology concern: redshift-dependent population evolution can be partially degenerate with cosmological dimming in the Hubble diagram whenever standardization is not strictly invariant \citep[e.g.,][]{Drell2000,Podsiadlowski2006,Riess2006}.
In this sense, a cosmic progenitor-channel crossover provides a concrete physical mechanism for an otherwise phenomenological ``evolution'' term \citep[e.g.,][]{Dominguez2001,Maoz2014,LivioMazzali2018}.
To make explicit where such a channel-crossover mechanism enters SN~Ia cosmology, Table~\ref{tab:taxonomy} summarizes key work in three themes.
Specifically, we classify redshift-dependent SN~Ia systematics by the statistical ``layer'' at which redshift dependence is introduced: 
(i) \emph{within-component evolution}---even within a single progenitor type, the standardized HR can show $z$ trends if standardized luminosity depends on progenitor conditions, such as age and metallicity;
(ii) \emph{between-component evolution}---even if multiple components are each standardizable on their own, an apparent evolution can arise as the mixture drifts with $z$, shifting the population-averaged HR; 
and (iii) \emph{survey-level bias}---even without (i) \& (ii), SN--host/environment correlations, coupled with $z$-dependent selection/measurement/analysis cuts, can reweight host/progenitor subpopulations with $z$, yielding a $z$-dependent mean HR.
This taxonomy is not meant to be mutually exclusive---all three effects can operate simultaneously---but it clarifies whether a claimed ``evolution'' originates in intrinsic standardization, population mixing, or observational reweighting.
Our model prediction of a cosmic progenitor-channel crossover is on the second layer, supplying a direct physical realization of the SN~Ia population mixing effect.

\section{Conclusion}
\label{sec:10}

\subsection{Summary}
\label{sec:10.1}

In \S\,\ref{sec:2}, we introduce a unified forward-modeling framework for synthesizing the cosmic SN~Ia population by coupling a cosmological hydrodynamic galaxy simulation with BPS.
Individual \texttt{IllustrisTNG} star particles are treated as simple stellar populations defined by their mass, age, and metallicity, and, for each particle, we generate binary systems from observationally motivated initial-parameter distributions.
These systems are then evolved with \texttt{COMPAS} through the SD and DD channels within a hybrid framework that combines the efficiency of rapid BPS with external constraints from detailed SD progenitor models.
We then construct an all-sky cosmic-volume realization of the \texttt{IllustrisTNG} snapshots, embedding the full galaxy population, including SN~Ia hosts, in a cosmological setting.
Using this cosmology--BPS pipeline, we resolve the spatial distributions of SNe~Ia within individual galaxies and connect host SFHs directly to the SN~Ia rate and progenitor-age distribution (\S\,\ref{sec:3}).
Across the simulation volume, the model reproduces key SN-related observations, including host-galaxy demographics (\S\,\ref{sec:4}), SN~Ia rates from galactic (\S\,\ref{sec:5}) to cosmic scale (\S\,\ref{sec:8}), as well as a progenitor-age step (\S\,\ref{sec:7}) suggested by the mass-step and sSFR-step signals in Hubble residuals.

Our main finding (\S\S\,\ref{sec:2.4} and \S\,\ref{sec:6}) is that the SN~Ia DTD is intrinsically non-universal: instead of a single global kernel, it is more naturally framed as an environment-conditioned response function, ${\rm DTD}(\tau;Z)$, whose integrated yield and effective power-law slope vary systematically with progenitor metallicity.
The canonical $\tau^{\alpha}$ description with $\alpha\,\simeq\,-1$ remains useful as a population-averaged approximation, but it can hide metallicity-driven variations tied to host and redshift.
Because progenitor metallicity is host- and redshift-dependent, this non-universality generically enforces a redshift-dependent re-weighting of progenitor ages and SD/DD demographics, thereby linking SN~Ia population evolution to potential cosmological systematics whenever standardized luminosities retain HR sensitivity to progenitor or host conditions in precision cosmology.

Another main result (\S\,\ref{sec:9}) is that we obtain a direct and self-consistent determination of SD and DD channel fractions across cosmic time.
By decomposing the cosmic SN~Ia rate into SD versus DD progenitors, we find that the SD channel---tightly coupled to the cosmic SFH---dominates at $z\gtrsim0.5$, while the DD contribution rises more gradually and takes over toward the present epoch.
The SN~Ia progenitor mixture is therefore intrinsically time-dependent---evolving from SD to DD dominance---with potential implications for redshift-dependent SN-luminosity evolution and for cosmological inference.
In future work, we will assign luminosities to synthetic SNe~Ia based on their progenitor properties, enabling a direct forward modeling of the Hubble diagram.
Forthcoming facilities and surveys will provide decisive tests: JWST, Roman, and Euclid will probe $1<z<3$, where the earliest SD progenitors may emerge in highly star-forming galaxies, while LSST will deliver millions of low-$z$ SNe~Ia, anchoring the DD contribution in massive, quenched hosts.

\subsection{Future Work}
\label{sec:10.2}

A natural direction for future work is to quantify the sensitivity of our conclusions to the use of a single, relatively small cosmological simulation volume. 
In particular, we will explicitly assess several sources of statistical uncertainty: (i) cosmic variance introduced by the finite simulated volume, (ii) subhalo-sampling noise arising from the discrete realization of the galaxy population, (iii) inter-snapshot covariance induced by galactic progenitor--descendant links, and (iv) the limited dynamic range for extremely under- and over-dense regions imposed by the finite box size.
Although these effects may introduce additional variance into the inferred SN~Ia population, their net impact is expected to remain modest given the size of the full catalogue, which contains $\sim$\,$10^{13}$ SNe~Ia identified from $\sim$\,$10^{14}$ star particles in $\sim$\,$10^{10}$ galaxies (Table~\ref{tab:snpm_stats}).
Nevertheless, larger-volume simulations are needed to determine the extent to which these effects may alter the inferred demographics of SNe~Ia and their hosts (and hence affect SN~Ia cosmology).
We will also investigate the dependence of our results on the adopted galaxy formation models, such as \texttt{EAGLE} \citep{Schaye2015}, \texttt{FIRE} \citep{Hopkins2018}, \texttt{SIMBA} \citep{Dave2019}, and \texttt{NEWHORIZON} \citep{Dubois2021}, thereby moving beyond a single realization (TNG100).
Because our framework is designed to ingest alternative mock galaxy samples flexibly, such tests can be carried out in a controlled manner.

Complementary to these cosmological-simulation-side robustness tests, we will quantify how sensitive our inferences are to the adopted \texttt{COMPAS} options and parameter choices.
More broadly, we will evaluate the extent to which the results depend on the underlying BPS framework, beyond our specific implementation adopted here (\texttt{COMPAS} together with an in-depth exploration of SD parameter spaces).
In practice, this can be achieved by re-running the same end-to-end pipeline while replacing \texttt{COMPAS} with other widely used BPS frameworks---e.g., \texttt{StarTrack} \citep{Belczynski2008}, \texttt{SeBa} \citep{Toonen2012}, \texttt{BSE/binary\_c} \citep{Hurley2002, Izzard2004}, the Brussels code \citep{Mennekens2010}, \texttt{BPASS} \citep{EldridgeStanway2016,Eldridge2017}, \texttt{COSMIC/MOBSE} \citep{Breivik2020, Giacobbo2018}, and \texttt{POSYDON} \citep{Fragos2023}, and related codes. 
The resulting cross-model scatter can then be folded into a systematic error budget, placing our conclusions on a firmer footing that is not tied to any single BPS realization.

The cosmology-facing contribution of this study is to lay the groundwork for a cosmology-level stress test of empirical standardization.
By generating synthetic event catalogues labeled by explosion channel and by progenitor and host properties, our model predicts the redshift evolution of progenitor-property distributions and their covariance with the host proxies used in modern analyses \citep[e.g.,][]{Kelly2010,Sullivan2010,Rigault_2013,Rigault2020,Kim2024}.
Forthcoming papers will analyze the synthetic sample with the empirical standardization pipeline used in cosmology, explicitly testing whether a single set of nuisance parameters remains valid under population drift and quantifying the induced $\Delta\mu(z)$ and its degeneracy with cosmological parameters.
In that companion work, recent claims that progenitor-age bias can mimic high-$z$ ``extra dimming'' \citep{Kang_2020,Lee2020,Zhang2021,Lee2022,Wang2023,Chung2023,Chung2025,Son2025} (but see \citealp[e.g.,][for an alternative dust-based explanation in which any age-driven signal is weak or already absorbed by standard corrections]{Popovic2024,Wiseman2026}) will be directly testable.
The same framework will also identify which observables (e.g., host spectroscopy, local-environment diagnostics, and/or sample-dependent re-training\footnote{Sample-dependent re-training refers to re-calibrating the SN~Ia standardization nuisance parameters for different subsamples, rather than adopting a single globally trained set, in order to account for variations in progenitor demographics and selection effects.}) are most effective at breaking the astrophysics--cosmology degeneracy.

While our discussion has focused primarily on demographic evolution in terms of the SD--DD balance, an evolving transition between near-$M_{\rm Ch}$ and sub-$M_{\rm Ch}$ explosions may be comparably important for SN~Ia cosmology. 
This is because sub-$M_{\rm Ch}$ progenitors likely contribute at a non-negligible, and potentially substantial, level to the population of SNe~Ia used for cosmology \citep[e.g.,][]{Scalzo2014a,Scalzo2014b,Shen2018,Polin_2019,Townsley2019,Rigault2020,Dhawan2025} and the explosion mass and density regime are more directly connected to the observables that enter luminosity standardization. 
We nevertheless do not attempt a quantitative prediction for the cosmic evolution of the near-$M_{\rm Ch}$/sub-$M_{\rm Ch}$ mixture in the present work, because this mapping is currently less robust than the broader SD--DD decomposition \citep{Bours2013,Toonen2014, RuiterSeitenzahl2025}. 
We therefore regard the redshift evolution of the near-$M_{\rm Ch}$/sub-$M_{\rm Ch}$ mixture, and its quantitative impact on inferred cosmology, as an important topic for future work.

\section*{Acknowledgments}

We are grateful to Young-Lo Kim, Jiwoo Kim, Seunghyun Park, Junhyuk Son, Heesue Kang, Jin-Hyung Lee, Do-Hyun Kim, and Chengpeng Zhang for insightful comments. 
This research was supported (a) by the Mid-career Researcher Program (RS-2024-00344283) through the NRF of Korea funded by the Ministry of Science and ICT and (b) by the Basic Science Research Program for the Center for Galaxy Evolution Research (RS-2022-NR070872, RS-2022-NR070525) through the NRF of Korea funded by the Ministry of Education.

\section*{Data Availability}

The \texttt{SN Ia Population Machine} framework, developed to synthesize galactic and cosmic SN~Ia populations by coupling cosmological hydrodynamic simulations with binary population synthesis, is publicly available at \url{https://doi.org/10.5281/zenodo.18603625}. 
(The site will be activated upon publication of the paper. 
In the meantime, the dataset is available via this \href{https://zenodo.org/records/18603625?preview=1&token=eyJhbGciOiJIUzUxMiJ9.eyJpZCI6IjBhYmIyZTlkLWM5ZWEtNGQ1Yi1iNTE1LWE1NDM5MTUzZTk2NCIsImRhdGEiOnt9LCJyYW5kb20iOiJlYjZlYjkzYzQ2MDcwMDFjMTdhNzg0MjQxZTJhZDE4MSJ9.GjyZAJV3PvwjOsO-4qAtFAMy9rVretFi8WmEBMgYNTc22Jo9dh4lN8-NPs-eTajoywnELC8SJsNrug81arruBw}{private link}.)
While this study utilizes the \texttt{IllustrisTNG} simulation as a primary input, the code is designed to be adaptable to other alternative cosmological simulation datasets.
The data underlying this article were derived from the \texttt{IllustrisTNG} simulation, available at \url{www.tng-project.org}, and the {\ttfamily COMPAS} rapid binary population synthesis code (version 03.29.00), which is available at \url{http://github.com/TeamCOMPAS/COMPAS}

\bibliographystyle{mnras}
\bibliography{Paper1}



\appendix

\section{Summary of Acronyms and Parameters/Variables}
\label{appendix:A}

Appendix\,\ref{appendix:A} summarizes the notation used throughout this paper for ease of reference. 
Table~\ref{tab:acronyms} compiles all acronyms and code/survey/simulation labels, while Table~\ref{tab:variables} lists the parameters and variables appearing in the main text and figures. 
Both tables are sorted alphabetically and provide one-line definitions to facilitate rapid lookup and to minimise ambiguity in terminology and notation across the cosmological hydrodynamic galaxy simulation, binary-evolution components, and cosmology.

\begin{table*}
\centering
\scriptsize
\caption{Acronyms used in this paper (alphabetical).}
\label{tab:acronyms}
\setlength{\tabcolsep}{4pt}
\renewcommand{\arraystretch}{1.05}
\begin{tabular}{@{}l p{0.42\textwidth} l p{0.42\textwidth}@{}}
\toprule
\textbf{Acronym} & \textbf{Definition} & \textbf{Acronym} & \textbf{Definition} \\
\midrule
AGN & Active galactic nucleus & MW & Milky Way \\
\textsc{AREPO} & Moving-mesh magneto-hydrodynamics code used for \texttt{IllustrisTNG} & NIR & Near-infrared \\
BPS & Binary population synthesis & PS1 & Pan-STARRS1 (Medium Deep Survey) \\
CANDELS & Cosmic Assembly Near-infrared Deep Extragalactic Legacy Survey & SD & Single-degenerate SN~Ia channel (accreting WD with a non-degenerate companion) \\
CLASH & Cluster Lensing And Supernova survey with Hubble & SDSS & Sloan Digital Sky Survey \\
CMR & Color--magnitude relation & SFH & SF history \\
\texttt{COMPAS} & BPS code: \emph{Compact Object Mergers: Population Astrophysics and Statistics} & SFR & Star-formation rate \\
DD & Double-degenerate SN~Ia channel (merging double white dwarfs) & sSFR & Specific star-formation rate ($\mathrm{sSFR}\equiv\mathrm{SFR}/M_*$) \\
DES & Dark Energy Survey (DES-SN in this context) & SN & Supernova (generic) \\
DTD & Delay-time distribution; SN~Ia response versus delay time after SF & SN~Ia & Type~Ia supernova (thermonuclear explosion of a C/O white dwarf in a binary) \\
FLRW & Friedmann--Lema\^{\i}tre--Robertson--Walker cosmology & SNLS & Supernova Legacy Survey \\
GFM & \texttt{IllustrisTNG} data-field prefix for the galaxy-formation model & SPAD & SN~Ia progenitor-age distribution \\
GOODS & Great Observatories Origins Deep Survey & SSP & Simple stellar population \\
GW & Gravitational wave(s) & \texttt{STARDUST} & STARDUST SN~Ia classification/analysis pipeline for CANDELS SN rates \\
HR & Hubble residual & \texttt{STARTRACK} & StarTrack binary population-synthesis code \\
HST & Hubble Space Telescope & \textsc{SUBFIND} & Algorithm to identify self-bound subhalos/galaxies in simulations \\
IMF & Initial mass function & TNG & \texttt{IllustrisTNG} cosmological hydrodynamic simulation suite \\
JWST & James Webb Space Telescope & TNG50 & \texttt{IllustrisTNG} run with $\sim 50\,{\rm Mpc}$ box \\
LSST & Legacy Survey of Space and Time (Rubin Observatory LSST) & TNG100 & \texttt{IllustrisTNG} run with $\sim 100\,{\rm Mpc}$ box \\
\bottomrule
\end{tabular}
\end{table*}

\begin{table*}
\centering
\scriptsize
\caption{Parameters and variables used in this paper (alphabetical).}
\label{tab:variables}
\setlength{\tabcolsep}{4pt}
\renewcommand{\arraystretch}{1.05}
\begin{tabular}{@{}l p{0.41\textwidth} l p{0.41\textwidth}@{}}
\toprule
\textbf{Symbol} & \textbf{Definition} & \textbf{Symbol} & \textbf{Definition} \\
\midrule
$a$ & Orbital semi-major axis & $N_{\rm SN}$ & Number of SN~Ia events in a sample/catalogue \\
$\alpha$ & DTD power-law slope; also stretch--luminosity coefficient in SN~Ia standardization & $\Omega_{\rm k}$ & Curvature density parameter \\
$\beta$ & Colour--luminosity coefficient in SN~Ia standardization & $\Omega_\Lambda$ & Dark-energy density parameter (cosmological constant) \\
$c$ & Speed of light; also SN~Ia colour parameter in standardization & $\Omega_{\rm m}$ & Matter density parameter \\
$\chi(z)$ & Line-of-sight comoving distance & $P_{\rm orb}$ & Binary orbital period \\
$D_{\rm M}(z)$ & Transverse (proper-motion) comoving distance ($D_{\rm M}=\chi$ for $\Omega_{\rm k}=0$) & $R(t)$ & SN~Ia rate at cosmic time $t$ (population-averaged) \\
${\rm DTD}(\tau)$ & Delay-time distribution (DTD) as a function of delay time & $R_{90}$ & Characteristic galaxy radius enclosing 90\% of the stellar mass \\
$\Delta_{\rm host}$ & Host-dependent correction term in standardization (e.g., a mass-step term) & $R_{\rm gal}$ & Galaxy-normalised SN~Ia rate (yr$^{-1}$ galaxy$^{-1}$) \\
$\Delta\mu(z)$ & Redshift-dependent distance-modulus bias & $R_{\rm half}$ & Stellar half-mass radius \\
$\Delta t$ & Observational time window / survey baseline used to select SN--host pairs & $R_{\rm vol}$ & Volumetric (comoving) SN~Ia rate (yr$^{-1}$ Mpc$^{-3}$) \\
$\Delta t_{\rm lb}$ & Look-back time & ${\rm SFR}(t)$ & Star-formation rate as a function of time \\
$\frac{{\rm d}V_{\rm c}}{{\rm d}z}$ & Differential comoving volume element & ${\rm sSFR}$ & Specific star-formation rate, ${\rm sSFR}\equiv {\rm SFR}/M_*$ \\
$G$ & Gravitational constant & ${\rm sSFR}_{\rm step}(z)$ & Redshift-dependent sSFR threshold defining the sSFR-step split \\
$\gamma$ & Host mass-step correction amplitude in standardized SN~Ia magnitudes & $t$ & Cosmic time (age of the Universe) \\
$h$ & Reduced Hubble constant, $h \equiv H_0/(100\,\mathrm{km\,s^{-1}\,Mpc^{-1}})$ & $t_0$ & Present epoch \\
$H_0$ & Hubble parameter at $z=0$ (Hubble constant) & $t_{\rm arrival}$ & Observer-frame arrival time \\
$H(z)$ & Hubble expansion rate at redshift $z$ & $t_{\rm explosion}$ & Source-frame explosion time \\
$L_{\rm box}$ & Simulation box side length (comoving) & $t_{\rm GW}$ & Gravitational-wave inspiral/merger timescale for a binary \\
$m_{\rm baryon}$ & Baryonic (gas/stellar) particle mass resolution in TNG100 & $T_*$ & Host-level star-particle-mass-weighted mean stellar age of a galaxy \\
$m_B$ & Observed peak apparent magnitude in rest-frame $B$ band (light-curve fit parameter) & $T_{\rm pro}$ & Host-level mean progenitor age of a galaxy \\
$M_*$ & Total stellar mass of a galaxy (sum of star-particle masses) & $T_{{\rm pro},i}$ & Event-level progenitor ages of individual SNe~Ia \\
$M_1$ & Initial primary mass in a binary system & $\tau$ & Delay time (progenitor age): time between SF and SN~Ia explosion \\
$M_{1}^{\rm WD}$ & Initial primary white-dwarf mass (in double-WD systems) & $V_{\rm box}$ & Simulation box volume, $V_{\rm box}=L_{\rm box}^3$ \\
$M_2$ & Initial secondary mass in a binary system & $V_{\rm shell}(\chi_0,\chi_1)$ & Comoving volume between radii $\chi_0$ and $\chi_1$ \\
$M_{2}^{\rm WD}$ & Initial secondary white-dwarf mass (in double-WD systems) & $w_{i,t}$ & Tile weight for tile $t$ in snapshot $i$ (used for weighted statistics) \\
$M_B$ & Peak absolute magnitude in the $B$ band for a normal SN~Ia & $x_1$ & Light-curve stretch parameter \\
$M_{\rm Ch}$ & Chandrasekhar mass & $X_j$ & Generic property of object $j$ used when computing weighted tile statistics \\
$M_{\rm p}$ & Primary mass, $M_{\rm p} \equiv \max(M_{1}^{\rm WD},M_{2}^{\rm WD})$ & $z$ & Redshift \\
$M_{\rm ref}$ & Reference SSP mass used to normalise star-particle-level SN~Ia populations & $z_{\rm crossover}$ & Redshift at which SD and DD contributions are equal (SD--DD demographic crossover) \\
$M_{\rm s}$ & Secondary mass, $M_{\rm s} \equiv \min(M_{1}^{\rm WD},M_{2}^{\rm WD})$ & $Z$ & Metallicity (used for progenitor or host metallicity depending on context) \\
$\mu$ & Distance modulus & $Z_*$ & Star-particle-mass-weighted mean stellar metallicity of a galaxy \\
$N_{\rm host}$ & Number of SN~Ia host galaxies in a sample/catalogue & $Z_{\rm pro}$ & Host-level mean progenitor metallicity of a galaxy \\
$N_{i,t}$ & Number of galaxies (or SN events) in tile $t$ of snapshot $i$ & $Z_{{\rm pro},i}$ & Event-level progenitor metallicities of individual SNe~Ia \\
\bottomrule
\end{tabular}
\end{table*}

\section{Examples of SD- \& DD-channel SN\MakeLowercase{e}~I\MakeLowercase{a}}
\label{appendix:B}

Each star particle can give rise to a diverse population of SNe~Ia, encompassing both SD and DD progenitors with a wide range of binary configurations and delay times. 
This diversity reflects the variety of evolutionary pathways encoded within each stellar population. 
In this Appendix\,\ref{appendix:B}, we present representative cases to illustrate the evolutionary pathway to SNe~Ia.

\begin{figure*}
\includegraphics[width=0.9\textwidth]{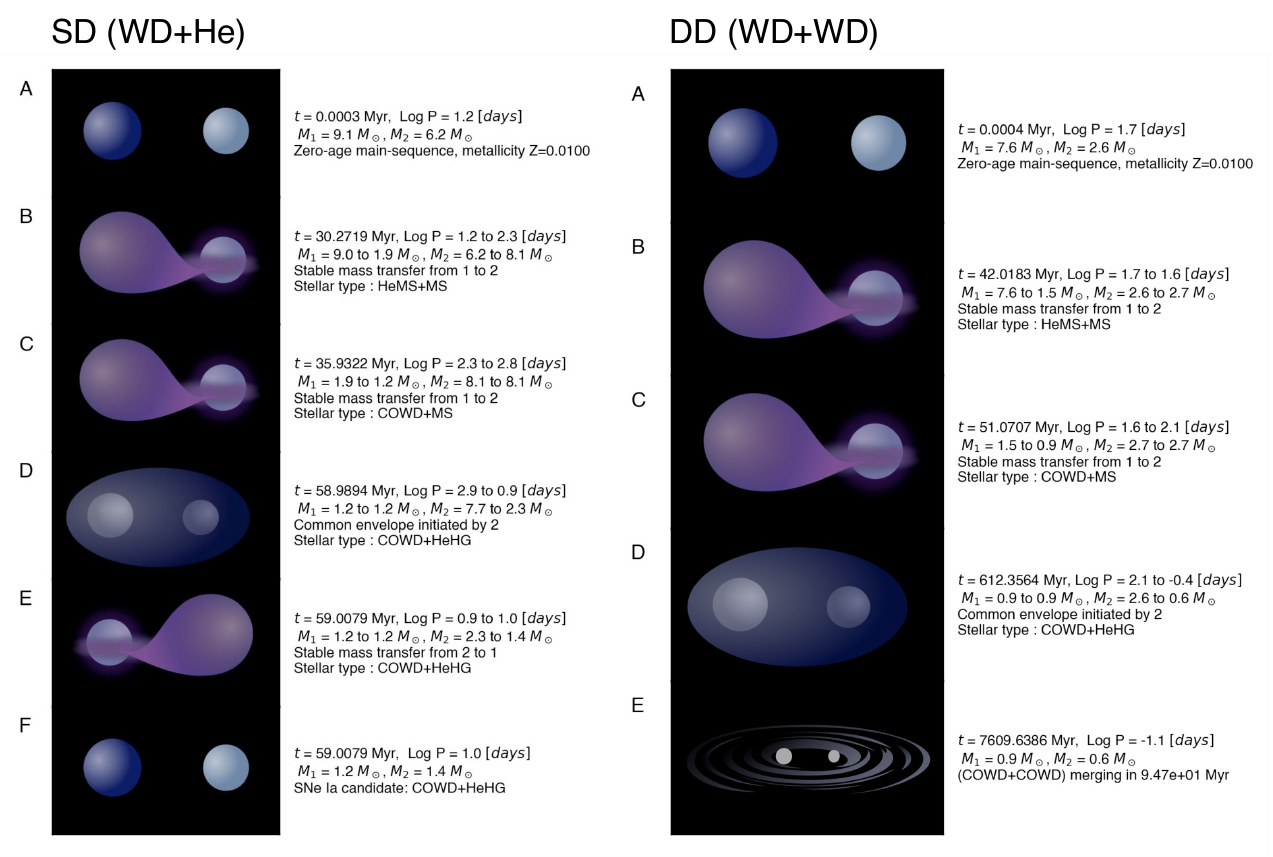}
\caption{
Schematic illustrations of binary evolutionary pathways for SN~Ia progenitors, generated with the evolution plotter from the \texttt{COMPAS} suite (Team COMPAS; \citealt{COMPASTeam2022a,COMPASTeam2025}).
\textbf{\textit{(Left)}} Example single-degenerate progenitor (a WD+He type). 
Circles denote the two components (primary on the left; secondary on the right). 
Each labeled stage (A--F) reports the time, orbital period, component masses, as well as stellar and interaction states.
\textbf{\textit{(Right)}} Analogous to the left panel, but for a double-degenerate progenitor (a CO+CO WD system).
}
\label{fig:Fig4}
\end{figure*}

Figure\,\ref{fig:Fig4} presents schematic illustrations of binary evolutionary pathways for SN~Ia progenitors, generated with the evolution plotter from the \texttt{COMPAS} suite (Team COMPAS; \citealt{COMPASTeam2022a,COMPASTeam2025}).
The left panel shows an example of an SD SN~Ia progenitor in the WD+He channel. 
Each row corresponds to a distinct evolutionary phase, showing the time, component masses, orbital period, and stellar and interaction states of the binary system. 
On the ZAMS (phase~A), the primary and secondary masses are $M_1 = 9.1\,\Msun$ and $M_2 = 6.2\,\Msun$, respectively, with an orbital period of $\sim$\,16\,days. 
At $t = 30.27$\,Myr (phase~B), Roche-lobe overflow occurs as the primary evolves off the MS and loses mass, forming a He~MS star. 
At phase~C, the primary has completed its evolution into a CO-WD and initiates a second episode of Roche-lobe overflow, while the secondary, still on the MS, has gained additional mass through accretion from the primary.
At this point, with $(M_{1}^{\rm WD}, M_2, \log P_{\rm orb}) = (1.2,\ 8.1,\ 2.8)$, the system lies outside the WD+MS parameter space depicted back in Figure\,\ref{fig:Fig3}, and thus continues to evolve. 
During this phase, the binary loses angular momentum, tightening the orbit and accelerating its rotation. 
The reduced separation enables renewed mass transfer from the secondary to the CO-WD, causing the donor’s mass to decrease (phases~D and~E). 
Eventually, in phase~F, the system reaches $(M_{1}^{\rm WD}, M_2, \log P_{\rm orb}) = (1.2,\ 1.4,\ 1.0)$. 
At this stage---approximately 60\,Myr after formation---the binary falls squarely within the WD+He parameter space shown in Fig.\,\ref{fig:Fig3} and is therefore identified as an SD SN~Ia progenitor.

The right panel is analogous to the left panel, but for a double-degenerate progenitor (a CO+CO WD system).
It shows an illustrative DD progenitor that follows an evolutionary sequence broadly similar to the SD case shown in the left panel.
However, by phase~D, the secondary star does not acquire enough mass to enter the WD+He parameter space. 
It instead evolves into a CO-WD, yielding a CO+CO WD binary whose total mass exceeds $M_{\rm Ch}$. 
For such systems, we define the delay time as the sum of the stellar evolutionary time and the GW inspiral timescale, $t_{\rm GW}$.

\section{Detailed distributions of SD and DD SN~I\MakeLowercase{a} events in clusters and their member galaxies}
\label{appendix:C}
\begin{figure*}
\includegraphics[width=0.90\textwidth]{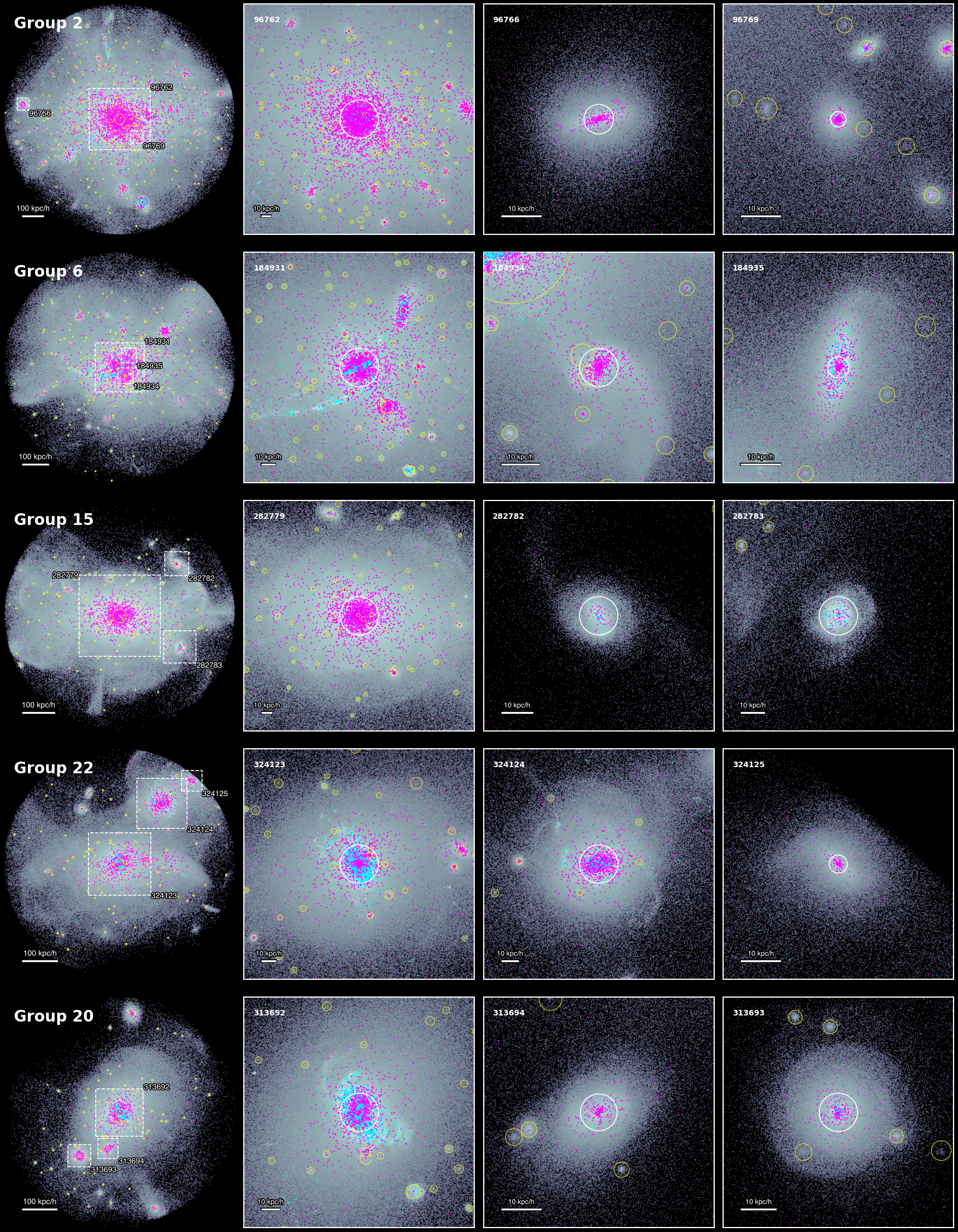}
\caption{
Extended version of Fig.\,\ref{fig:Fig_intro}($d$) for the other five groups/clusters of galaxies from Fig.\,\ref{fig:Fig_intro}($a$), along with SNe~Ia (SD and DD) identified in a mock survey with an observational time window of $\Delta t = 10^5$~yr.
Each group/cluster is labeled by its TNG50 halo catalogue index.
The groups/clusters are arranged in descending stellar mass of their central halos.
}
\label{fig:Combined_5x4_Part1}
\end{figure*}

\begin{figure*}
\includegraphics[width=0.90\textwidth]{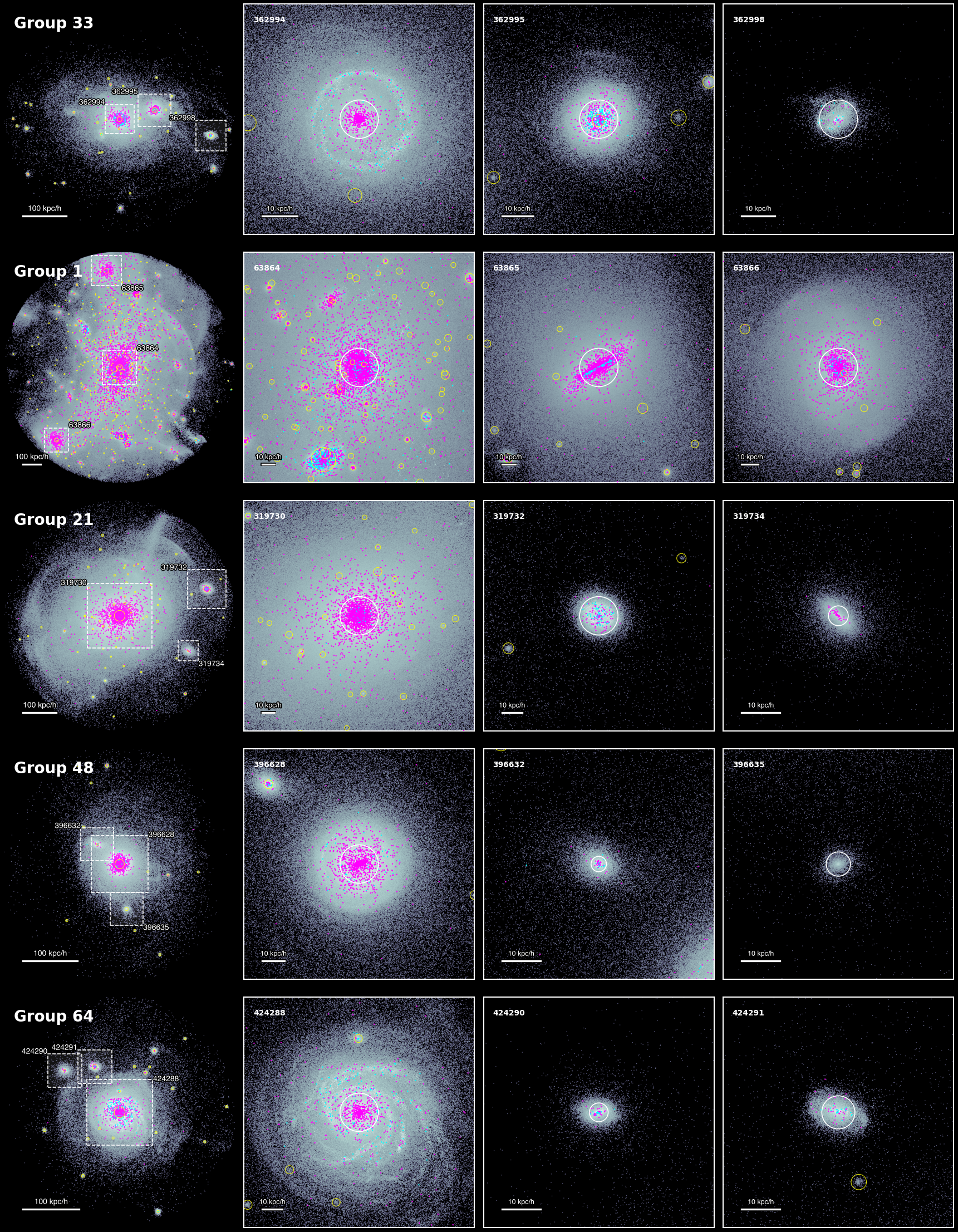}
\caption{
Same as Fig.\,\ref{fig:Combined_5x4_Part1}, but for the different five groups/clusters from Fig.\,\ref{fig:Fig_intro}($a$).
}
\label{fig:Combined_5x4_Part2}
\end{figure*}

In Figures\,\ref{fig:Combined_5x4_Part1} and \ref{fig:Combined_5x4_Part2}, we show the SN~Ia locations on galactic scales and the detailed distributions of SD and DD events in groups/clusters and their member galaxies, using the same presentation style as Fig.\,\ref{fig:Fig_intro}($d$).

\section{Various sets of pairwise parameter combinations}
\label{appendix:D}

\begin{figure*}
\includegraphics[width=\textwidth]{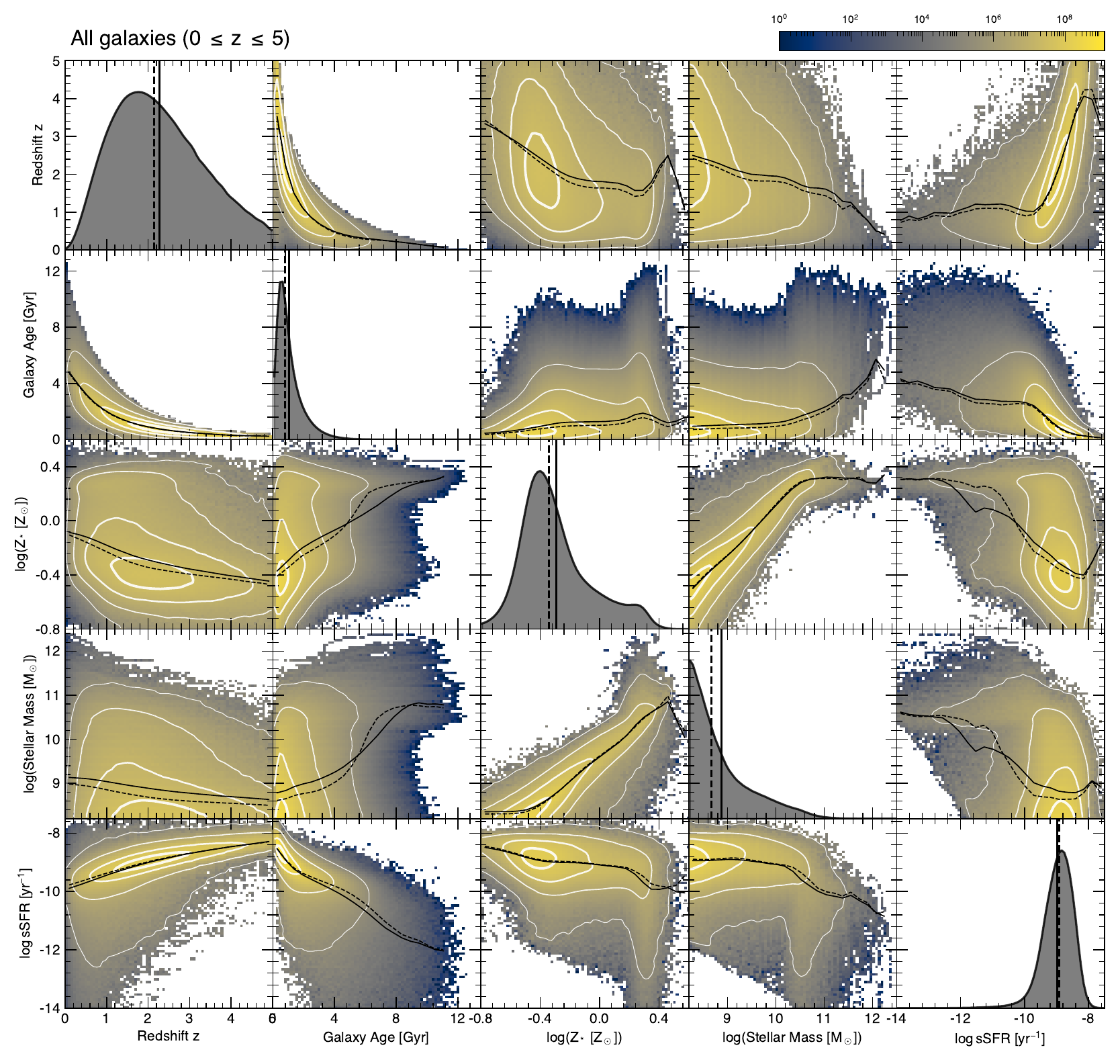}
\caption{
Various sets of pairwise parameter combinations for \emph{all} galaxies in the \emph{full catalogue} ($0 \leq z \leq 5$).
The parameters used are redshift, stellar-particle-mass-weighted mean stellar age, stellar-particle-mass-weighted mean stellar metallicity, total stellar mass, and sSFR of the full population of galaxies.
Symbols and lines follow the conventions of Figs.\,\ref{fig:Demo_z3}, \ref{fig:Demo_z0p1_and_z0p55}, \ref{fig:Host_Prog_age}, and \ref{fig:Prog_age_event}.
The diagonal panels, running from the upper left to the lower right, show the corresponding one-dimensional parameter density distributions.
All curves are normalized to unity at their respective maxima to facilitate direct comparison of their shapes.
The y-axis range is fixed to [0,\,1.2] in each diagonal panel.
}
\label{fig:corner_all_z0to5}
\end{figure*}

\begin{figure*}
\includegraphics[width=\textwidth]{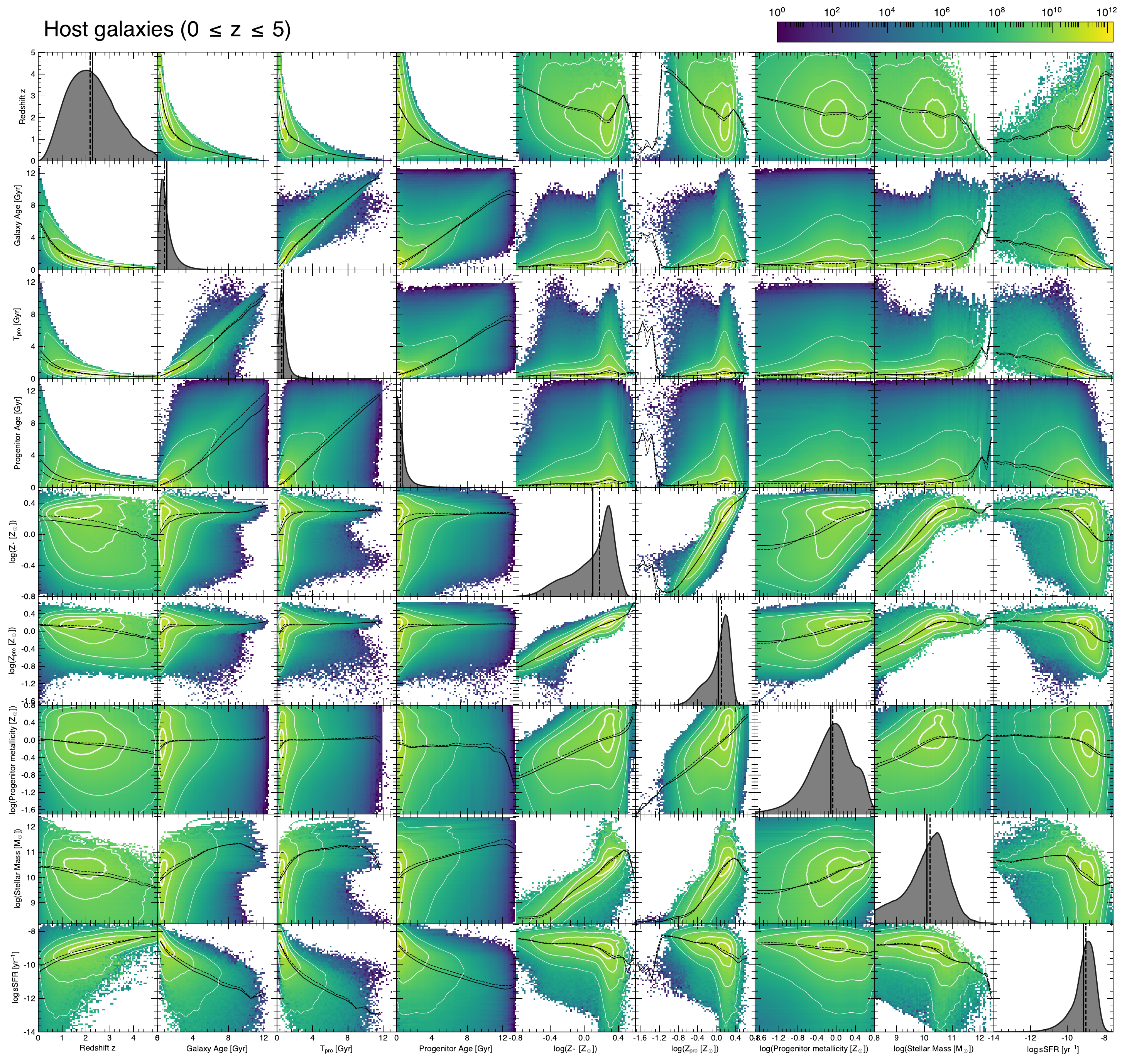}
\caption{
Various sets of pairwise parameter combinations for SN~Ia \emph{host} galaxies in the \emph{full catalogue} ($0 \leq z \leq 5$).
The parameters used are redshift, stellar-particle-mass-weighted mean stellar age, host-level event-weighted mean progenitor age, individual progenitor age, stellar-particle-mass-weighted mean stellar metallicity, host-level event-weighted mean progenitor metallicity, individual progenitor metallicity, total stellar mass, and sSFR of the SN~Ia host galaxies.
Symbols and lines follow the conventions of Figs.\,\ref{fig:Demo_z3}, \ref{fig:Demo_z0p1_and_z0p55}, \ref{fig:Host_Prog_age}, and \ref{fig:Prog_age_event}.
Diagonal panels show the corresponding 1D parameter distributions.
}
\label{fig:corner_host_z0to5}
\end{figure*}

\begin{figure*}
\includegraphics[width=\textwidth]{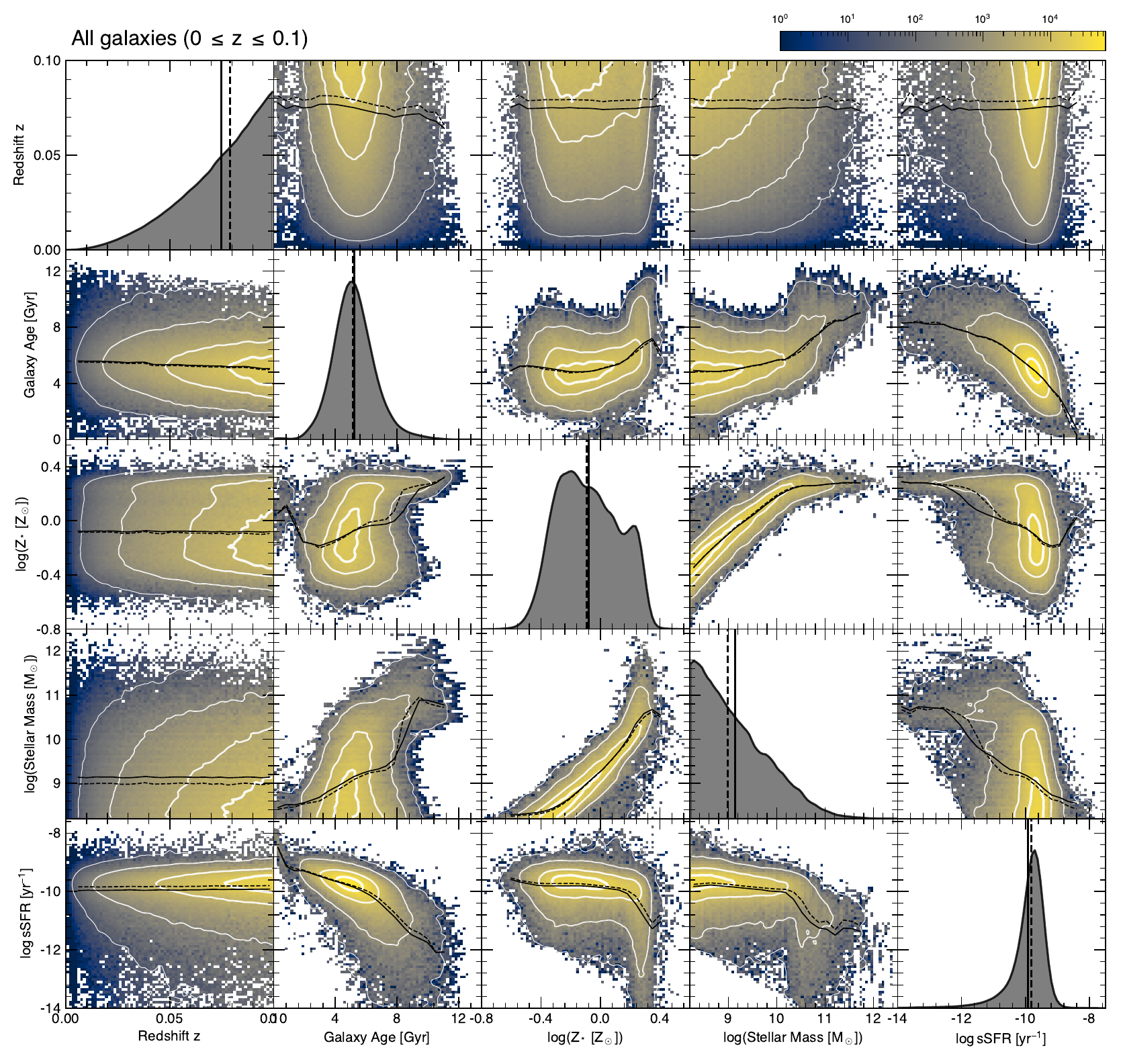}
\caption{The same as Figure\,\ref{fig:corner_all_z0to5} (i.e., for \emph{all} galaxies), but in the \emph{local} ($0 \leq z \leq 0.1$) sample.}
\label{fig:corner_all_z0}
\end{figure*}

\begin{figure*}
\includegraphics[width=\textwidth]{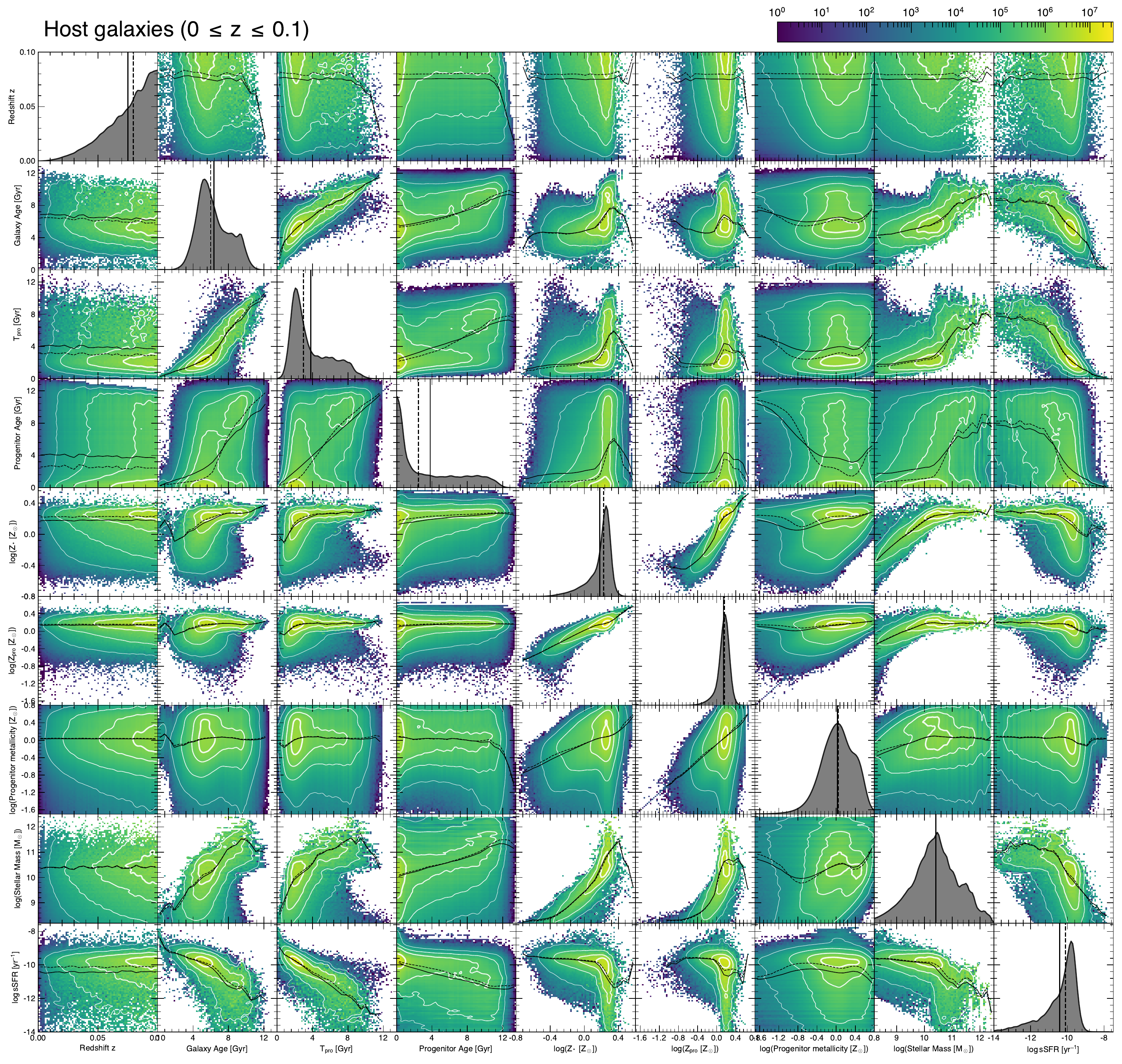}
\caption{The same as Figure\,\ref{fig:corner_host_z0to5} (i.e., for SN~Ia \emph{host} galaxies), but in the \emph{local} ($0 \leq z \leq 0.1$) sample.}
\label{fig:corner_host_z0}
\end{figure*}

This appendix presents a suite of pairwise parameter combinations, contrasting the distributions of all galaxies and SN~Ia host galaxies for both the full simulation volume ($0 \leq z \leq 5$) and local-volume ($0 \leq z \leq 0.1$) samples.
Figures\,\ref{fig:corner_all_z0to5} and \ref{fig:corner_host_z0to5} show the all-galaxy and host-galaxy comparisons, respectively, for the full simulation volume sample, while Figures\,\ref{fig:corner_all_z0} and \ref{fig:corner_host_z0}
present the analogous comparisons for the local-volume sample.
These panels are included to provide a comprehensive view of how SN~Ia host selection manifests across the parameter spaces, beyond the representative combinations highlighted in \S\,\ref{sec:4}.

\section{Observer-frame Demographics of SN\MakeLowercase{e}~I\MakeLowercase{a} Host Galaxies over }
\label{appendix:E}

In the main text, we report all main results in the \emph{source frame} in order to highlight the intrinsic predictions of the framework rather than the observer-frame realization of a survey-selected sample.
For completeness, Appendix\,\ref{appendix:E} presents observer-frame demographics of SN~Ia hosts over $0 \le z \le 5$.
A mock-observed catalogue is constructed by applying the selection interval
$t_0-\Delta t_{\rm lb}(z)-\Delta t/(1+z) \le t_{\rm explosion} \le t_0-\Delta t_{\rm lb}(z)$,
where $\Delta t$ is the mock observational time window and the factor $1/(1+z)$ arises from cosmological time dilation.
This transformation incorporates cosmological time dilation and produces distributions that are more directly comparable to detected-sample demographics reported by observational surveys, although direct comparison with any individual survey catalogue would still require survey-specific modeling of its selection function.
Table~\ref{tab:snpm_catalogue} gives the quantities required to convert the intrinsic source-frame catalogue to its observer-frame counterpart.

\begin{figure*}
\includegraphics[width=\textwidth]{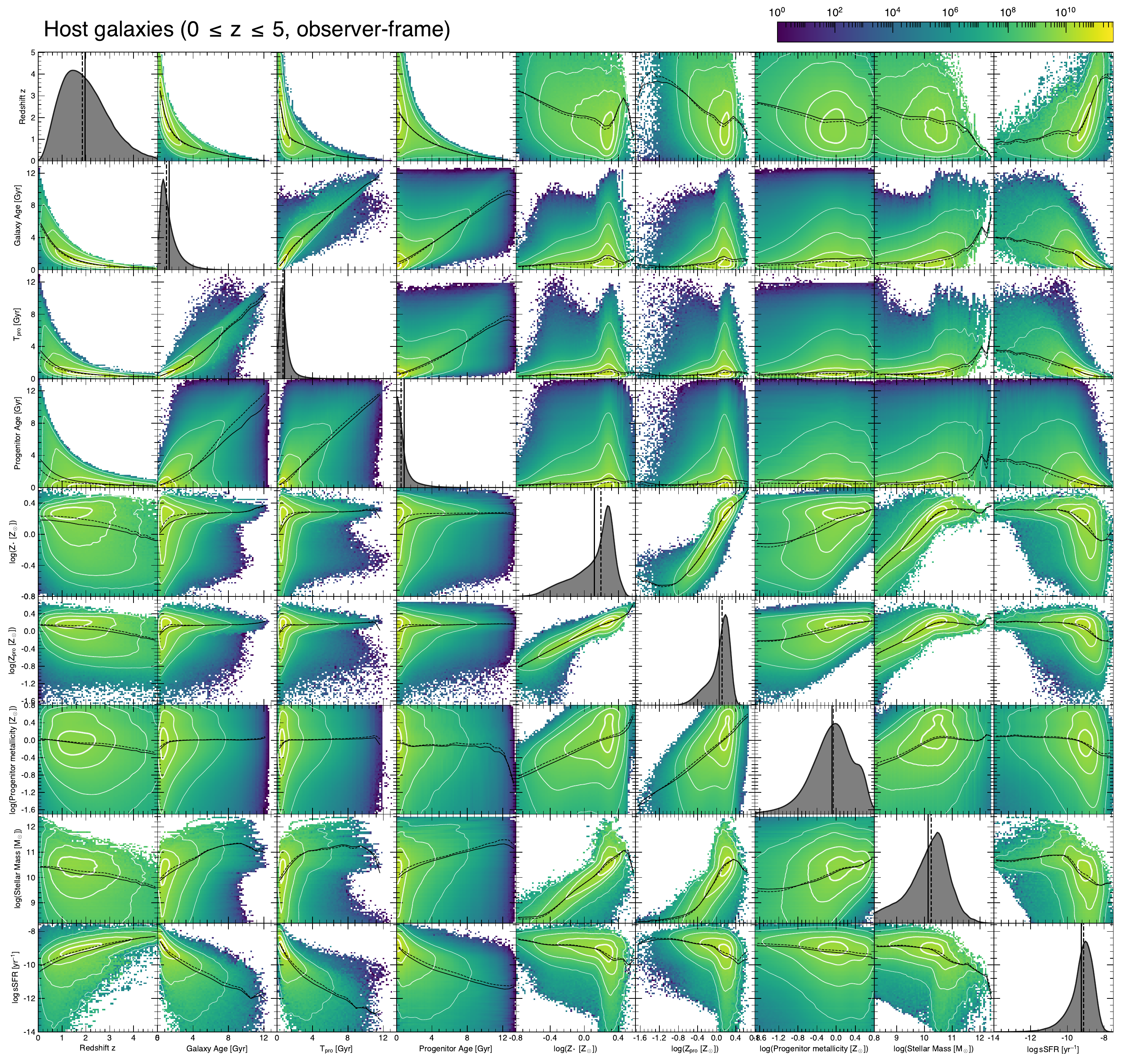}
\caption{The same as Figure\,\ref{fig:corner_host_z0to5} (i.e., host galaxies in the source-frame), but host galaxies for the observer-frame survey-selected sample.}
\label{fig:corner_host_observerframe}
\end{figure*}

\bsp	
\label{lastpage}
\end{document}